\documentclass[review]{elsarticle}
\pdfoutput=1
\usepackage{lineno,hyperref}
\usepackage{multirow}
\usepackage{amsmath}
\usepackage{caption}
\usepackage[utf8]{inputenc}
\usepackage{verbatim}
\usepackage{longtable}
\usepackage{graphicx}
\usepackage[colorinlistoftodos]{todonotes}
\modulolinenumbers[5]

\journal{Applied Radiation and Isotopes}

\begin{document}

\begin{frontmatter}

\title{Optimization of $^{99m}$Tc proton-induced production by means of cross-section measurements of technetium radioisotopes from molybdenum targets}

\author[tera,lhep]{Gaia Dellepiane \corref{mycorrespondingauthor}}
\ead{gaia.dellepiane@unibe.ch}
\author[tera,lhep,pavia]{Sofia D'Uva}
\author[lhep]{Lars Eggimann}
\author[lhep]{Alexander Gottstein}
\author[lhep]{Samuel~Juillerat}
\author[lhep]{Eva Kasanda}
\author[lhep]{Isidre Mateu}
\author[lhep,napoli]{Paola Scampoli}
\author[ukr,ukr1]{Ihor Kadenko}
\author[lhep]{Saverio Braccini}

\address[tera]{Tera-Care Foundation, Chemin du Coq-d'Inde 8a, Cologny, Switzerland}
\address[lhep]{Albert Einstein Center for Fundamental Physics (AEC), Laboratory for High Energy Physics (LHEP), University of Bern, Sidlerstrasse 5, Bern, Switzerland}
\address[pavia]{Deparment of Nuclear and Theoretical Physics, University of Pavia, Pavia, Italy}
\address[napoli]{Department of Physics “Ettore Pancini", University of Napoli Federico II, Complesso Universitario di Monte S. Angelo, Napoli, Italy}
\address[ukr]{Department of Nuclear and High Energy Physics, Faculty of Physics, Taras Shevchenko National University of Kyiv, Kyiv, Ukraine}   
\address[ukr1]{International Nuclear Safety Center of Ukraine of Taras Shevchenko National University of Kyiv, Kyiv, Ukraine}

\cortext[mycorrespondingauthor]{Corresponding author}

\begin{abstract}
$^{99m}$Tc [t$_{1/2}$=6.0072~h, IT:~99.9963\%, E$_{\gamma}$=140.51~keV~(89\%)] is currently the most widely used radionuclide for SPECT imaging. It is mainly obtained from $^{99}$Mo/$^{99m}$Tc generators, where $^{99}$Mo [t$_{1/2}$=65.924~h, $\beta^{-}$:~100\%] is produced through the fission of $^{235}$U targets in nuclear reactors. However, the gradual aging of nuclear reactors dedicated to $^{99}$Mo production, combined with the increasing frequency and duration of maintenance shutdowns, poses a significant risk to the global supply of $^{99}$Mo/$^{99m}$Tc generators and highlights the need for alternative production strategies. Among the several accelerator-based approaches proposed, the direct production of $^{99m}$Tc from enriched $^{100}$Mo targets using relatively low-energy proton beams ($<$18 MeV) is one of the most promising routes. Within a collaboration between the Laboratory for High Energy Physics (LHEP) of the University of Bern, Switzerland, and the Department of Nuclear and High Energy Physics of the Taras Shevchenko National University of Kyiv, Ukraine, we investigated the production of $^{99m}$Tc with medical cyclotrons. The cross sections of the reactions leading to the formation of technetium radioisotopes were measured by irradiating solid molybdenum targets. This data enables an accurate assessment of the Tc impurities produced during irradiation and provide essential information for optimizing production parameters to maximize both production yield and radionuclidic purity. Based on these results, a proton energy of about 16~MeV and a highly enriched $^{100}$Mo target with a thickness of approximately 0.5~mm were identified as suitable conditions for $^{99m}$Tc production. For a 10 hour irradiation, this configuration provides a calculated $^{99m}$Tc yield of about 1.7~GBq/$\mu$A, while maintaining the radionuclidic purity within the limits of the European Pharmacopoeia. Several production tests were performed to validate the cross-section measurements and confirm the feasibility of the investigated production route.
\end{abstract}

\end{frontmatter}

\section{Introduction}
$^{99m}$Tc is the most widely used radionuclide in nuclear medicine, particularly for planar and SPECT (Single Photon Emission Computed Tomography) imaging. Its favorable nuclear properties, including a relatively short half-life (t$_{1/2}$ = 6.0072~h), the emission of a high-abundance $\gamma$-ray at 140.51~keV~(89\%), and negligible particulate radiation, make it an ideal diagnostic radionuclide, allowing high-quality imaging while minimizing the dose delivered to the patient. \\
Currently, $^{99m}$Tc is mainly obtained from $^{99}$Mo/$^{99m}$Tc generators, where $^{99}$Mo (t$_{1/2}$ = 65.924~h) is produced by the fission of $^{235}$U targets in research nuclear reactors. However, this supply chain experienced a major disruption in 2009 due to temporary and permanent shutdowns of the main production facilities, resulting in a worldwide shortage of $^{99m}$Tc. This critical situation highlighted the vulnerability of the current production system and stimulated extensive research into alternative production routes capable of ensuring a reliable and sustainable supply of this essential medical radionuclide. Among the accelerator-based approaches proposed as alternatives, the direct production of $^{99m}$Tc from enriched $^{100}$Mo targets using low-energy cyclotrons (up to 18~MeV) is one of the most promising strategies. \\
The development of reliable accelerator-based production routes for medical radionuclides is part of the research activities promoted by the Tera-Care Foundation~\cite{teracare}, which supports the development of advanced accelerator-based technologies for cancer care in collaboration with CERN and other academic and research institutions. The present study was carried out within a broader research collaboration between the Laboratory for High Energy Physics (LHEP) of the University of Bern, Switzerland, and the Department of Nuclear and High Energy Physics of the Taras Shevchenko National University of Kyiv, Ukraine, aimed at investigating the production of $^{99m}$Tc using low-energy proton beams. Within this framework, the cross section of the $^{100}$Mo(p,2n)$^{99m}$Tc nuclear reaction was accurately measured through the irradiation of solid enriched $^{100}$Mo targets. Although this reaction has been extensively investigated in the literature, the significant discrepancies among the reported cross-section values motivated further experimental studies. In addition, the cross sections of the reactions leading to the production of technetium radioisotopes up to 18 MeV were determined. This data is essential for optimizing irradiation parameters and evaluating the feasibility of producing $^{99m}$Tc with the required activity and radionuclidic purity for clinical applications. \\
In this paper, we report on the cross-section measurements obtained by irradiating natural Mo, enriched $^{97}$Mo, and enriched $^{100}$Mo targets. The main physical characteristics of the investigated Tc radioisotopes are summarized in Table~\ref{tab:Tc_radioisotopes}. The contribution of each nuclear reaction leading to the production of technetium radioisotopes in the energy range up to 18 MeV was determined using a method based on the inversion of a linear system of equations. Some of the cross-section data presented here are reported for the first time. \\
Based on these measurements, the $^{99m}$Tc production yield and radionuclidic purity were evaluated to identify optimal irradiation conditions. Further tests were performed to validate the proposed production strategies.

\begin{table}[h]
\centering
\caption{Decay properties of technetium radioisotopes and $^{99}$Mo~\cite{IAEA}. The values in parentheses are the uncertainties referred to the last digits of the value.}
\resizebox{0.9\textwidth}{!}{%
\begin{tabular}{c  c  c  c  c }
    \hline
  Radioisotope & t$_{1/2}$ & Decay mode:~[$\%$] & E$_{\gamma}$~[keV]& BR~[$\%$] \\
\hline
  $^{93}$Tc & 2.75(5) h & ec + $\beta^{+}$: 100  & 1362.94(7) & 66.2(6)\\
  \noalign{\smallskip}
  \multirow{2}*{$^{93m}$Tc} & \multirow{2}*{43.5(10) min} & IT: 77.4 &  391.83(8) & 58.3(9)\\
          & & ec + $\beta^{+}$: 22.6 & 943.7(5) & 2.92(24)\\ 
  \noalign{\smallskip}
  $^{94}$Tc  &  293(1) min & ec + $\beta^{+}$: 100 & 702.67(7) & 99.6(18)\\
  \noalign{\smallskip}
  $^{94m}$Tc  & 52.0(10) min &  ec + $\beta^{+}$: 100 & 1868.69(8) & 5.7(3)\\
  \noalign{\smallskip}
  $^{95}$Tc  &  20.0(1) h & ec + $\beta^{+}$: 100 & 765.789(9) & 93.8(3)\\
  \noalign{\smallskip}
  \multirow{2}*{$^{95m}$Tc} & \multirow{2}*{61(2) d} & ec + $\beta^{+}$: 96.12 &  204.117(2) & 63.2(8)\\
          & & IT: 3.88 & 38.9(1) & 7.5E-5(6)\\ 
  \noalign{\smallskip}
  $^{96}$Tc  &  4.28(7) d & ec + $\beta^{+}$: 100 & 812.54(4) & 82(3)\\
  \noalign{\smallskip}
  \multirow{2}*{$^{96m}$Tc} & \multirow{2}*{51.5(10) min} & IT: 98.0 &  34.20(5) & 0.0259(4)\\
          & & ec + $\beta^{+}$: 2.0 & 778.22(4) & 1.9(5)\\ 
  \noalign{\smallskip}
  $^{97}$Tc  &  4.21E+6(16) y & ec: 100 & - & -\\          
  \noalign{\smallskip}
  \multirow{2}*{$^{97m}$Tc} & \multirow{2}*{91.0(6) d} & IT: 96.06 &  96.5(1) & 0.321(3)\\
          & & ec + $\beta^{+}$: 3.94 & - & -\\         
  \noalign{\smallskip}
  $^{99}$Tc  & 2.111E+5(12) y & $\beta^{-}$: 100 & 89.5(2) & 0.00065(15)\\  
  \noalign{\smallskip}
  \multirow{2}*{$^{99m}$Tc} & \multirow{2}*{6.0072(9) h} & IT: 99.9963 &  140.511(1) & 89(4)\\ 
          & & $\beta^{-}$: 0.0037 & 89.6(3) &  0.00104(19)\\ 
  \noalign{\smallskip}
  $^{101}$Tc  & 14.2(1) min & $\beta^{-}$: 100 & 306.83(3) & 89(4)\\  
  \noalign{\smallskip}
  $^{99}$Mo  & 65.924(6) h & $\beta^{-}$: 100 & 739.500(17) & 12.2(2)\\            
  \hline
\end{tabular}
}
\label{tab:Tc_radioisotopes}
\end{table}

\section{Material and methods}
The Bern medical cyclotron laboratory is located at the Bern University Hospital (Inselspital)~\cite{Bern_cyclotron} and hosts an IBA Cyclone 18/18 high-current cyclotron, providing proton beams of 18~MeV, beam currents ranging from a few pA to 150~$\mu$A~\cite{Auger2015}, and eight extraction ports. The main feature of the facility is the presence of a second bunker with independent access, enabling multidisciplinary research activities to be performed in parallel with routine radiopharmaceutical production. The beam is transported to the second bunker through a 6-m-long external Beam Transfer Line (BTL), which was used for all measurements presented in this paper. The BTL is equipped with beam focusing and diagnostic systems for adjusting and monitoring the beam position and profile, including the nondestructive two-dimensional beam profiler UniBEaM, developed by LHEP and commercialized by D-Pace~\cite{unibeam,DPACE_1}. The beam energy extracted from the BTL was characterized using several independent measurements performed under differing cyclotron operating conditions. The nominal beam energy was determined as the weighted average of the experimental measurements and was found to be 18.24(25)~MeV~\cite{Nesteruk2018, Nesteruk2019, CAMPAJOLA2019, Haffner2019, Gottstein2026}.\\
The produced radionuclide activities were measured using an N-type High Purity Germanium (HPGe) detector (Canberra GR2019), coupled to a preamplifier and to a Lynx digital signal analyzer. Gamma-ray spectra were acquired using Genie2k~\cite{Genie2k} software for individual measurements and the Excel2Genie Microsoft Excel application~\cite{Excel2Genie} for repeated measurements over time. Spectral analysis was performed with the InterSpec~\cite{interspec} software package, developed by the Sandia National Laboratories. The HPGe detector efficiency calibration was performed according to international standards~\cite{HPGe_cal} using a multi-peak $\gamma$-ray reference source containing $^{57}$Co, $^{60}$Co, $^{85}$Sr, $^{88}$Y, $^{109}$Cd, $^{113}$Sn, $^{137}$Cs, and $^{241}$Am. The calibration source had the same geometry as the irradiated targets and was measured at distances up to 10~cm from the detector using a custom-designed plexiglass ladder with equally spaced positions at 1~cm intervals. At each level, the measured efficiency values as a function of the $\gamma$-ray energy were fitted using a parametric least-squares approach. The calibration procedure was validated through metrological studies~\cite{DURAN2022110507}, resulting in efficiency uncertainties below 3\%~\cite{Juget2023}.\\
To determine the cross sections of the reactions leading to the production of Tc radioisotopes in the energy range of interest, natural and enriched molybdenum were used as target materials. Natural Mo, and 89.86\% enriched $^{97}$Mo were supplied by Goodfellow and Isoflex, respectively, while 92.43\% enriched $^{100}$Mo was provided by the University of Kyiv. For the production tests, 99.73\% enriched $^{100}$Mo was purchased from Isoflex. All materials were independently characterized by Inductively Coupled Plasma Mass Spectrometry (ICP-MS) at the Department of Chemistry, Biochemistry and Pharmacy (DCBP) of the University of Bern. The resulting isotopic compositions are reported in Table~\ref{tab:Mo_mat}.

\begin{table}[h]
\centering
\caption{Isotopic abundances measured by ICP-MS for natural Mo purchased from Goodfellow~\cite{GoodFellow}, enriched $^{97}$Mo and enriched $^{100}$Mo (a) supplied by Isoflex~\cite{isoflex} and enriched $^{100}$Mo (b) provided by the University of Kyiv. The values in parentheses are the uncertainties referred to the last digits of the value.}
\resizebox{\textwidth}{!}{
\begin{tabular}{ l | c  c  c  c  c  c  c}
   & $^{92}$Mo~[\%] & $^{94}$Mo~[\%] & $^{95}$Mo~[\%] & $^{96}$Mo~[\%] & $^{97}$Mo~[\%] & $^{98}$Mo~[\%] & $^{100}$Mo~[\%] \\
  \hline
  nat.Mo & 12.10(4) & 8.35(3) & 15.09(9) & 16.60(4) & 10.01(3) & 26.47(9) & 11.38(6) \\
  \hline
  enr.$^{97}$Mo & 0.49 & 0.39 & 0.90 & 2.44 & 89.65(4) & 5.46 & 0.67 \\
  \hline
  enr.$^{100}$Mo (a) & 0.01 & 0.01 & 0.01 & 0.01 & 0.03 & 0.19 & 99.73(3) \\
  \hline
  enr.$^{100}$Mo (b) & 0.75 & 0.54 & 1.08 & 1.25 & 0.85 & 3.10 & 92.43(6)\\
\end{tabular}
}
\label{tab:Mo_mat}
\end{table}

\subsection{Cross-section measurements}
Natural molybdenum was provided in the form of a 25~$\mu$m thick foil. Square samples of approximately 2$\times$2~mm$^{2}$ were cut and placed into a cylindrical pocket (4.2 mm diameter, 0.8 mm depth) at the center of an aluminum disc (22.8~mm diameter, 2~mm thickness). The mass of each foil was measured using an analytical balance (Shimadzu AUW120D), with a readability of 0.01~mg and a repeatability of 0.02~mg. The pocket was sealed with a 13-$\mu$m-thick aluminum foil, fixed onto the disc with adhesive tape to prevent material loss before and during irradiation. \\
Enriched $^{97}$Mo and enriched $^{100}$Mo~(b) were available in the form of metallic powder. A few milligrams of material were suspended in absolute ethanol and deposited into the pocket. After complete evaporation of the ethanol, the target mass was measured using the balance, and the pocket was sealed following the same procedure adopted for the natural case. This procedure resulted in target thicknesses below 15~$\mu$m, allowing the proton energy loss in the irradiated material to be considered negligible within the uncertainty. \\
Each target was irradiated following the same experimental procedure used in our previous studies on cross-section measurements~\cite{Dellepiane2021b}. This method is based on the irradiation of the full mass of a thin target by a proton beam with uniform surface distribution and has the advantage that the target does not have to be necessarily homogeneous in thickness, provided that the proton energy can be considered constant within its mass. \\
The beam was defocused by the optical elements of the BTL to achieve a flat distribution at the target position and was monitored using the UniBEaM detector. The beam current delivered to the target was measured using a custom-designed target station connected to an electrometer (B2985A Keysight). The station provides a beam of controlled diameter thanks to a 6-mm-diameter collimator and is equipped with an electron suppressor ring biased at negative voltage to minimize the contribution of secondary electrons generated during irradiation.\\
To perform irradiations below 18~MeV, the beam energy was degraded using aluminum attenuator discs placed upstream the target. The resulting beam energy was determined by means of Monte Carlo simulations performed with the FLUKA code~\cite{FLUKA}, using an imported geometry of the irradiation setup and the measured beam characteristics~\cite{eggimann2026}.\\  
Following irradiation, each target was measured with the HPGe detector, and the cross sections of the produced Tc radioisotopes as a function of the proton energy were determined according to:
\begin{equation}
    \sigma(E) = A(t_{EOB}) \cdot \dfrac{q}{dI/dS} \cdot \dfrac{1}{m}\cdot \dfrac{m_{mol}}{N_A \epsilon \eta} \cdot f_C
    \label{eq:cs}
\end{equation}
where $A(t_{EOB})$ is the activity at the End of Beam (EOB), $q$ is the elementary charge, $dI/dS$ is the current on target per unit surface, $m$ is the target material mass, $m_{mol}$ is the molar mass of the material, $N_A$ is the Avogadro number, $\epsilon$ is the isotopic abundance of the target nucleus and $\eta$ its stoichiometric number. \\
The correction factor 
\begin{equation}
    f_C = \dfrac{\int^{t_{EOB}}_{t_{SOB}} I(t') dt'}{e^{-\lambda t_{i}} \int^{t_{EOB}}_{t_{SOB}} e^{\lambda t'}I(t') dt'}
    \label{eq:corr_factor}
\end{equation}
accounts for the decay of the radionuclides (with decay constant $\lambda$) during the irradiation time $t_i = t_{EOB} - t_{SOB}$, where $t_{SOB}$ is the time at the Start of Beam (SOB) and $I(t')$ is the current on target at time $t'$.\\
For irradiations performed with a constant beam current, this correction factor reduces to
\begin{equation}
    f_C = \dfrac{\lambda t_i}{1-e^{-\lambda t_i}}
    \label{eq:easy_corr_factor}
\end{equation}
The general expression of $f_{C}$ is required for short-lived radionuclides, such as $^{101}$Tc, for which fluctuations in beam current during irradiation can significantly affect the activity produced. \\
For Tc radioisotopes whose production involves multiple reaction pathways, the contributions of the individual nuclear reactions were separated using an inversion method based on a linear system of equations. This approach, described in our previous publications~\cite{BRACCINI2022, DELLEPIANE202311}, requires measurements performed with different target materials having distinct isotopic compositions, with the number of independent materials matching the number of reaction channels contributing to the production of the radionuclide. Considering two materials, A and B, with different known isotopic compositions, and two reactions (p,x) and (p,y), involving the $i$-th and $j$-th target isotope, respectively, the following linear system holds for a given proton energy:
\begin{equation}
\begin{cases}
\sigma_{TOT}^{A} = \beta_{i}^{A} \cdot \sigma(p,x) + \beta_{j}^{A} \cdot \sigma(p,y)\\
\sigma_{TOT}^{B} = \beta_{i}^{B} \cdot \sigma(p,x) + \beta_{j}^{B} \cdot \sigma(p,y)
\label{eq:lin_syst}
\end{cases}
\end{equation}
where the experimentally measured production cross sections appear on the left-hand side, while the individual reaction cross sections are the unknown quantities on the right-hand side. The coefficients $\beta_{i,j}^{A,B}$ represent the isotopic abundances of the target isotope in the corresponding material. \\

\subsection{Study of $^{99m}$Tc production yield and radionuclidic purity}
To identify the optimal irradiation conditions for the cyclotron production of $^{99m}$Tc, the Thick Target Yield (TTY) and the resulting radionuclidic purity were evaluated based on the measured cross sections. The TTY represents the activity produced per unit current and irradiation time (GBq/$\mu$Ah) when the incident protons are completely stopped in the target. As a function of the incident proton energy $E$, it is given by
\begin{equation}
    TTY(E) = f_{C} \cdot \dfrac{N_{A}\cdot \eta }{m_{mol}\cdot q} \int_{E_{th}}^{E} {\dfrac{\sigma(E')}{S_{p}(E')} dE'}
    \label{eq:tty}
\end{equation}
where $f_C$ is the correction factor defined in Eq.~\ref{eq:corr_factor}, $\sigma(E')$ is the cross section as a function of the proton energy $E'$, $S_{p}(E')$ is the mass stopping power of the target material, and $E_{th}$ is the threshold energy of the considered reaction. The mass stopping power was calculated using the SRIM-2013 Monte Carlo code~\cite{SRIM}.\\
For thin targets, where the incident protons are only partially slowed down within the target material, the production yield $Y(E)$ can be expressed as 
\begin{equation}
    Y(E) = TTY(E) - TTY(E_{out})
    \label{eq:prod_yield}
\end{equation}
where $E_{out}$ is the proton energy after traversing the target, determined using FLUKA simulations.\\
The theoretical production yield derived from the measured cross sections was compared with the experimental yield obtained from the production tests carried out by irradiating enriched $^{100}$Mo. The experimental yield was calculated from the activity measured at the EOB, $A(t_{EOB})$, according to 
\begin{equation}
    Y_{exp}(E) = \dfrac{A(t_{EOB})}{Q}
    \label{eq:y_exp}
\end{equation}
where $Q$ is the total charge delivered to the target during the irradiation, obtained by integrating the beam current over the irradiation time\\
\begin{equation}
    Q = \int_{t_{SOB}}^{t_{EOB}} I(t')\,dt'
\end{equation}
The radionuclidic purity was evaluated by considering the activities of all technetium radioisotopes produced during irradiation. For a sample containing N radioisotopes, the radionuclidic purity of the radionuclide of interest, X, is defined as
\begin{equation}
P_{X} = \dfrac{A_{X}}{\sum_{i}^{N}A_{i}}  
\label{eq:purity}
\end{equation}
where $A_{i}$ is the activity of the $i$-th radionuclide.

\subsection{Production tests}
For the production tests, two targets were prepared by compressing approximately 75~mg of enriched $^{100}$Mo (a) powder and 82~mg of enriched $^{100}$Mo (b) powder under an axial load of about 4$\cdot$10$^{4}$~N. The pellets had a diameter of 0.62(2)~cm. Their densities were determined from the measured mass and thickness and were found to be 5.2(3)~g/cm$^{3}$ and 5.4(3)~g/cm$^{3}$, respectively. These values were used in all subsequent calculations and simulations.\\
Each pellet was placed in a dedicated target capsule, referred to as the coin, developed by our group and previously employed for the production of several radionuclides~\cite{Dellepiane2021b}. 
The coin containing the enriched $^{100}$Mo pellet was mounted in a dedicated target holder and installed in the irradiation station used for cross-section measurements. In this configuration, an 8-mm-diameter collimator was employed. Since the station was not equipped with an active cooling system, the maximum beam current was limited to avoid excessive target heating. 

\section{Data analysis and results}
\subsection{Cross-section measurements}
The natural Mo, enriched $^{97}$Mo, and enriched $^{100}$Mo targets were irradiated for average durations of 14, 18, and 15~min, respectively, at mean beam currents of 10, 11, and 12~nA. After the EOB, the produced activities were measured with the HPGe detector. In all measurements, the count frequency was sufficiently low to limit pile-up effects, with an average dead time below 3\%.\\
The main contribution to the experimental uncertainty in the measured cross sections  depended on the radionuclide considered. For $^{93g}$Tc and $^{93m}$Tc, the dominant contribution was the statistical uncertainty associated with the $\gamma$-ray peak fitting ($\sim$9\% and 6\%, respectively), resulting from the relatively low activities produced. A similarly relevant contribution was observed for $^{94m}$Tc, for which the main contributions were associated with the $\gamma$-ray peak fitting (7\%) and the uncertainty on the branching ratio (5\%). In the case of $^{101}$Tc, the statistical component was even larger, reaching up to 13\%, and representing the leading source of uncertainty. To reduce the impact of low-statistics measurements, the same targets were measured repeatedly at different times after EOB, and the final cross-section values were obtained by fitting the activity as a function of time, as described in detail later in this section. For $^{97m}$Tc, only a single long-duration measurement was performed, and the statistical uncertainty was the largest contribution, reaching up to 17\%. For $^{96g}$Tc, the main contribution to uncertainty arose from the activity determined through the Bateman fit (about 20\%).\\
Additional systematic sources of uncertainty affecting all measurements include the flatness of the beam (5\%), the beam current integration (1\%), the HPGe detector efficiency (2.5-4.5\%, depending on $\gamma$-ray energy), and the target mass determination (below 5\%).\\
The individual uncertainty contributions were considered independent and propagated according to Gaussian error propagation to obtain the total experimental uncertainty for each measured cross section.\\
The uncertainties of the nuclear cross sections obtained by solving the linear system of equations were estimated by propagating the upper and lower bounds of the measured production cross sections through the solution of the linear system. \\
The measured cross sections are discussed below for each technetium radionuclide and $^{99}$Mo, together with comparisons with the TENDL nuclear data library and previously published experimental results. Numerical values of the measured cross sections are provided in the Appendix. 

\begin{figure}
    \centering
    \includegraphics[width=1\textwidth]{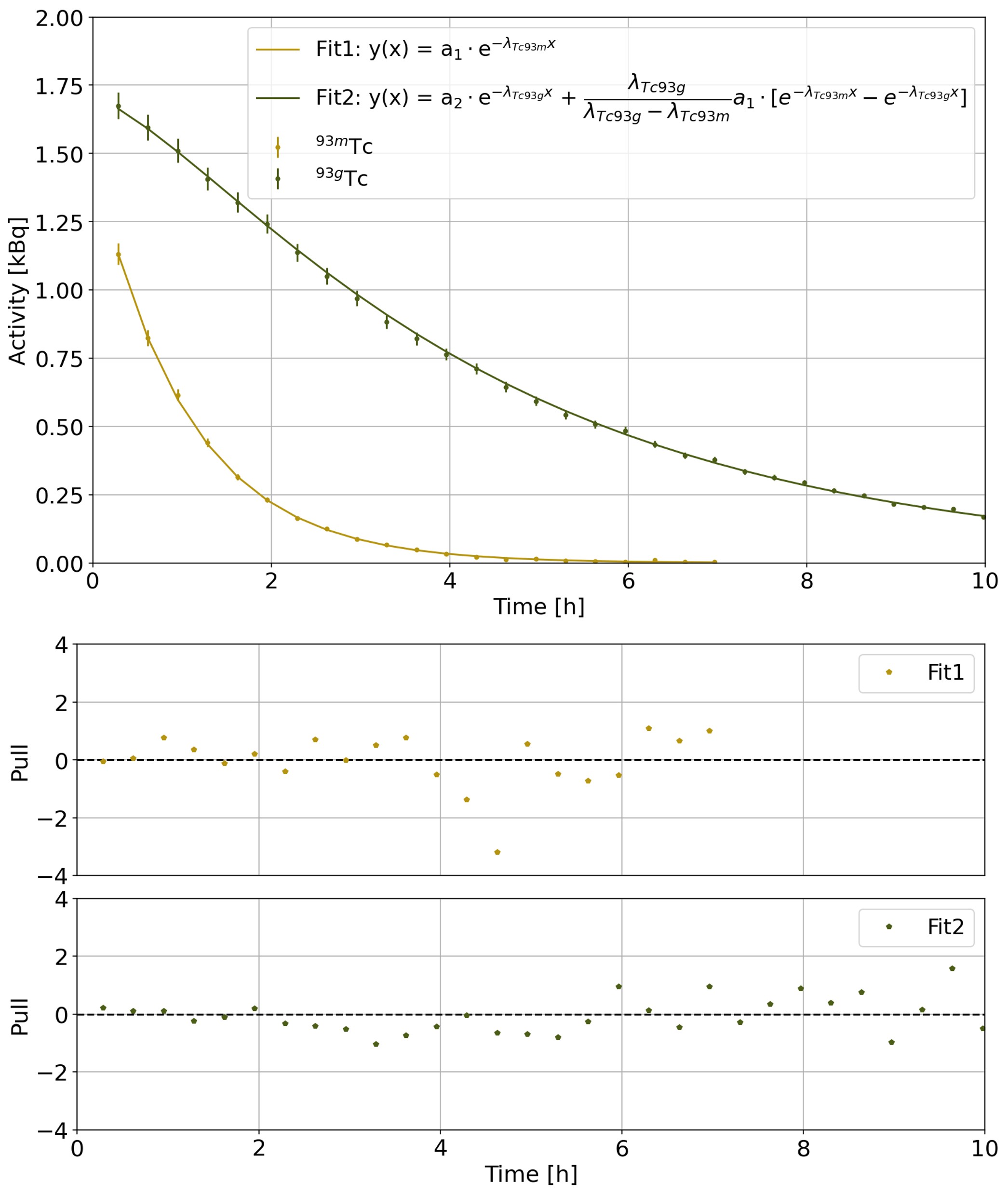}
    \caption{Activity of $^{93g}$Tc and $^{93m}$Tc measured over time and fit curves with the corresponding pulls.}
    \label{fig:fits_Tc93}
\end{figure}

\paragraph{Technetium-93}
Within the investigated energy range, the production of $^{93}$Tc proceeds through two reactions: $^{92}$Mo(p,$\gamma$)$^{93}$Tc and $^{94}$Mo(p,2n)$^{93}$Tc. Irradiation of Mo targets populates both the ground state ($^{93g}$Tc, t$_{1/2}$ = 2.75~h) and the metastable state ($^{93m}$Tc, t$_{1/2}$ = 43.5~min). Since $^{93m}$Tc decays to $^{93g}$Tc through isomeric transition with a branching ratio of 77.4\%, its contribution must be taken into account when determining the activity of the ground state. The activities of both states were therefore monitored over time through repeated measurements. The EOB activity of $^{93m}$Tc was determined by fitting the measured activity values with the exponential decay function
\begin{equation}
    A(t) = A(t_{EOB})\cdot e^{-\lambda t}
\label{eq:exponential_fit}
\end{equation}
where $A(t_{EOB})$ was treated as the only free parameter, while the decay constant $\lambda$ was fixed to the literature value. \\
The resulting $A(t_{EOB})$ value for $^{93m}$Tc was then used as a fixed input parameter in the Bateman equation describing the activity of $^{93g}$Tc:
\begin{equation}
    A_{D}(t) = A_{D}(t_{EOB})\cdot e^{-\lambda_{D}t} + \dfrac{\lambda_{D}}{\lambda_{D} - \lambda_{P}} A_{P}(t_{EOB}) (e^{-\lambda_{P} t} -  e^{-\lambda_{D} t})
\end{equation}
where $A_{D}$ and $A_{P}$ denote the activities of the daughter ($^{93g}$Tc) and parent ($^{93m}$Tc) radionuclides, respectively.\\
The experimental activity data together with the corresponding fits are shown in Figure~\ref{fig:fits_Tc93}. The extracted $A(t_{EOB})$ values were subsequently used in Eq.~\ref{eq:cs} to determine the production cross sections. 

\begin{figure}
    \centering
    \includegraphics[width=0.7\textwidth]{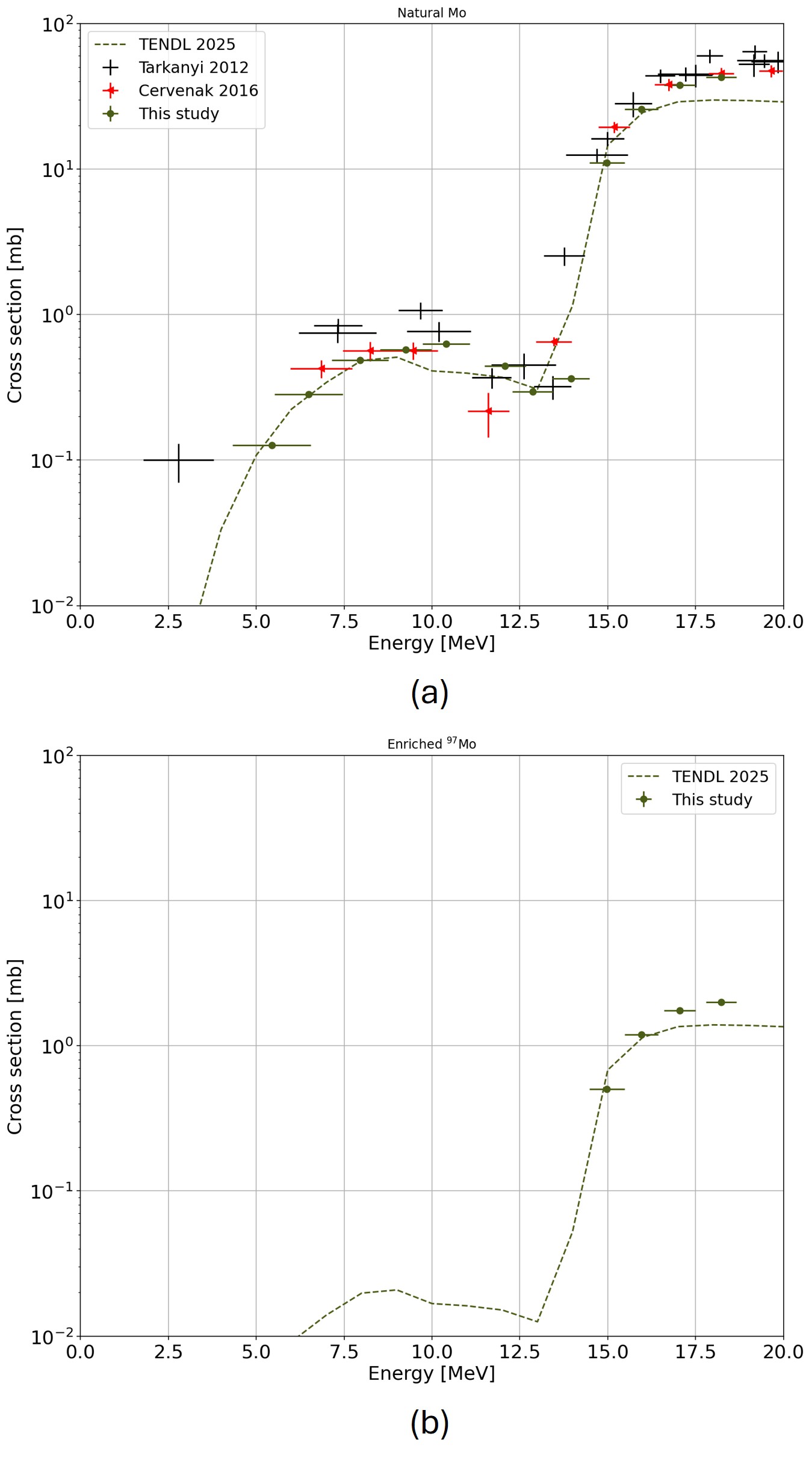}
    \caption{$^{93g}$Tc production cross sections from natural Mo (a) and enriched $^{97}$Mo (b), whose isotopic composition is reported in Table~\ref{tab:Mo_mat}. The dots are the experimental data, the dashed lines are the TENDL-2025 calculations.}
    \label{fig:cs_prod_Tc93}
\end{figure}

\begin{figure}
    \centering
    \includegraphics[width=0.7\textwidth]{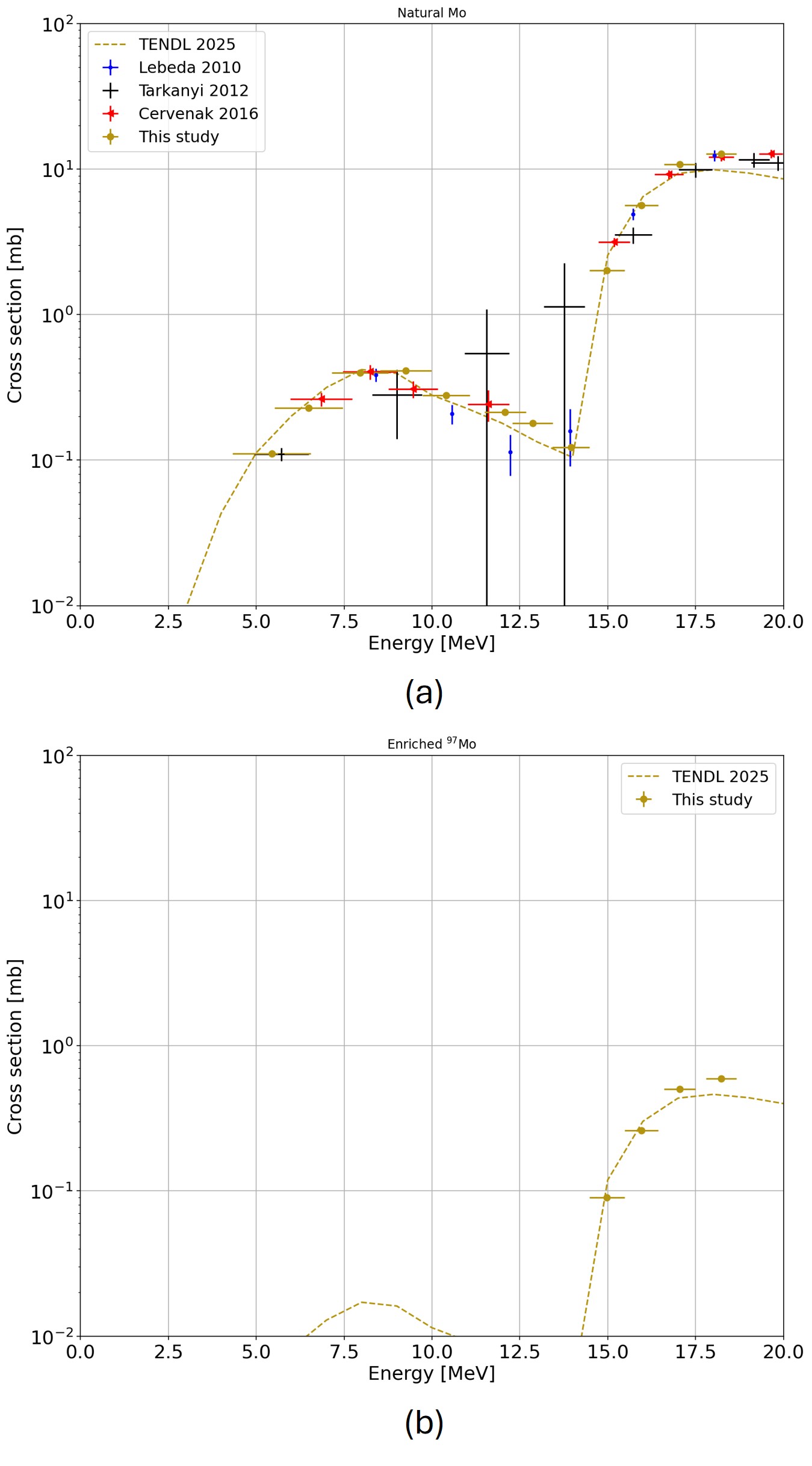}
    \caption{$^{93m}$Tc production cross sections from natural Mo (a) and enriched $^{97}$Mo (b), whose isotopic composition is reported in Table~\ref{tab:Mo_mat}. The dots are the experimental data, the dashed lines are the TENDL-2025 calculations.}
    \label{fig:cs_prod_Tc93m}
\end{figure}

\begin{figure}
    \centering
    \includegraphics[width=0.7\textwidth]{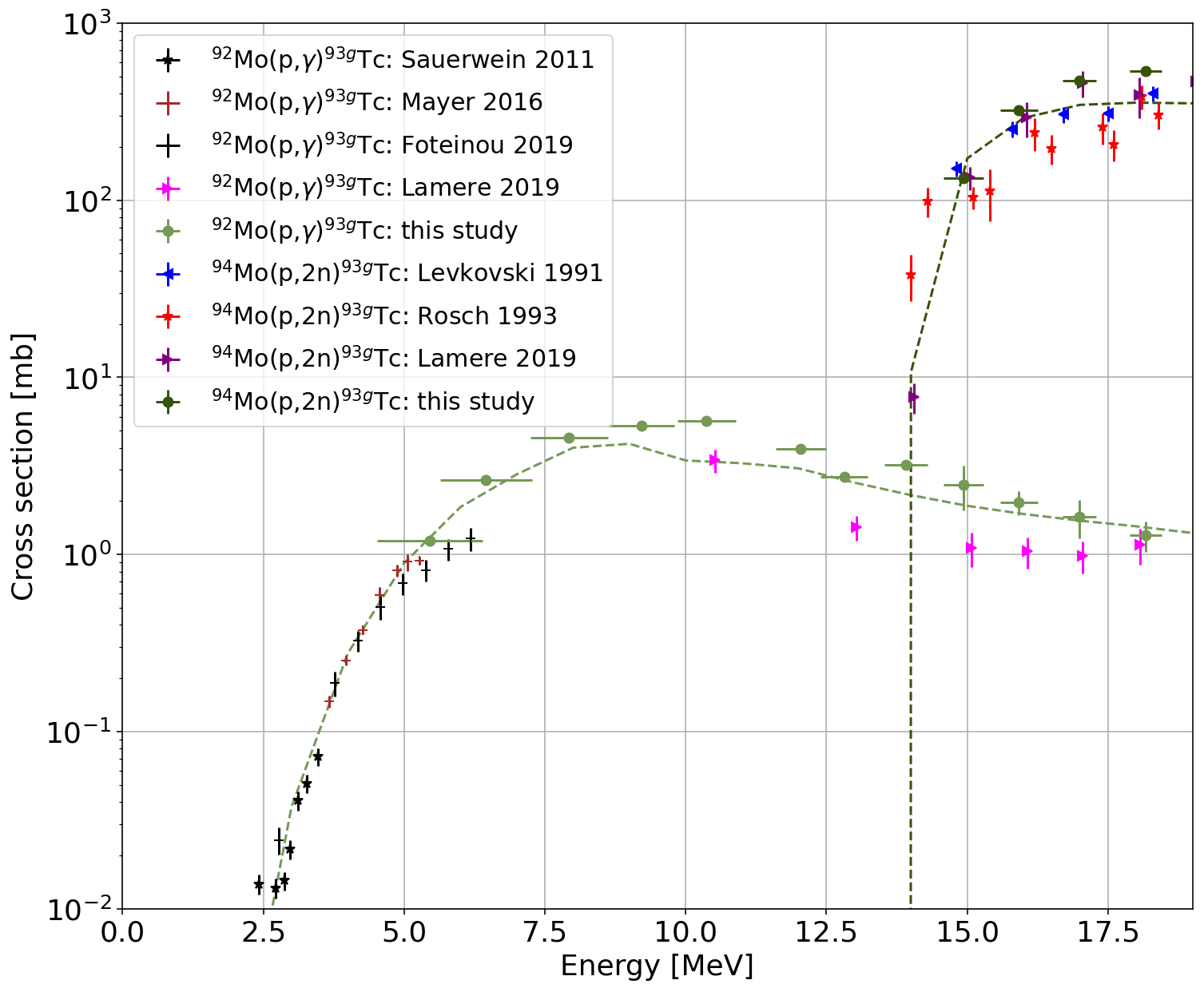}
    \caption{$^{92}$Mo(p,$\gamma$)$^{93g}$Tc and $^{94}$Mo(p,2n)$^{93g}$Tc nuclear cross sections. The dots are the experimental data, the dashed lines are the TENDL-2025 calculations.}
    \label{fig:cs_nucl_Tc93}
\end{figure}

\begin{figure}
    \centering
    \includegraphics[width=0.7\textwidth]{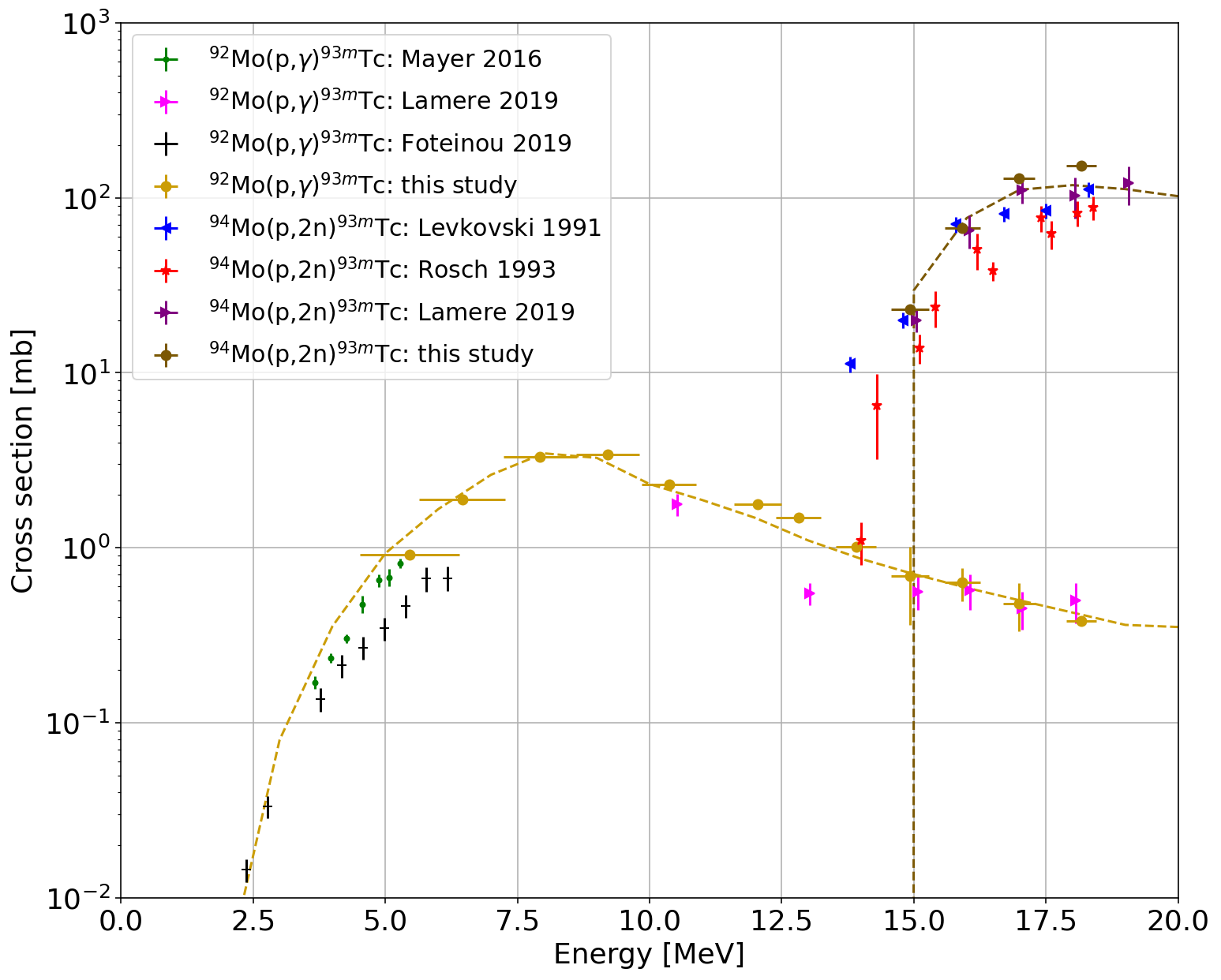}
    \caption{$^{92}$Mo(p,$\gamma$)$^{93m}$Tc and $^{94}$Mo(p,2n)$^{93m}$Tc nuclear cross sections. The dots are the experimental data, the dashed lines are the TENDL-2025 calculations.}
    \label{fig:cs_nucl_Tc93m}
\end{figure}

\noindent Production cross sections were measured using natural Mo and enriched $^{97}$Mo targets. The corresponding results for $^{93}$Tc and $^{93m}$Tc are presented in Figures~\ref{fig:cs_prod_Tc93} and~\ref{fig:cs_prod_Tc93m}, respectively, together with TENDL predictions and the experimental data available in the literature~\cite{TARKANYI2012, CERVENAK2016}. \\
The individual reaction cross sections, extracted using the linear system method, are shown in Figures~\ref{fig:cs_nucl_Tc93} and \ref{fig:cs_nucl_Tc93m} for $^{93g}$Tc and $^{93m}$Tc, respectively, together with TENDL predictions and the available experimental data from the literature~\cite{Sauerwein2011, Mayer2016, Foteinou2019, Lamere2019, Levkowskij1991, Rosch1993}. Overall, a good agreement is observed. 


\begin{figure}
    \centering
    \includegraphics[width=0.7\textwidth]{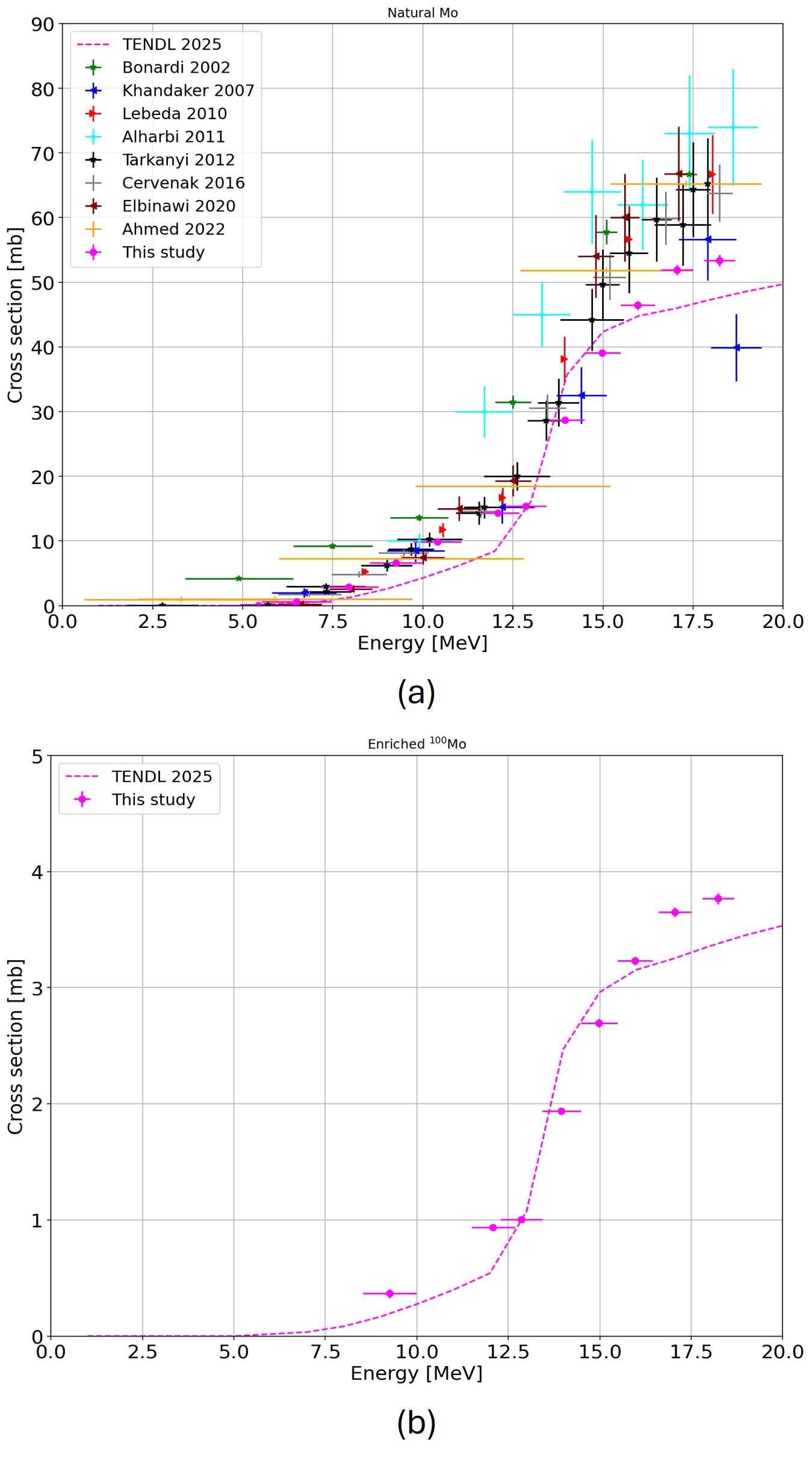}
    \caption{$^{94g}$Tc production cross sections from natural Mo (a) and enriched $^{100}$Mo (b), whose isotopic composition is reported in Table~\ref{tab:Mo_mat}. The dots are the experimental data, the dashed lines are the TENDL-2025 calculations.}
    \label{fig:cs_prod_Tc94}
\end{figure}

\begin{figure}
    \centering
    \includegraphics[width=0.7\textwidth]{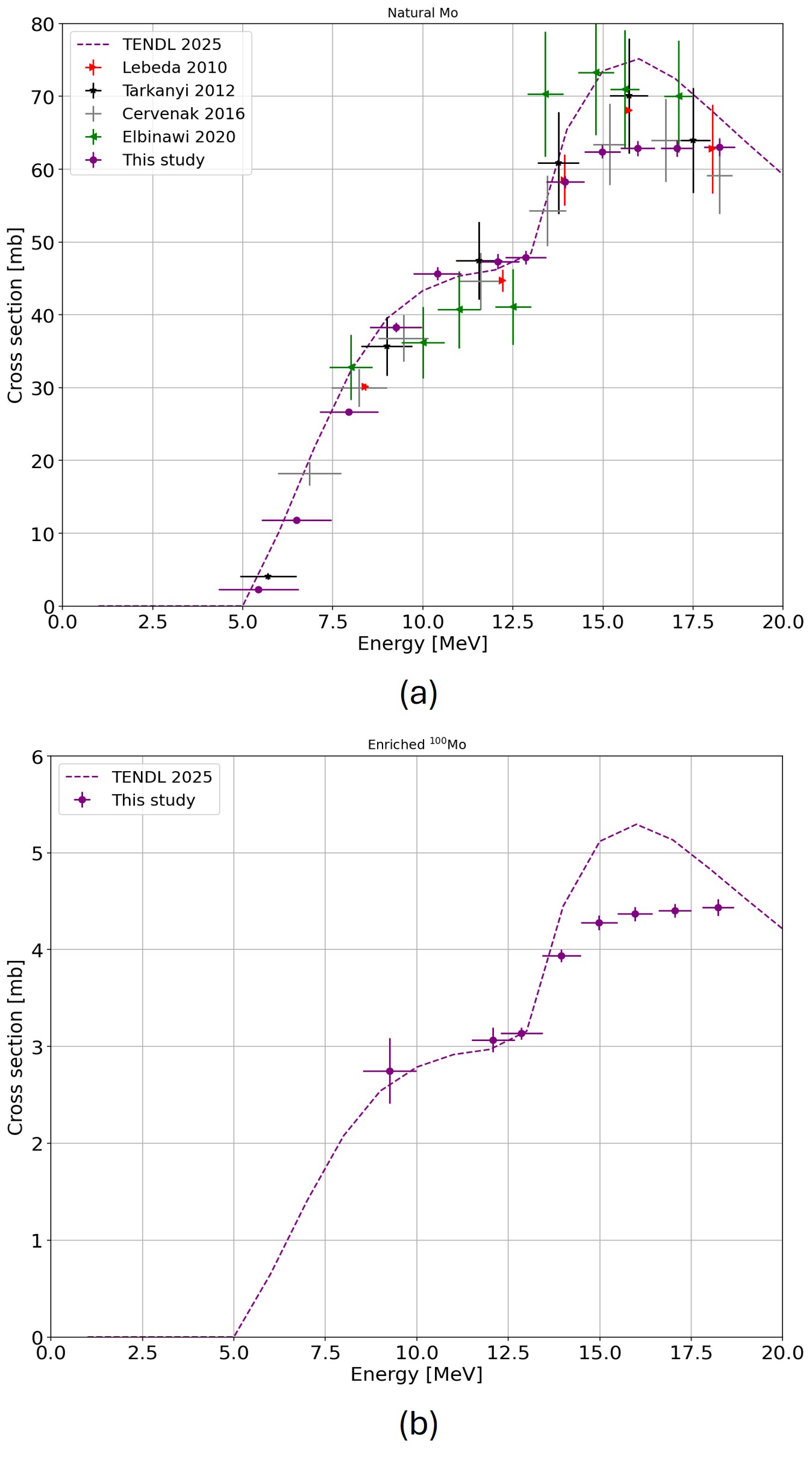}
    \caption{$^{94m}$Tc production cross sections from natural Mo (a) and enriched $^{100}$Mo (b), whose isotopic composition is reported in Table~\ref{tab:Mo_mat}. The dots are the experimental data, the dashed lines are the TENDL-2025 calculations.}
    \label{fig:cs_prod_Tc94m}
\end{figure}

\begin{figure}
    \centering
    \includegraphics[width=0.7\textwidth]{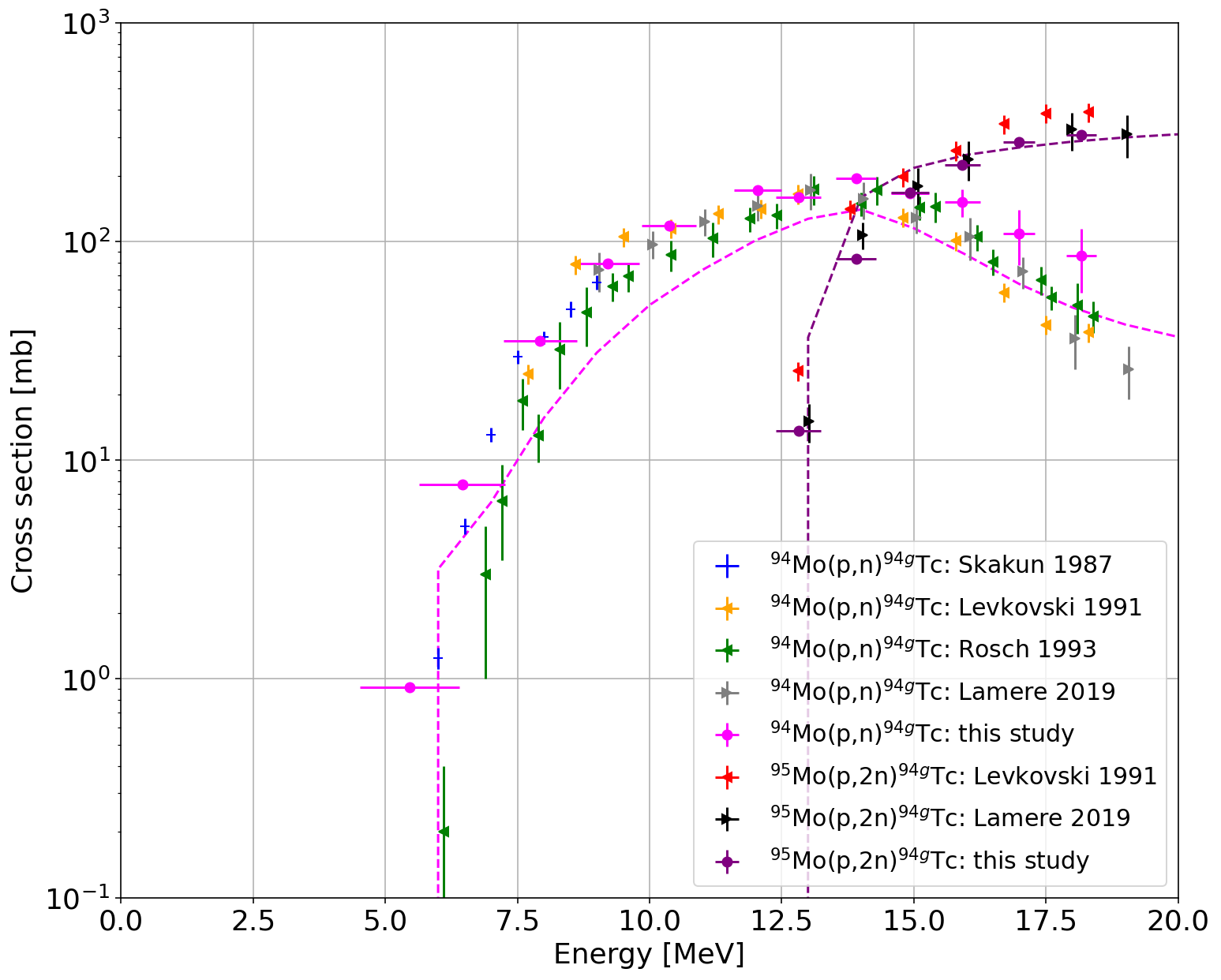}
    \caption{$^{94}$Mo(p,n)$^{94g}$Tc and $^{95}$Mo(p,2n)$^{94g}$Tc nuclear cross sections. The dots are the experimental data, the dashed lines are the TENDL-2025 calculations.}
    \label{fig:cs_nucl_Tc94}
\end{figure}

\begin{figure}
    \centering
    \includegraphics[width=0.7\textwidth]{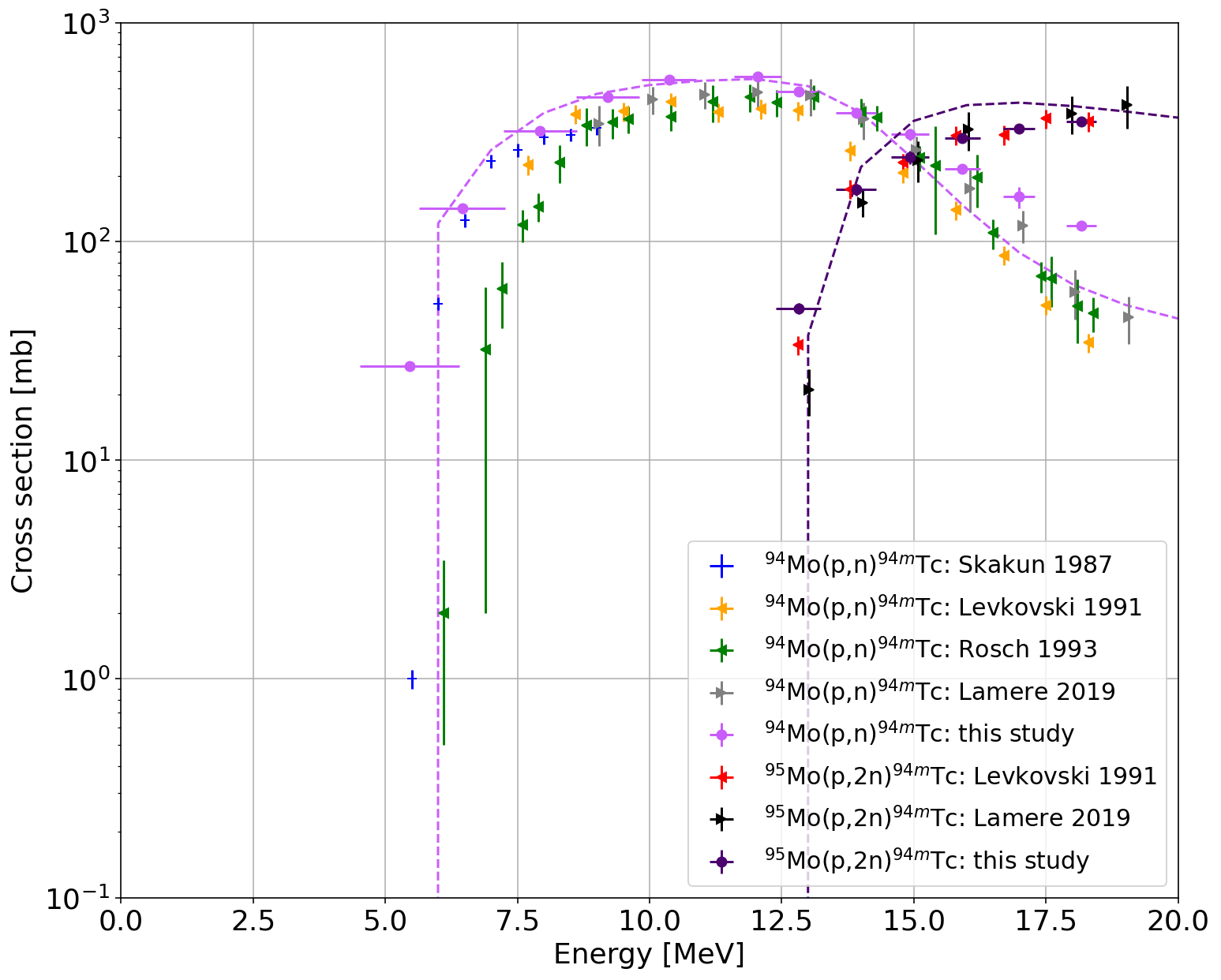}
    \caption{$^{94}$Mo(p,n)$^{94m}$Tc and $^{95}$Mo(p,2n)$^{94m}$Tc nuclear cross sections. The dots are the experimental data, the dashed lines are the TENDL-2025 calculations.}
    \label{fig:cs_nucl_Tc94m}
\end{figure}

\paragraph{Technetium-94}
    In the energy range of interest, $^{94}$Tc is produced via the $^{94}$Mo(p,n) and $^{95}$Mo(p,2n) reactions, populating the ground state $^{94g}$Tc (t$_{1/2}$ = 293 min) and the metastable state $^{94m}$Tc (t$_{1/2}$ = 52 min). The metastable state decays exclusively via EC + $\beta^{+}$ emission to stable $^{94}$Mo and therefore does not contribute to the activity of the ground state. Consequently, the activities of the two states were treated independently. Production cross sections were measured using natural Mo and enriched $^{100}$Mo targets and are presented in Figures~\ref{fig:cs_prod_Tc94} and~\ref{fig:cs_prod_Tc94m} for $^{94g}$Tc and $^{94m}$Tc, respectively, together with TENDL predictions and the available experimental data from the literature~\cite{BONARDI2002, KHANDAKER2007, Alharbi2011, TARKANYI2012, CERVENAK2016, Elbinawi2020, AHMED2022}. For $^{94g}$Tc, the present production cross sections are consistent, within the experimental uncertainties, with the available experimental data, while the TENDL tends to underestimate the measured values. For $^{94m}$Tc, the present results show good agreement with both the experimental data and the predictions at low proton energies, whereas at higher energies the latter tend to overestimate the measured cross sections.\\
    The individual reaction cross sections, obtained by disentangling the contributions of (p,n) and (p,2n) reactions using the linear system method, are shown in Figures~\ref{fig:cs_nucl_Tc94} and~\ref{fig:cs_nucl_Tc94m}, for $^{94g}$Tc and $^{94m}$Tc, respectively, together with TENDL predictions and with literature data~\cite{Skakun1987, Levkowskij1991, Rosch1993, Lamere2019}. The present results are generally consistent with the available experimental data, although some deviations from the model predictions are observed.

\begin{figure}
    \centering
    \includegraphics[width=1\textwidth]{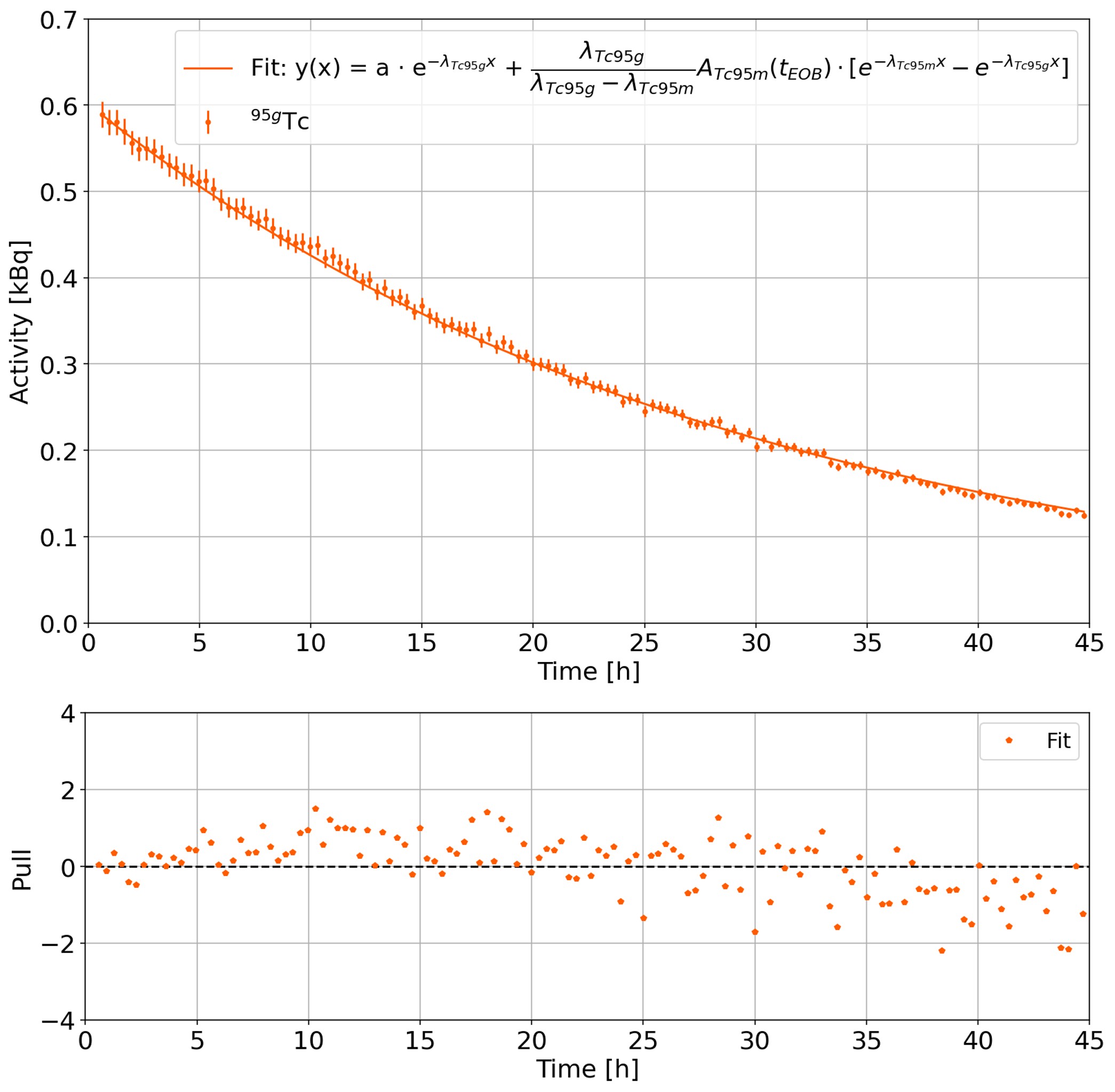}
    \caption{Activity of $^{95}$Tc measured over time and fit curve with the corresponding pull.}
    \label{fig:fits_Tc95}
\end{figure}

\paragraph{Technetium-95}
In the investigated energy range, $^{95}$Tc is produced through the $^{95}$Mo(p,n)$^{95}$Tc and $^{96}$Mo(p,2n)$^{95}$Tc reactions, populating both the ground state $^{95g}$Tc (t$_{1/2}$ = 20 h) and the metastable state $^{95m}$Tc (t$_{1/2}$ = 61 d). The metastable state decays to the ground state via isomeric transition with a branching ratio of 3.88\%, while the remaining fraction decays directly to $^{95}$Mo. Owing to its long half-life, the activity of $^{95m}$Tc was determined from a long-duration measurement performed several days after the EOB. The corresponding EOB activity was then used as a fixed input parameter in the Bateman fit applied to the $^{95g}$Tc activity data (Figure~\ref{fig:fits_Tc95}). The extracted $A(t_{EOB})$ values were subsequently used in Eq.~\ref{eq:cs} to calculate the production cross sections. 

\begin{figure}
    \centering
    \includegraphics[width=0.7\textwidth]{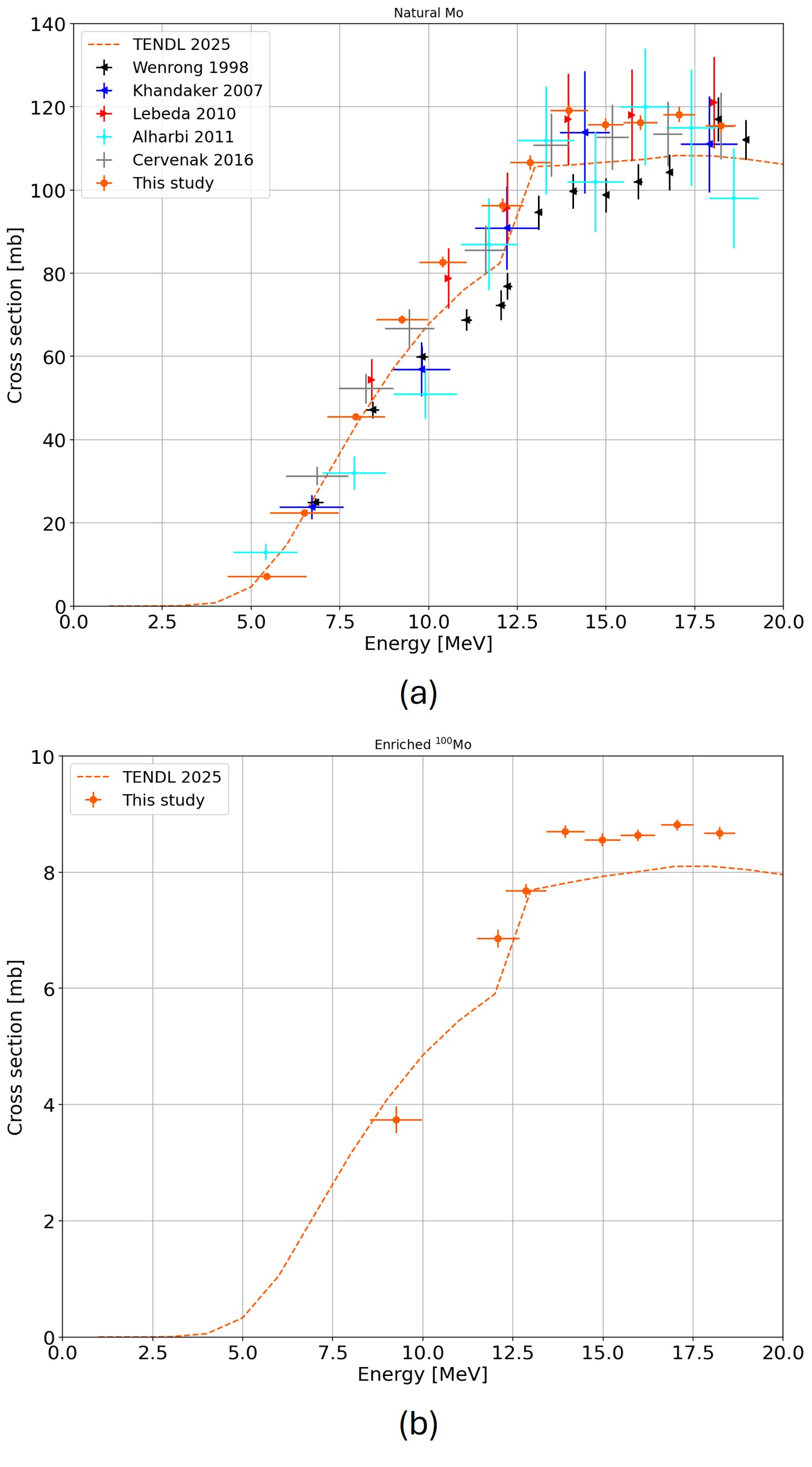}
    \caption{$^{95g}$Tc production cross sections from natural Mo (a) and enriched $^{100}$Mo (b), whose isotopic composition is reported in Table~\ref{tab:Mo_mat}. The dots are the experimental data, the dashed lines are the TENDL-2025 calculations.}
    \label{fig:cs_prod_Tc95}
\end{figure}

\begin{figure}
    \centering
    \includegraphics[width=0.7\textwidth]{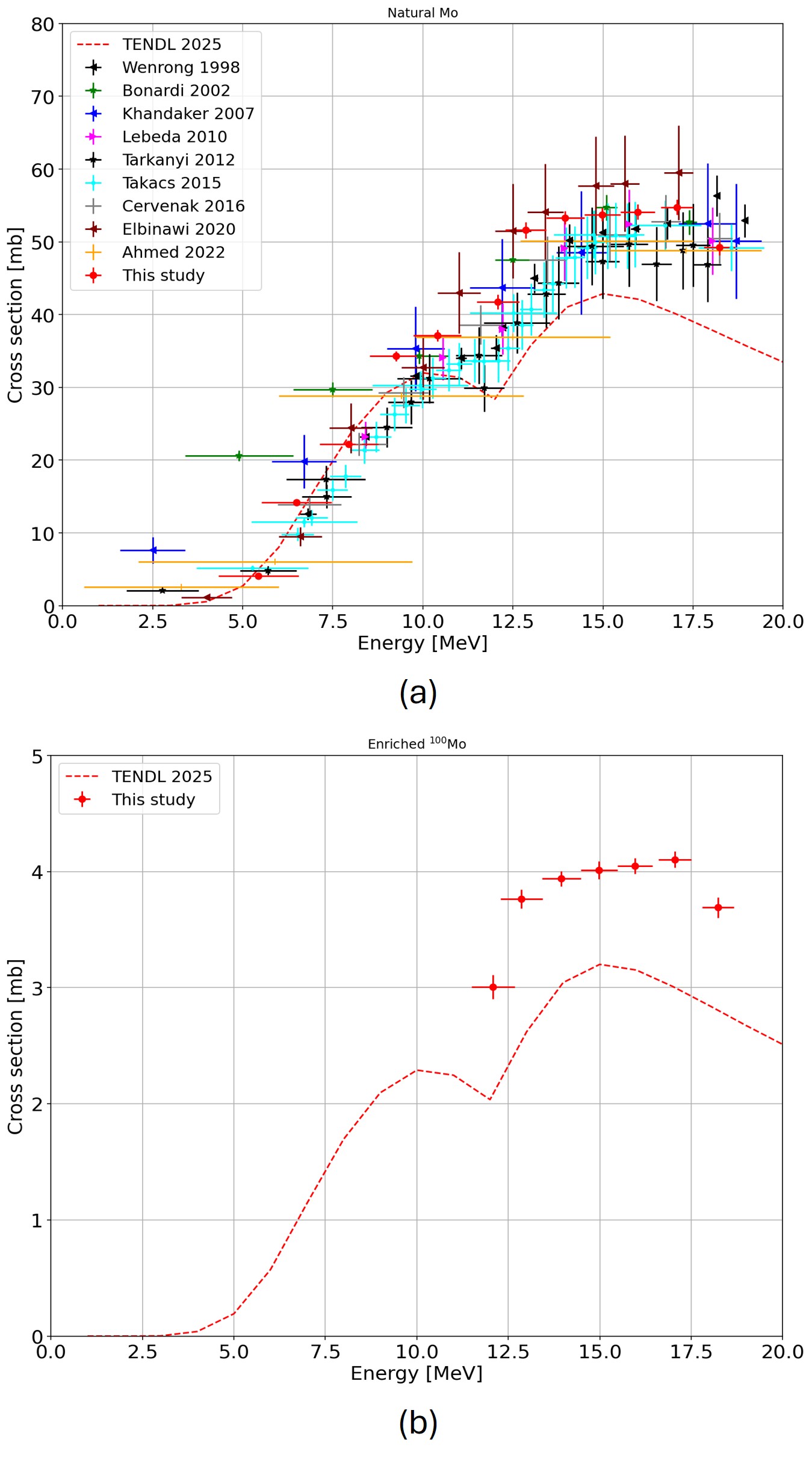}
    \caption{$^{95m}$Tc production cross sections from natural Mo (a) and enriched $^{100}$Mo (b), whose isotopic composition is reported in Table~\ref{tab:Mo_mat}. The dots are the experimental data, the dashed lines are the TENDL-2025 calculations.}
    \label{fig:cs_prod_Tc95m}
\end{figure}

\begin{figure}
    \centering
    \includegraphics[width=0.7\textwidth]{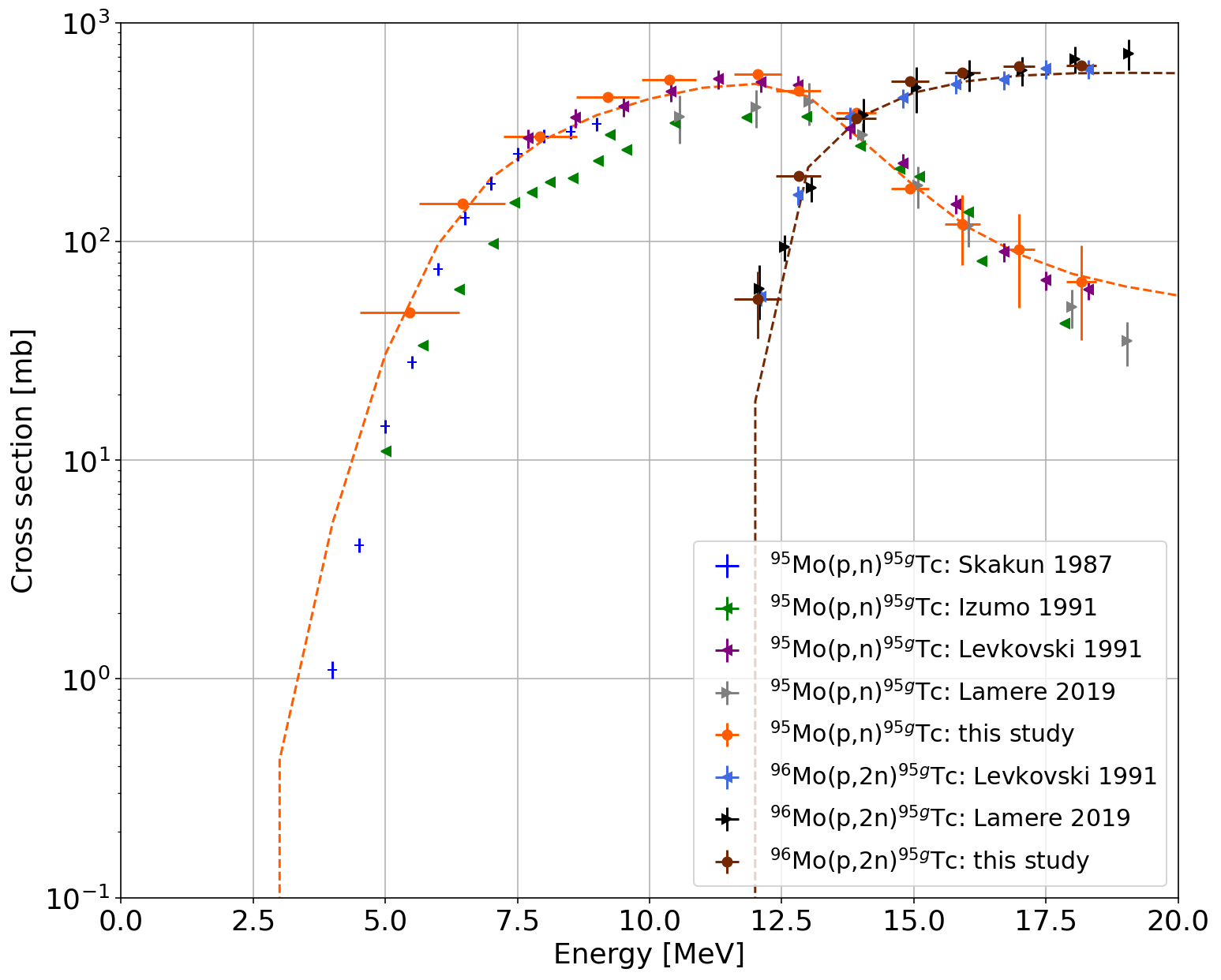}
    \caption{$^{95}$Mo(p,n)$^{95g}$Tc and $^{96}$Mo(p,2n)$^{95g}$Tc nuclear cross sections. The dots are the experimental data, the dashed lines are the TENDL-2025 calculations.}
    \label{fig:cs_nucl_Tc95}
\end{figure}

\begin{figure}
    \centering
    \includegraphics[width=0.7\textwidth]{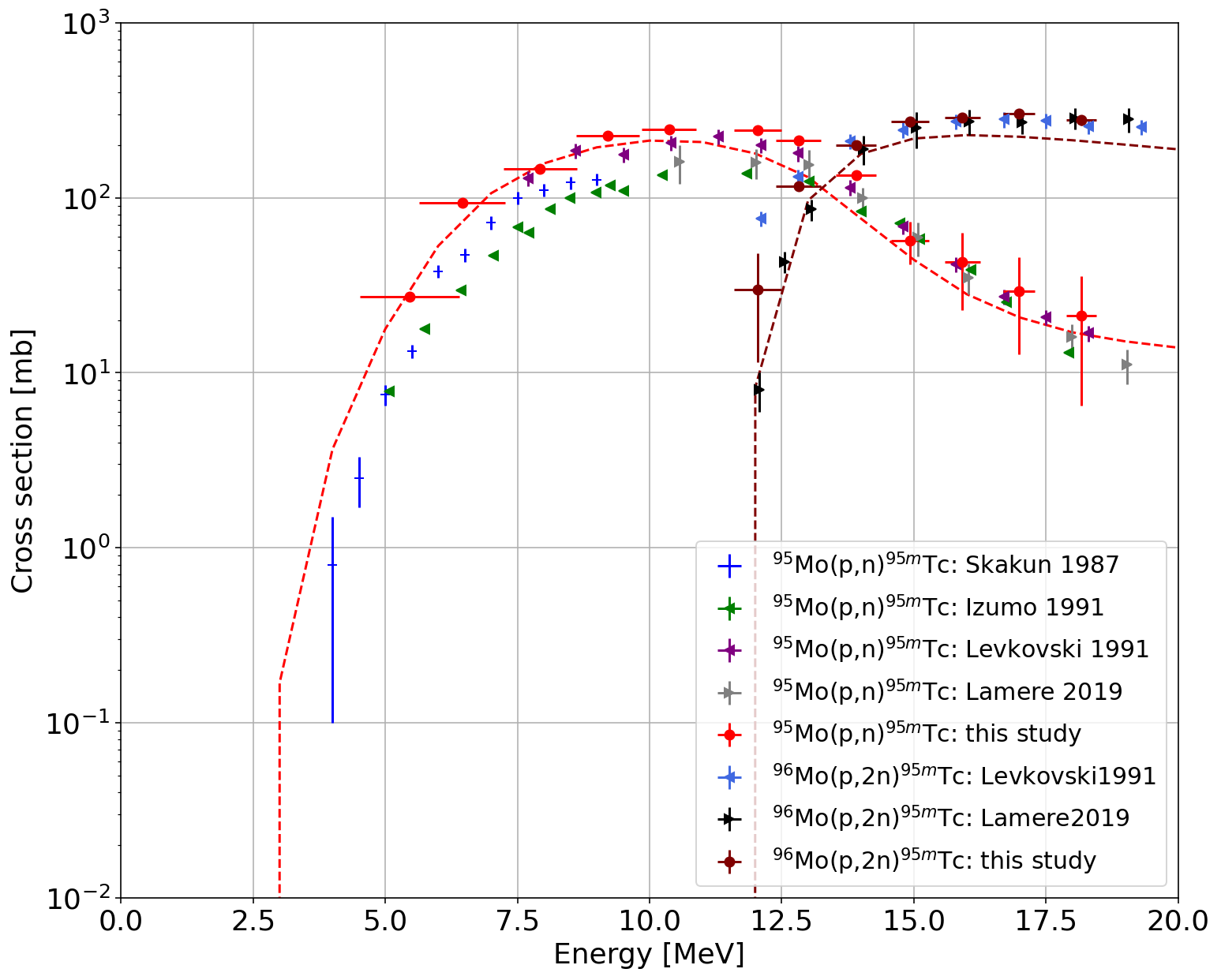}
    \caption{$^{95}$Mo(p,n)$^{95m}$Tc and $^{96}$Mo(p,2n)$^{95m}$Tc nuclear cross sections. The dots are the experimental data, the dashed lines are the TENDL-2025 calculations.}
    \label{fig:cs_nucl_Tc95m}
\end{figure}

\noindent Cross sections measured with natural Mo and enriched $^{100}$Mo targets are presented in Figures~\ref{fig:cs_prod_Tc95} and~\ref{fig:cs_prod_Tc95m} for $^{95g}$Tc and $^{95m}$Tc, respectively.\\
The individual reaction cross sections, extracted using the linear system method, are shown in Figures~\ref{fig:cs_nucl_Tc95} and~\ref{fig:cs_nucl_Tc95m} for the ground and metastable states, respectively.
The present results are compared with the TENDL predictions and the available experimental data from the literature~\cite{Wenrong1998, BONARDI2002, KHANDAKER2007, LEBEDA2010, Alharbi2011,TARKANYI2012, TAKACS2015, CERVENAK2016, Elbinawi2020, AHMED2022, Skakun1987, IZUMO1991, Levkowskij1991, Lamere2019}. For the production cross sections, the present results show good agreement with the available experimental data, while the TENDL calculations tend to underestimate the measured values at higher proton energies. The individual reaction cross sections show overall good agreement with both the experimental data and the model predictions over the investigated energy range.

\begin{figure}
    \centering
    \includegraphics[width=1\textwidth]{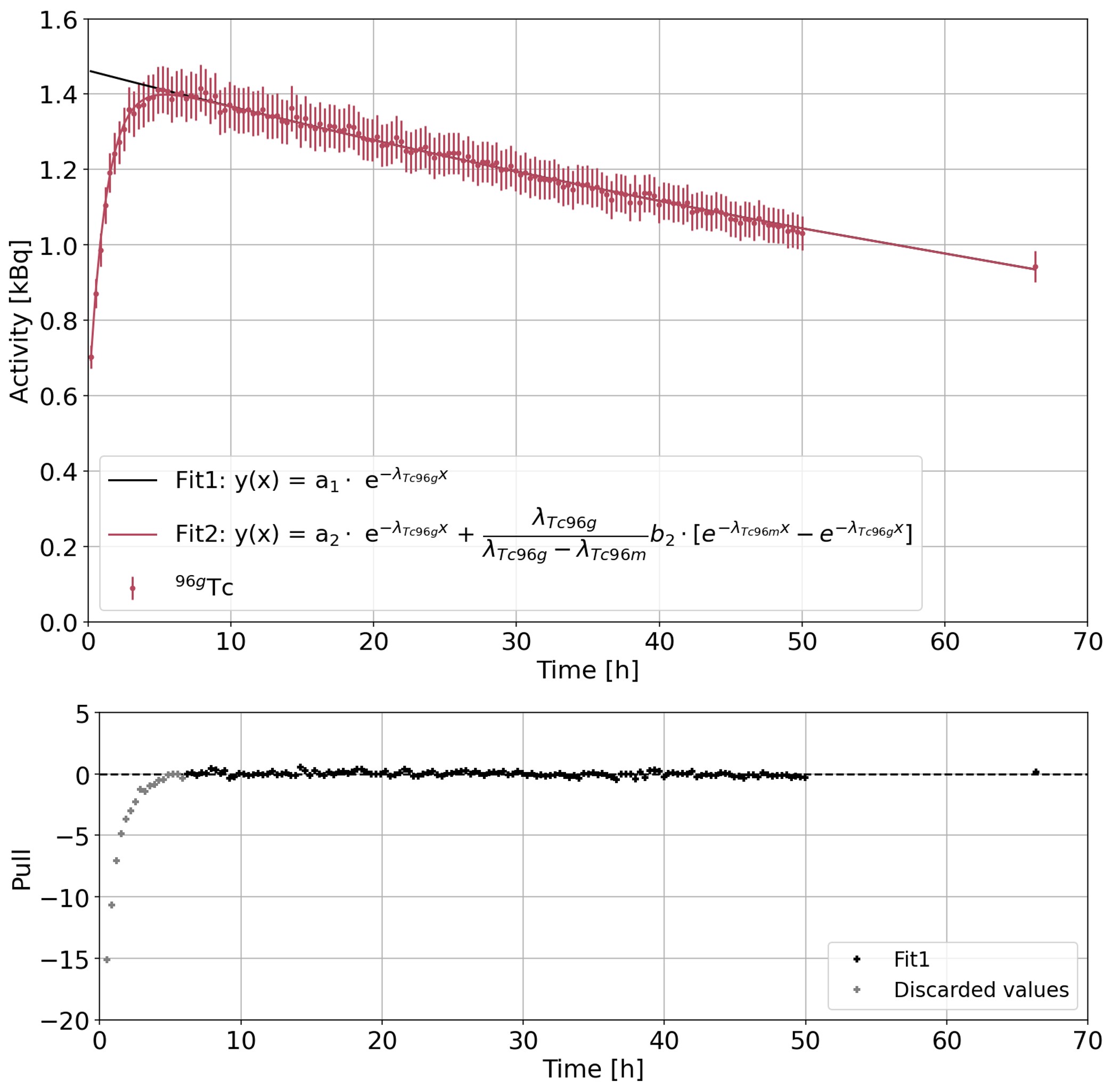}
    \caption{Activity of $^{96}$Tc measured over time and fit curve with the corresponding pull.}
    \label{fig:fits_Tc96}
\end{figure}

\begin{figure}
    \centering
    \includegraphics[width=0.7\textwidth]{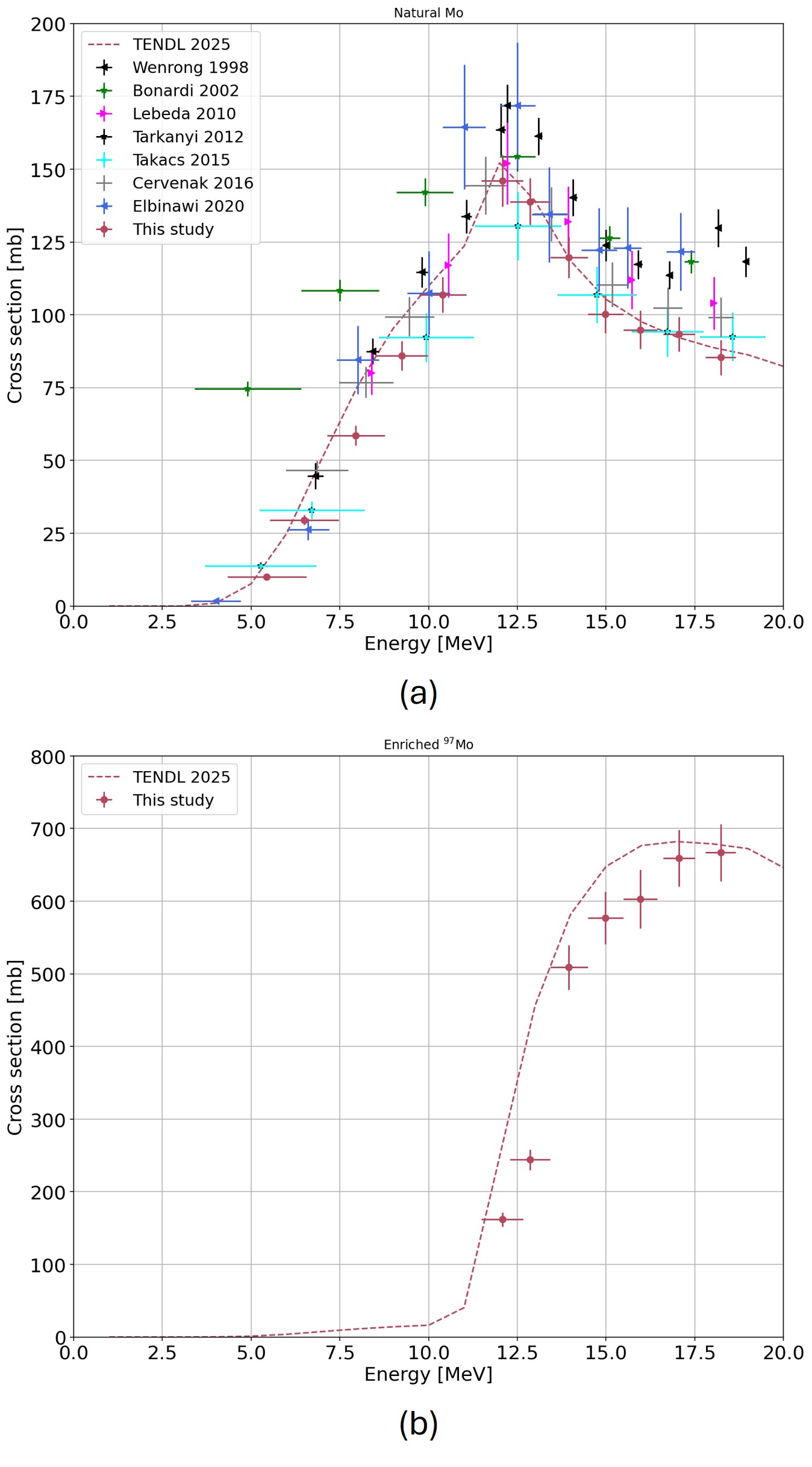}
    \caption{$^{96}$Tc production cross sections from natural Mo (a) and enriched $^{97}$Mo (b), whose isotopic composition is reported in Table~\ref{tab:Mo_mat}. The dots are the experimental data, the dashed lines are the TENDL-2025 calculations.}
    \label{fig:cs_prod_Tc96_cum}
\end{figure}

\begin{figure}
    \centering
    \includegraphics[width=0.7\textwidth]{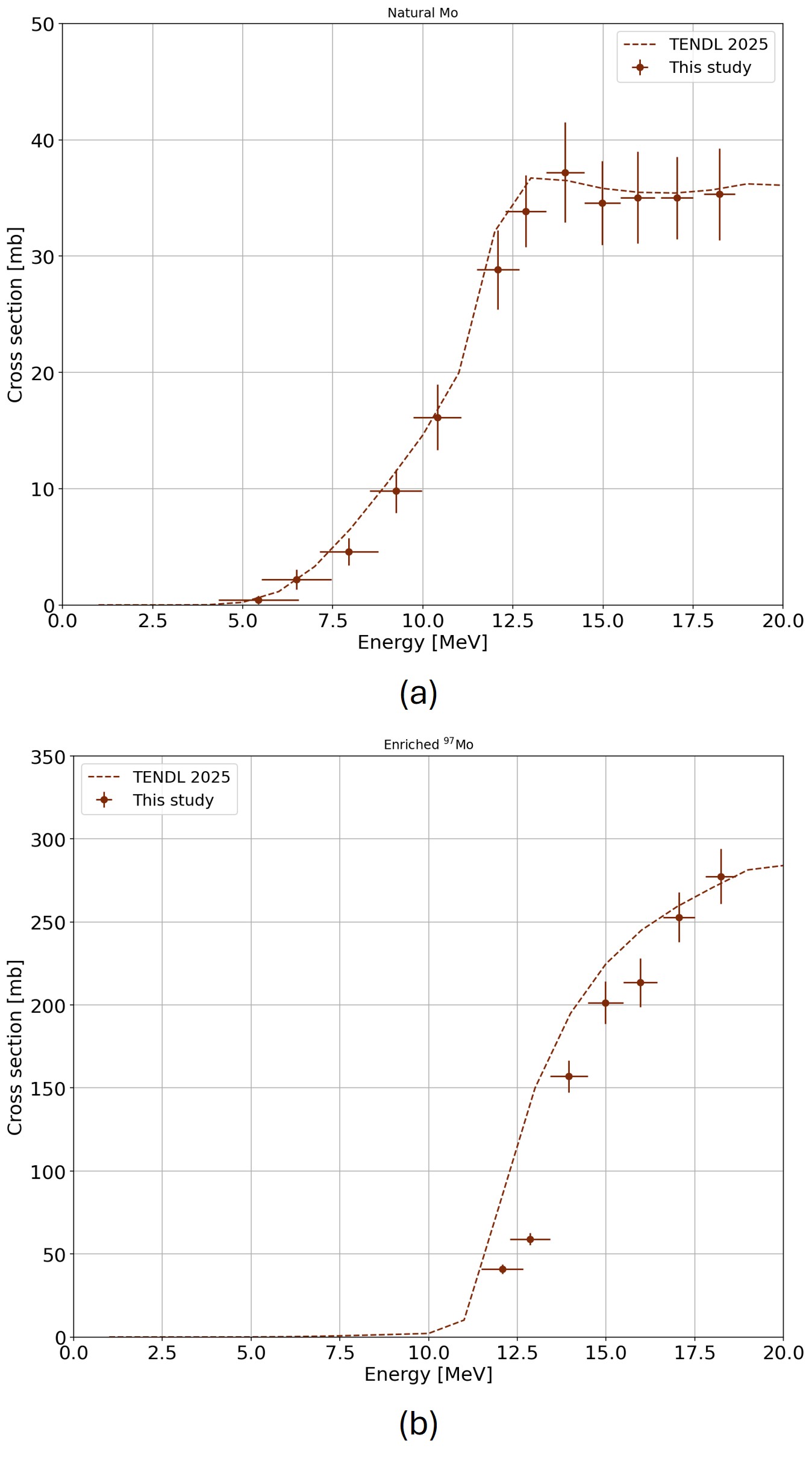}
    \caption{$^{96g}$Tc production cross sections from natural Mo (a) and enriched $^{97}$Mo (b), whose isotopic composition is reported in Table~\ref{tab:Mo_mat}. The dots are the experimental data, the dashed lines are the TENDL-2025 calculations.}
    \label{fig:cs_prod_Tc96g}
\end{figure}

\begin{figure}
    \centering
    \includegraphics[width=0.7\textwidth]{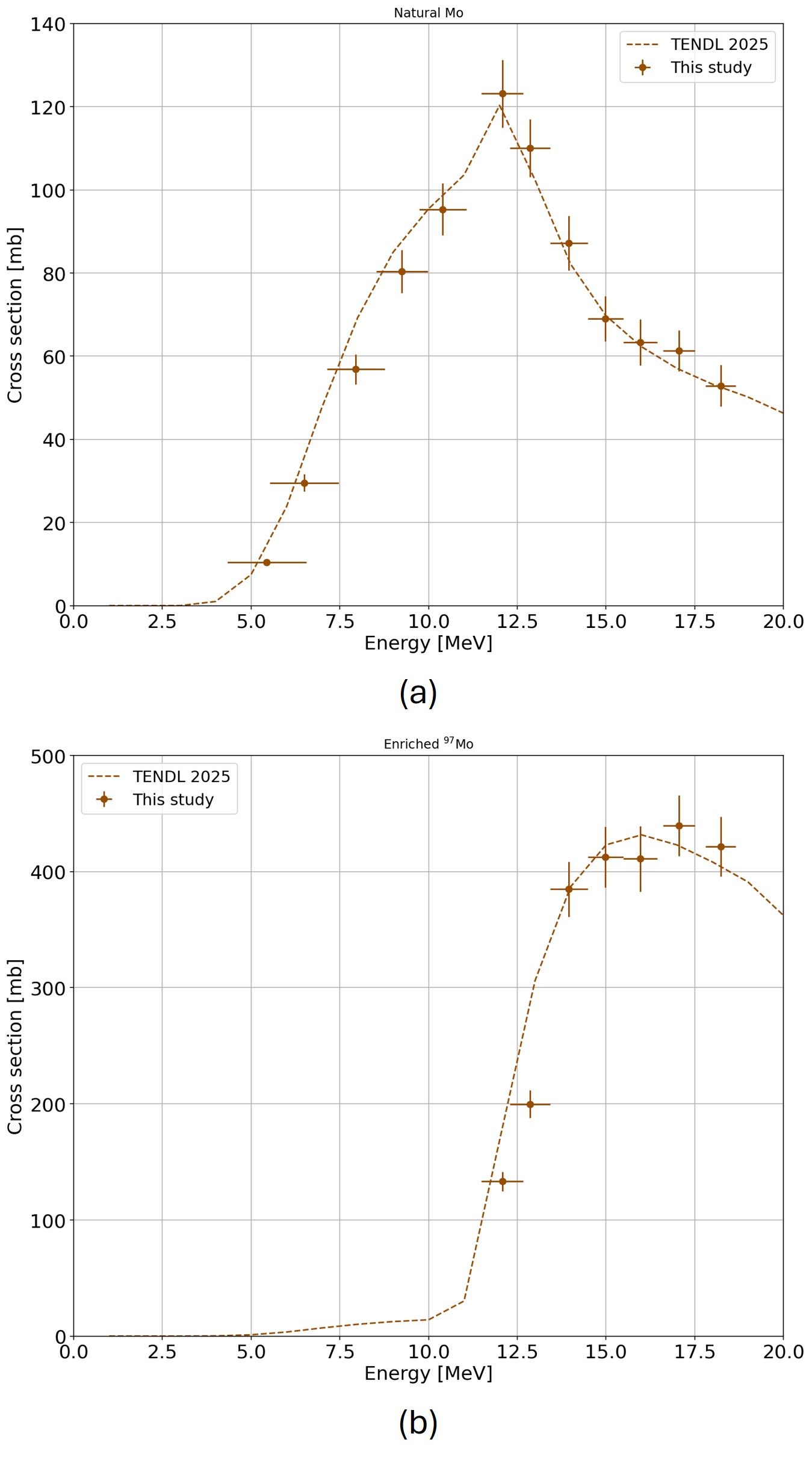}
    \caption{$^{96m}$Tc production cross sections from natural Mo (a) and enriched $^{97}$Mo (b), whose isotopic composition is reported in Table~\ref{tab:Mo_mat}. The dots are the experimental data, the dashed lines are the TENDL-2025 calculations.}
    \label{fig:cs_prod_Tc96m}
\end{figure}

\begin{figure}
    \centering
    \includegraphics[width=0.7\textwidth]{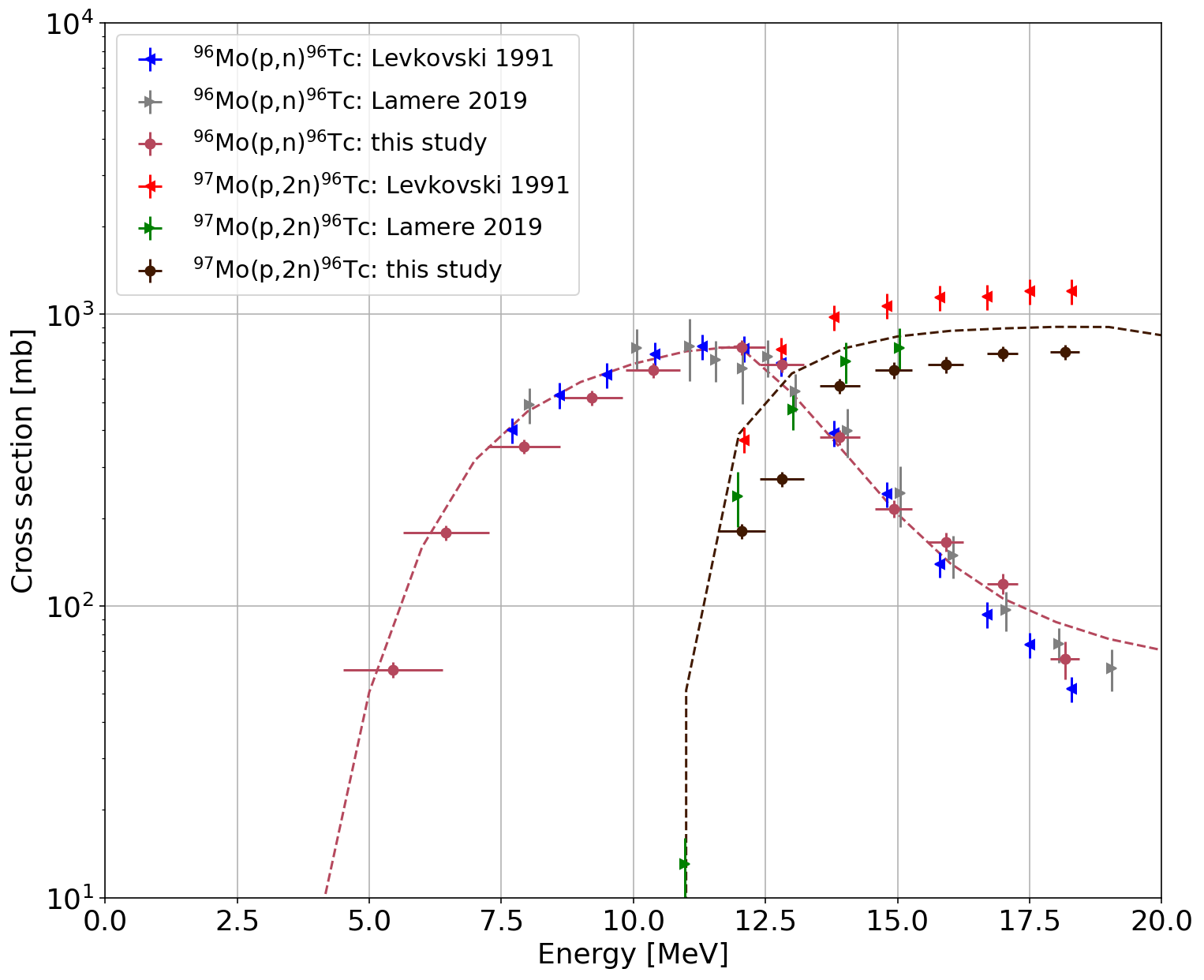}
    \caption{$^{96}$Mo(p,n)$^{96}$Tc and $^{97}$Mo(p,2n)$^{96}$Tc nuclear cross sections. The dots are the experimental data, the dashed lines are the TENDL-2025 calculations.}
    \label{fig:cs_nucl_Tc96}
\end{figure}

\begin{figure}
    \centering
    \includegraphics[width=0.7\textwidth]{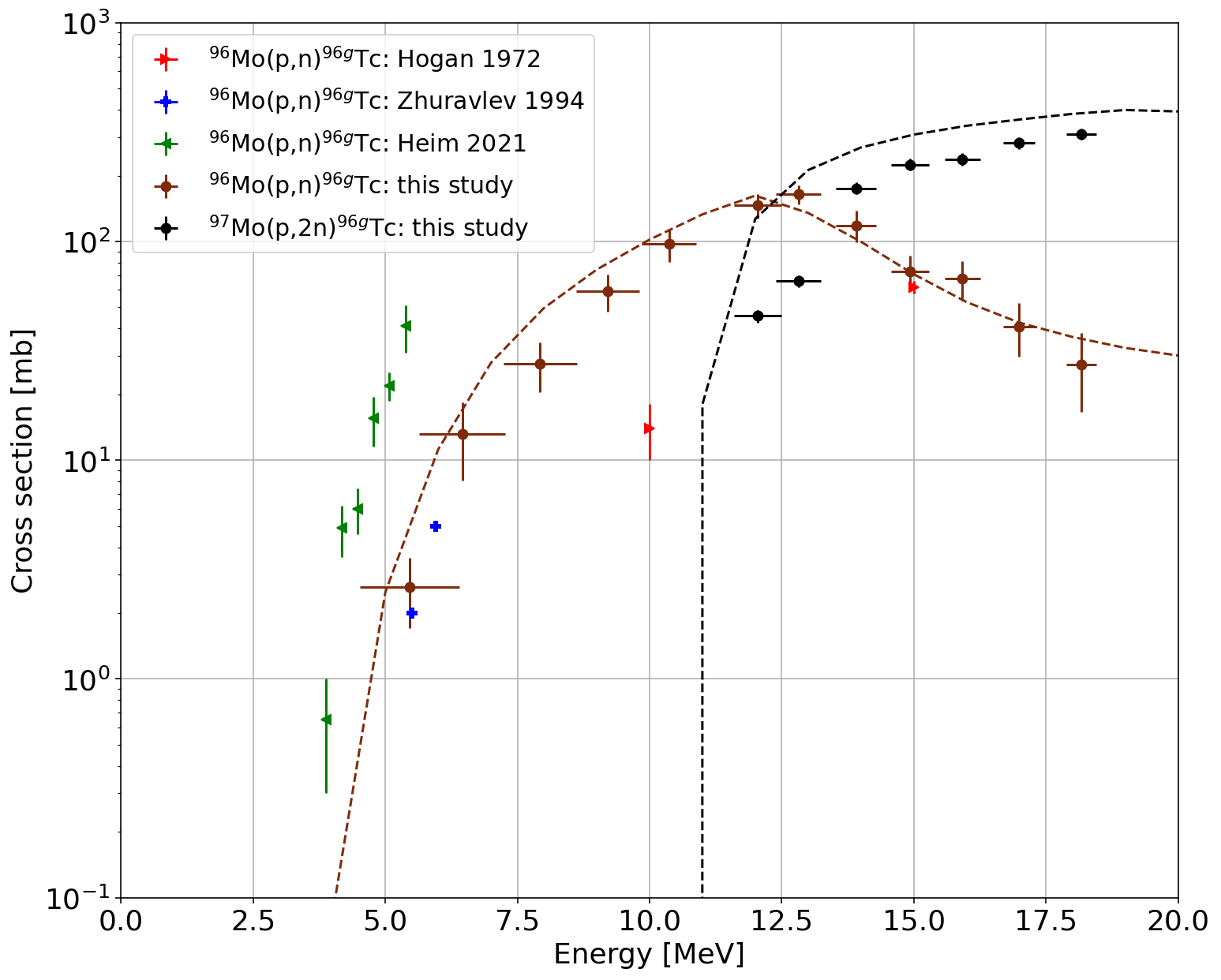}
    \caption{$^{96}$Mo(p,n)$^{96g}$Tc and $^{97}$Mo(p,2n)$^{96g}$Tc nuclear cross sections. The dots are the experimental data, the dashed lines are the TENDL-2025 calculations.}
    \label{fig:cs_nucl_Tc96g}
\end{figure}

\begin{figure}
    \centering
    \includegraphics[width=0.7\textwidth]{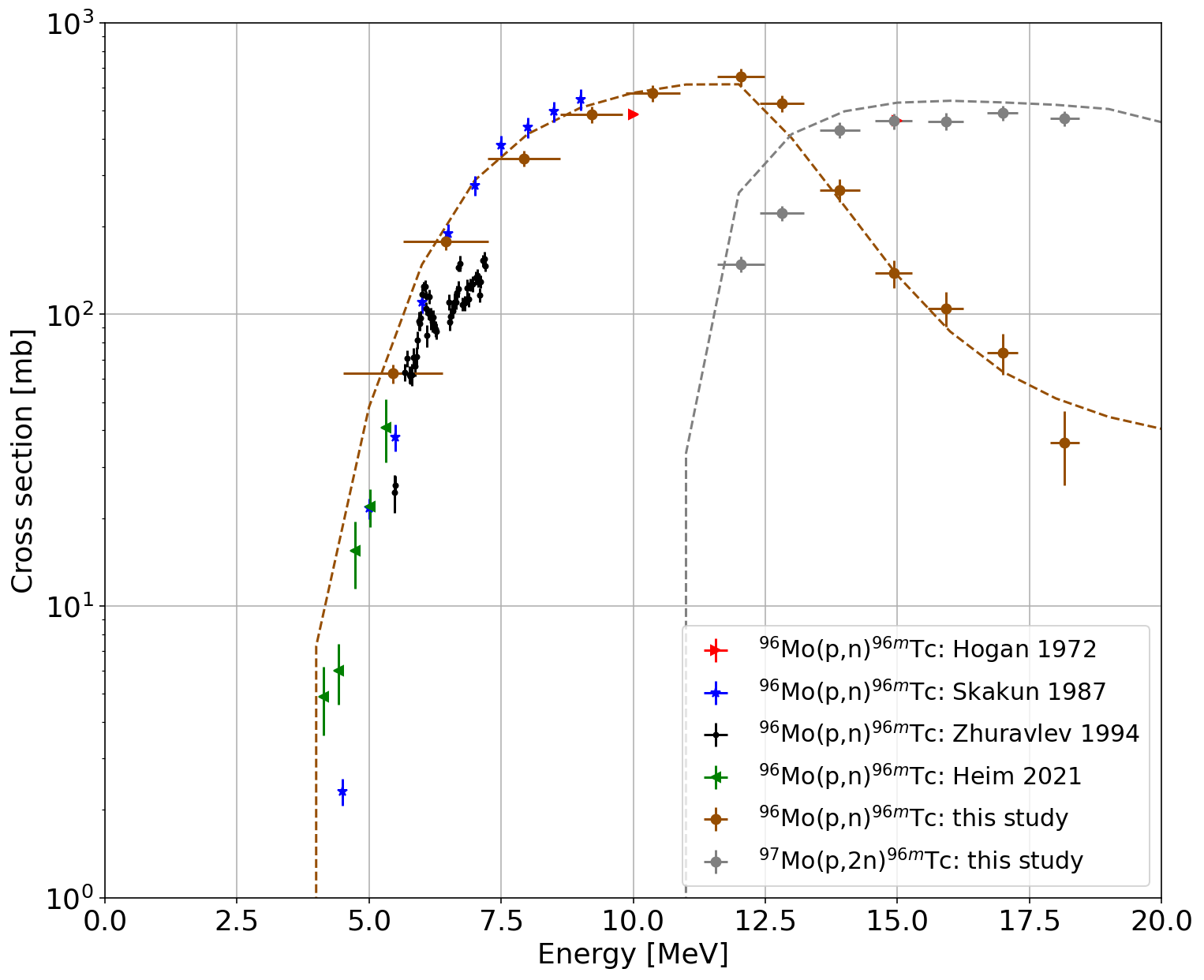}
    \caption{$^{96}$Mo(p,n)$^{96m}$Tc and $^{97}$Mo(p,2n)$^{96m}$Tc nuclear cross sections. The dots are the experimental data, the dashed lines are the TENDL-2025 calculations.}
    \label{fig:cs_nucl_Tc96m}
\end{figure}

\paragraph{Technetium-96}
Two reactions produce $^{96}$Tc in the energy range of interest, namely $^{96}$Mo(p,n) and $^{97}$Mo(p,2n), which lead to the population of both the ground state $^{96g}$Tc and the metastable state $^{96m}$Tc. The metastable state decays mainly via isomeric transition (98\%) to the ground state, resulting in an increase in the $^{96g}$Tc activity. Due to the absence of an intense $\gamma$-ray emission from $^{96m}$Tc, its contribution is difficult to quantify directly. Therefore, two complementary approaches were adopted. First, exploiting the significantly shorter half-life of $^{96m}$Tc compared with $^{96g}$Tc, the activity associated with the 812~keV $\gamma$-ray peak of $^{96}$Tc (see Table~\ref{tab:Tc_radioisotopes}) was measured after a sufficiently long cooling time, when the decay of the metastable state could be considered complete. The resulting cross section corresponds to the cumulative production cross section of $^{96g}$Tc, including both the directly produced ground state and the contribution from the decay of $^{96m}$Tc. Second, the independent production cross sections of cyclotron-produced $^{96g}$Tc and $^{96m}$Tc were determined by fitting the time evolution of the $^{96}$Tc activity using a two-parameter Bateman model (Figure~\ref{fig:fits_Tc96}), where the two fitting parameters correspond to the EOB activities of the two states. \\
For both natural Mo and enriched $^{97}$Mo, an increase in activity was observed during the first 5~h after EOB, due to the decay of $^{96m}$Tc. At longer cooling times, the activity reached a maximum and subsequently decreased according to the decay constant of $^{96g}$Tc. The equilibrium condition was verified by fitting the data measured after the equilibrium time, t$_{eq}$, with the exponential function described in Eq.~\ref{eq:exponential_fit}. The EOB activity was used as a fitting parameter, while the decay constant $\lambda$ was fixed to the literature value. The extrapolated $A(t_{EOB})$ was found to be compatible, within the experimental uncertainties, with the value obtained from the cumulative approach described above.
\noindent The production cross sections measured using natural Mo and enriched $^{97}$Mo are shown in Figures~\ref{fig:cs_prod_Tc96_cum},~\ref{fig:cs_prod_Tc96g} and~\ref{fig:cs_prod_Tc96m} for the cumulative $^{96}$Tc, $^{96g}$Tc and $^{96m}$Tc production, respectively. \\
The contributions of the (p,n) and (p,2n) nuclear reactions were disentangled using the linear system method and are shown in Figures~\ref{fig:cs_nucl_Tc96},~\ref{fig:cs_nucl_Tc96g} and~\ref{fig:cs_nucl_Tc96m} for $^{96}$Tc, $^{96g}$Tc and $^{96m}$Tc, respectively. Overall, a good agreement is observed with TENDL and, where available, with the experimental data from the literature~\cite{Wenrong1998, BONARDI2002, LEBEDA2010, TARKANYI2012, TAKACS2015, CERVENAK2016, Elbinawi2020, Hogan1972, Zhuravlev1994, Heim2021}. To the best of our knowledge, no experimental data are reported for the direct production of $^{96g}$Tc and $^{96m}$Tc from natural Mo, nor for the $^{97}$Mo(p,2n) reaction leading to either state.

\begin{figure}
    \centering
    \includegraphics[width=0.7\textwidth]{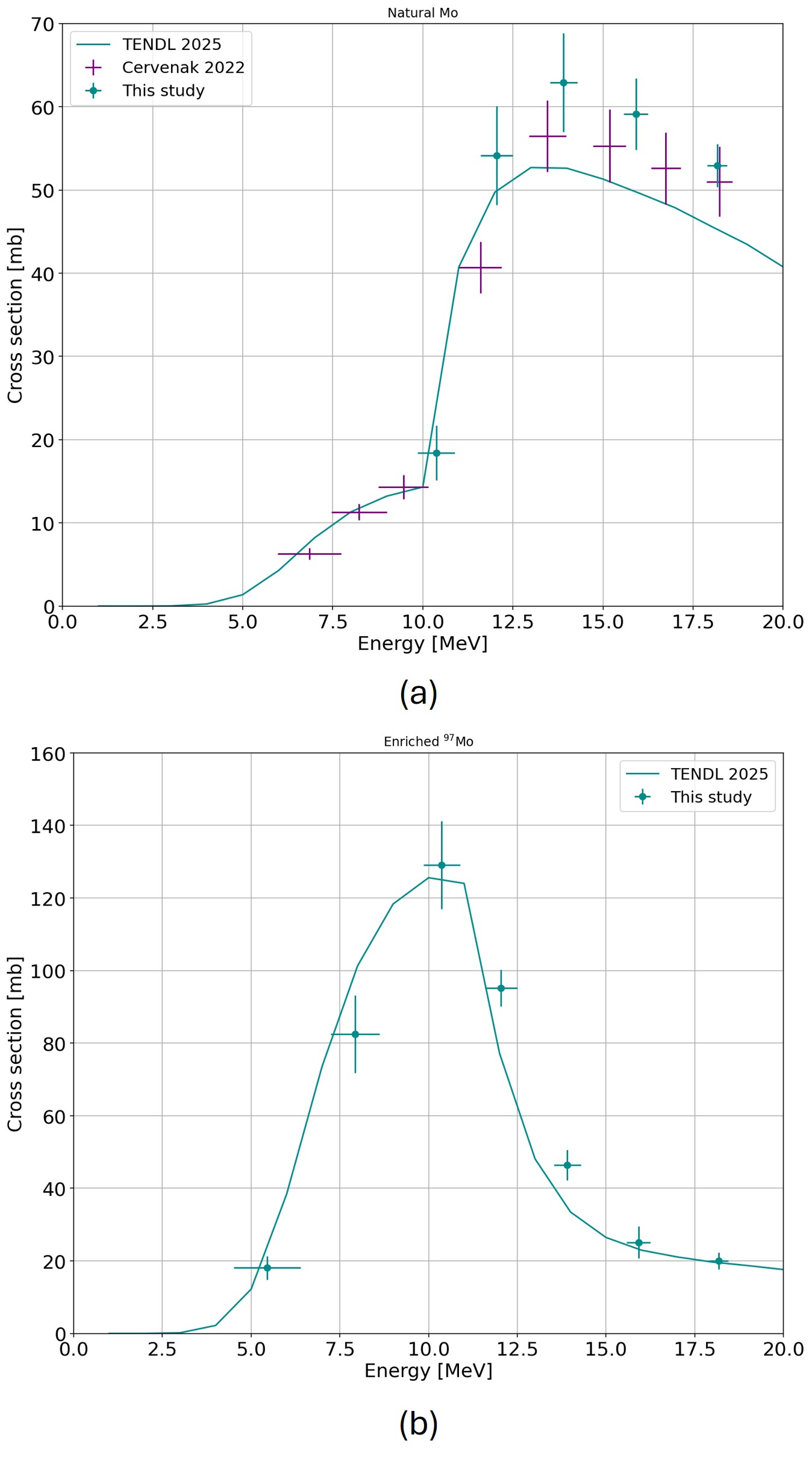}
    \caption{$^{97m}$Tc production cross sections from natural Mo (a) and enriched $^{97}$Mo (b), whose isotopic composition is reported in Table~\ref{tab:Mo_mat}. The dots are the experimental data, the dashed lines are the TENDL-2025 calculations.}
    \label{fig:cs_prod_Tc97m}
\end{figure}

\begin{figure}
    \centering
    \includegraphics[width=0.7\textwidth]{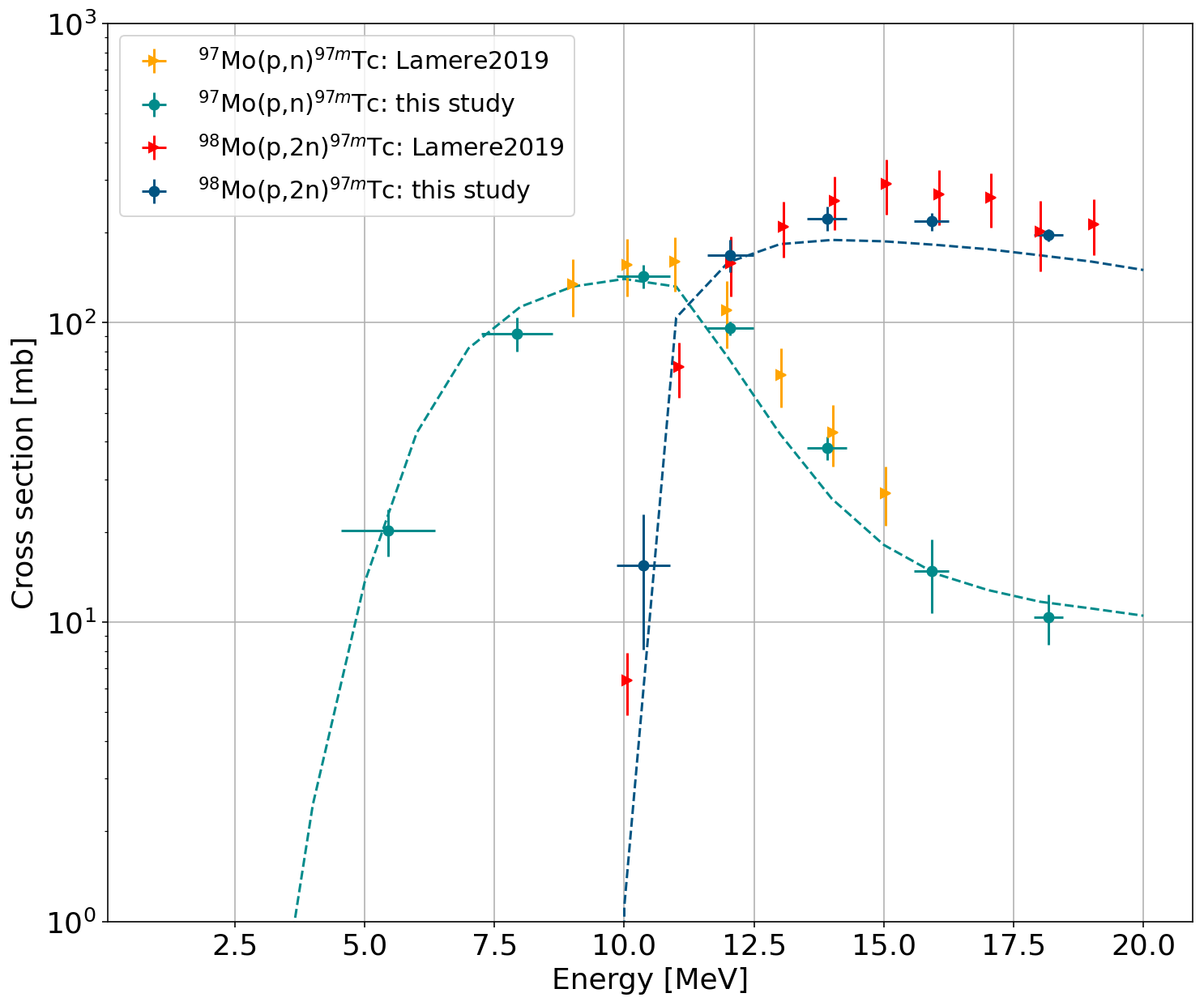}
    \caption{$^{97}$Mo(p,n)$^{97m}$Tc and $^{98}$Mo(p,2n)$^{97m}$Tc nuclear cross sections. The dots are the experimental data, the dashed lines are the TENDL-2025 calculations.}
    \label{fig:cs_nucl_Tc97m}
\end{figure}

\paragraph{Technetium-97m}
Within the investigated energy range, $^{97}$Tc is formed through the $^{97}$Mo(p,n) and $^{98}$Mo(p,2n) reactions, leading to the population of both the ground state $^{97g}$Tc (t$_{1/2}$ = 4.21$\cdot$10$^{6}$ y) and the metastable state $^{97m}$Tc (t$_{1/2}$ = 91.0 d). The extremely long half-life of the ground state, together with the absence of an intense $\gamma$-ray emission, prevents its activity from being quantified within the time scale of the present measurements; its production cross section was therefore not determined. Further measurements using appropriate experimental techniques would be required.  \\
The production of $^{97m}$Tc was investigated using natural Mo and enriched $^{97}$Mo targets. Due to the long half-life of $^{97m}$Tc, a single long-duration measurement was performed for each target. The resulting cross sections are presented in Figure~\ref{fig:cs_prod_Tc97m}, together with the TENDL predictions and the available experimental data from the literature~\cite{CERVENAK2016}. The present results show good agreement with the available experimental data, whereas the TENDL predictions tend to underestimate the measured cross sections at higher proton energies. \\
The individual reaction cross sections were extracted using the linear system method and are shown in Figure~\ref{fig:cs_nucl_Tc97m}. Our results are in good agreement with the experimental data reported by Lamere et al.~\cite{Lamere2019}, while TENDL tends to underestimate the measured cross sections.

\begin{figure}
    \centering
    \includegraphics[width=1\textwidth]{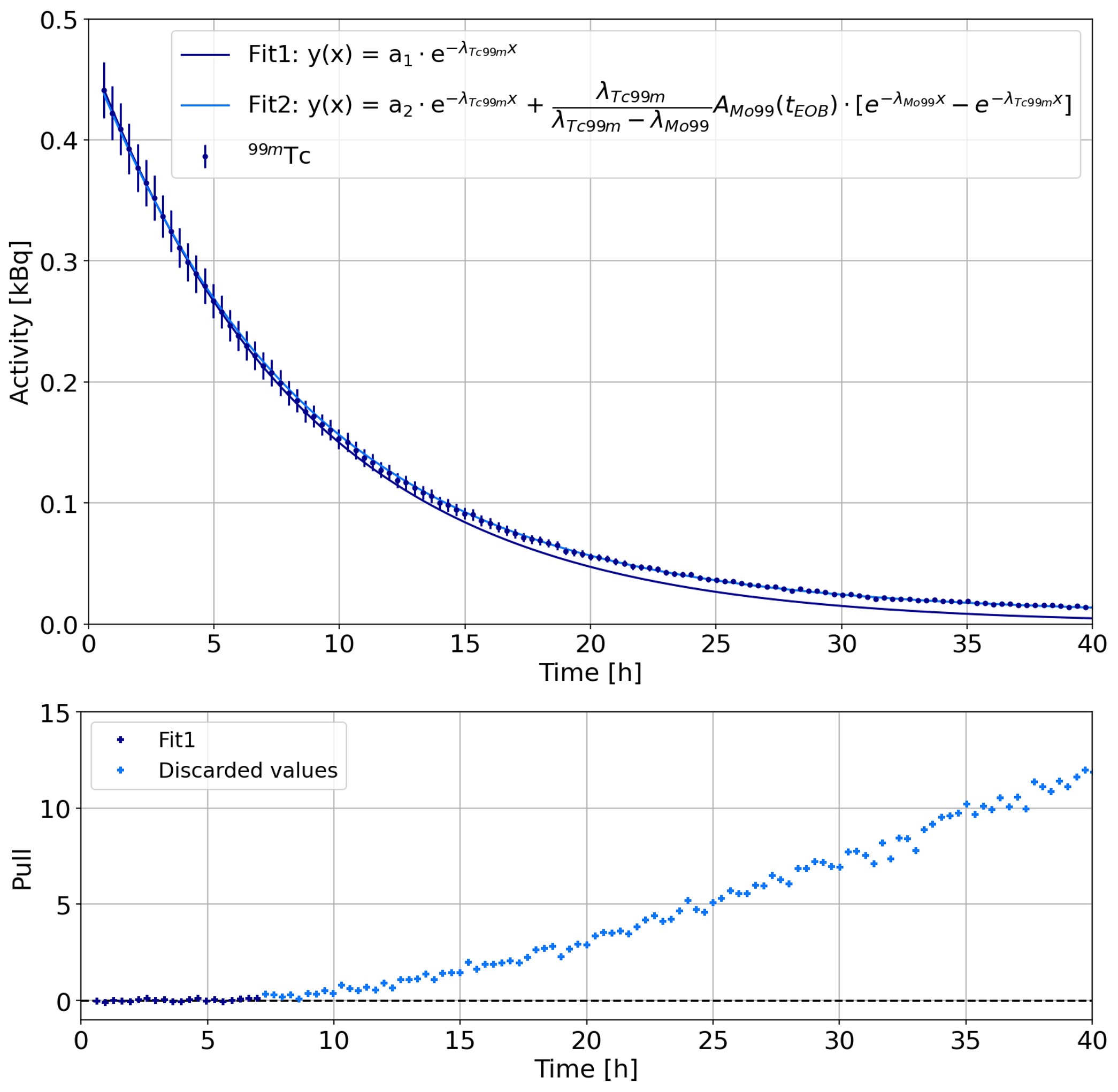}
    \caption{Activity of $^{99m}$Tc measured over time and fit curve with the corresponding pull.}
    \label{fig:fits_Tc99m}
\end{figure}

\begin{figure}
    \centering
    \includegraphics[width=0.7\textwidth]{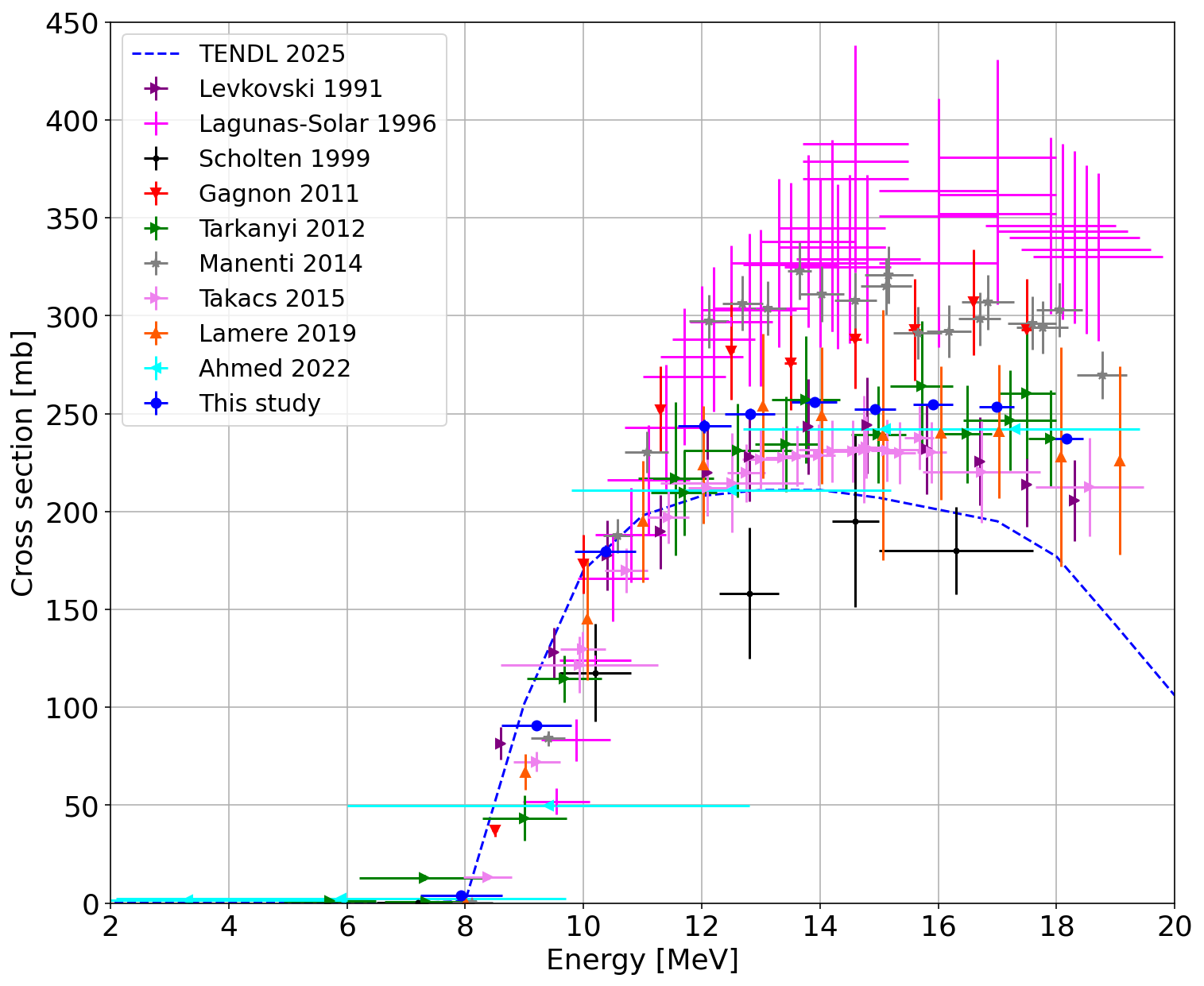}
    \caption{$^{100}$Mo(p,2n)$^{99m}$Tc nuclear cross sections. The dots are the experimental data, the dashed lines are the TENDL-2025 calculations.}
    \label{fig:cs_nucl_Tc99m}
\end{figure}

\paragraph{Technetium-99m}
During the irradiation of Mo targets, $^{99m}$Tc is mainly produced directly through the $^{100}$Mo(p,2n)$^{99m}$Tc reaction and indirectly through the decay of $^{99}$Mo, produced via the $^{100}$Mo(p,pn)$^{99}$Mo reaction. The $^{98}$Mo(p,$\gamma$)$^{99m}$Tc reaction can also lead to the direct production of $^{99m}$Tc; however, its contribution is negligible in the investigated energy range due to the low abundance of $^{98}$Mo in the targets and the small corresponding cross sections.  The production of $^{99}$Mo and its subsequent decay to $^{99m}$Tc occur both during irradiation and after the EOB, throughout the cooling and measuring times. Given the difference in half-life between $^{99m}$Tc and $^{99}$Mo, short irradiation and cooling times were adopted to minimize the indirect contribution and allow the direct production cross section to be determined. In addition, the $^{99m}$Tc activity was monitored as a function of time and analyzed using both an exponential decay function and a Bateman equation (Figure~\ref{fig:fits_Tc99m}). For all samples, the measured activity followed a simple exponential decay for approximately 7~h after the EOB, indicating that the contribution from $^{99}$Mo was negligible over this time interval. The $^{99m}$Tc EOB activities were therefore obtained by fitting the data measured during the first 7~h after the EOB with an exponential decay function and were used to calculate the direct production cross section according to Eq.~\ref{eq:cs}.\\
Cross sections were measured using both natural Mo and enriched $^{100}$Mo targets. For completeness, the individual experimental results are reported in Table~\ref{appx:prod_cs_Tc99m} of the Appendix. The final cross-section values, obtained as the weighted average of the two measurements, are shown in Figure~\ref{fig:cs_nucl_Tc99m} together with the TENDL predictions and the experimental data available in the literature~\cite{Levkowskij1991, Lagunas1996, SCHOLTEN1999, GAGNON2011, MANENTI2014, TAKACS2015, Lamere2019, AHMED2022}. Despite the substantial spread of the published experimental values, the present results are consistent with the overall distribution of the available data. Furthermore, the present measurements exhibit the smallest experimental uncertainties among the datasets currently reported in the literature.

\begin{figure}
    \centering
    \includegraphics[width=0.7\textwidth]{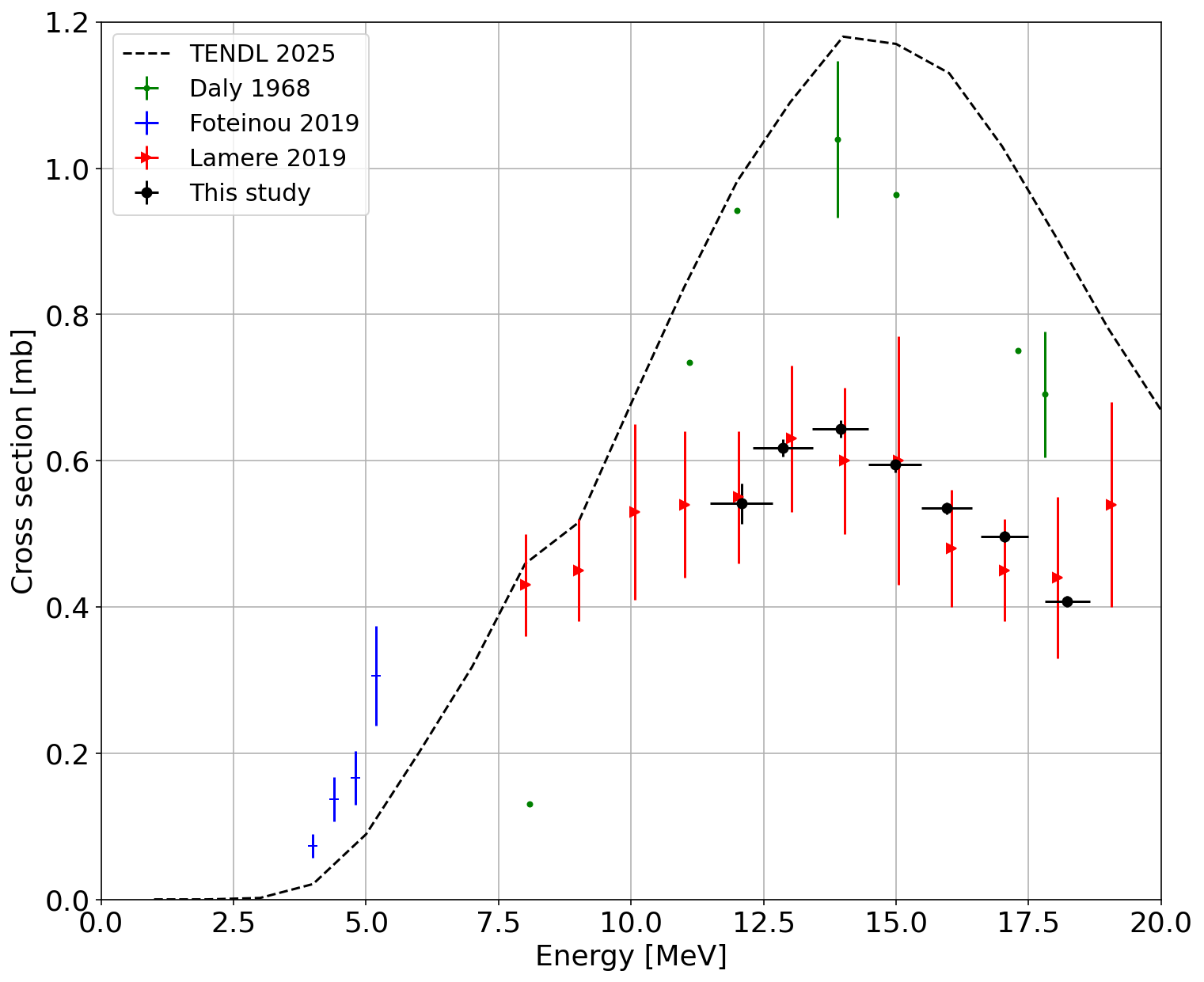}
    \caption{$^{100}$Mo(p,$\gamma$)$^{101}$Tc nuclear cross sections. The dots are the experimental data, the dashed lines are the TENDL-2025 calculations.}
    \label{fig:cs_nucl_Tc101}
\end{figure}

\paragraph{Technetium-101}
Within the investigated energy range, $^{101}$Tc is produced through the $^{100}$Mo(p,$\gamma$)$^{101}$Tc reaction. The corresponding cross sections, measured using enriched $^{100}$Mo targets, are shown in Figure~\ref{fig:cs_nucl_Tc101}. The present results are in good agreement with the experimental data reported by Lamere et al.~\cite{Lamere2019}, whereas both the TENDL predictions and the other available experimental datasets~\cite{DALY1968, Foteinou2019} tend to overestimate the measured cross sections.

\begin{figure}
    \centering
    \includegraphics[width=0.7\textwidth]{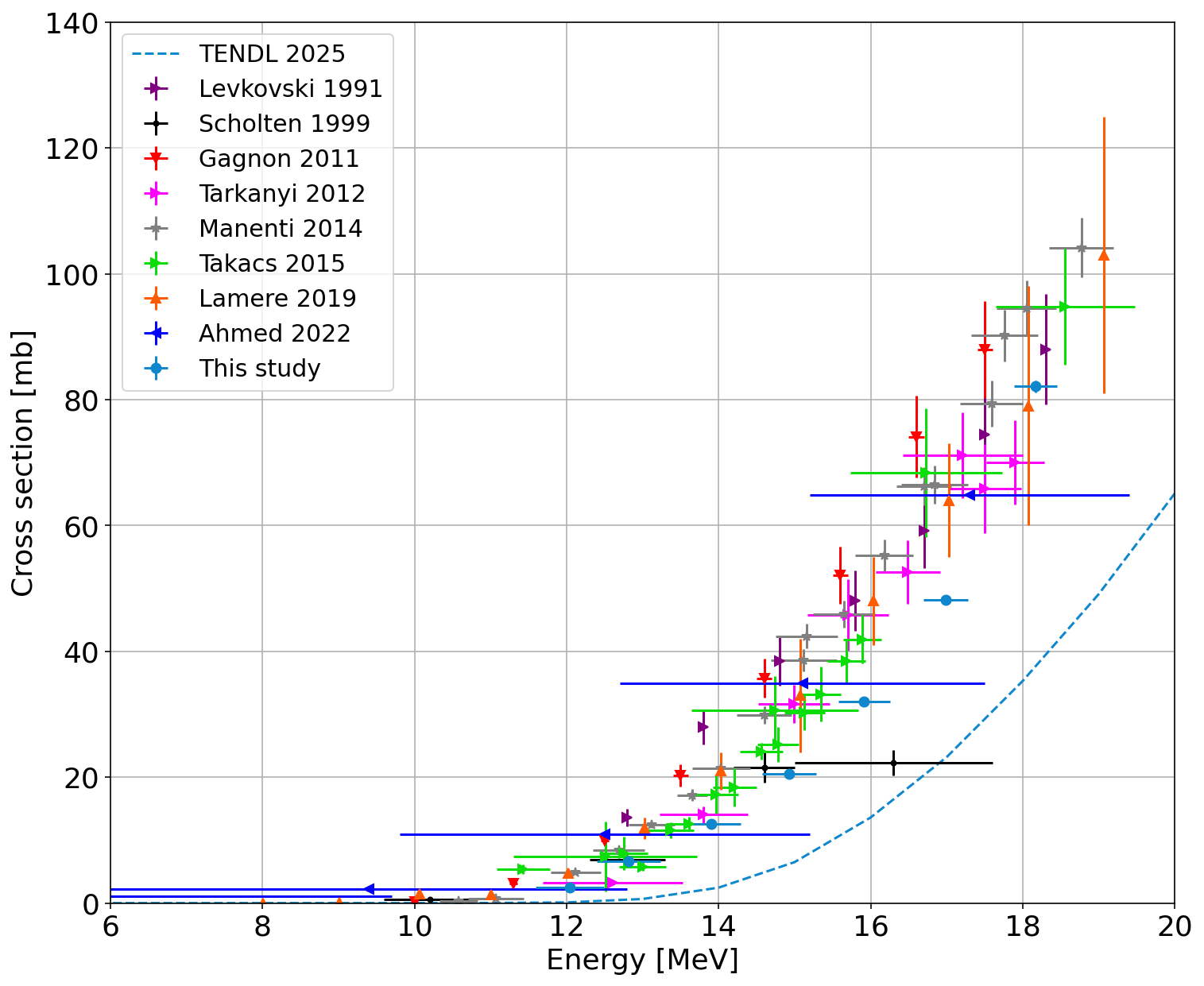}
    \caption{$^{100}$Mo(p,pn)$^{99}$Mo nuclear cross sections. The dots are the experimental data, the dashed lines are the TENDL-2025 calculations.}
    \label{fig:cs_nucl_Mo99}
\end{figure}

\paragraph{Molybdenum-99}
Up to 18 MeV, $^{99}$Mo is produced through the $^{100}$Mo(p,pn)$^{99}$Mo reaction. Cross-section measurements were performed using both natural Mo and enriched $^{100}$Mo targets, and the final cross-section values were obtained as the weighted average of the two measurements. The results are presented in Figure~\ref{fig:cs_nucl_Mo99}, together with TENDL predictions and the available data from the literature~\cite{Levkowskij1991, SCHOLTEN1999, GAGNON2011, TARKANYI2012, MANENTI2014, TAKACS2015, Lamere2019, AHMED2022}. The present results follow the trend of the available experimental data, although slightly lower values are observed in the 16--17~MeV energy range. In contrast, TENDL  predictions tend to underestimate the measured cross sections.

\subsection{$^{99m}$Tc production optimization}
\begin{figure}
    \centering
    \includegraphics[width=0.7\textwidth]{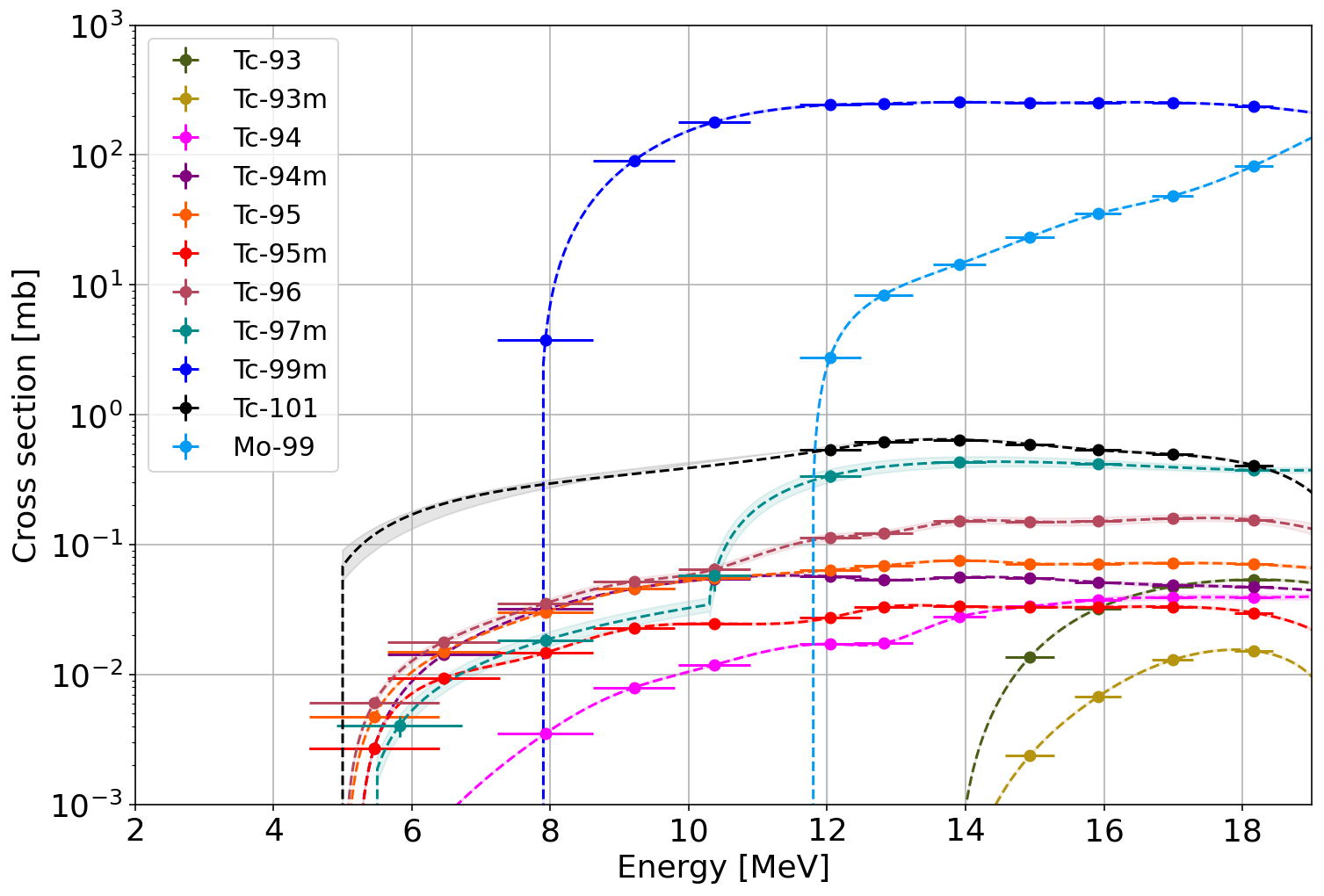}
    \caption{Production cross sections of Tc radioisotopes and $^{99}$Mo obtained by irradiating a 99.73\% enriched $^{100}$Mo target (see Table~\ref{tab:Mo_mat}).}
    \label{fig:prod_cs_pelletA}
\end{figure}

\begin{figure}
    \centering
    \includegraphics[width=0.7\textwidth]{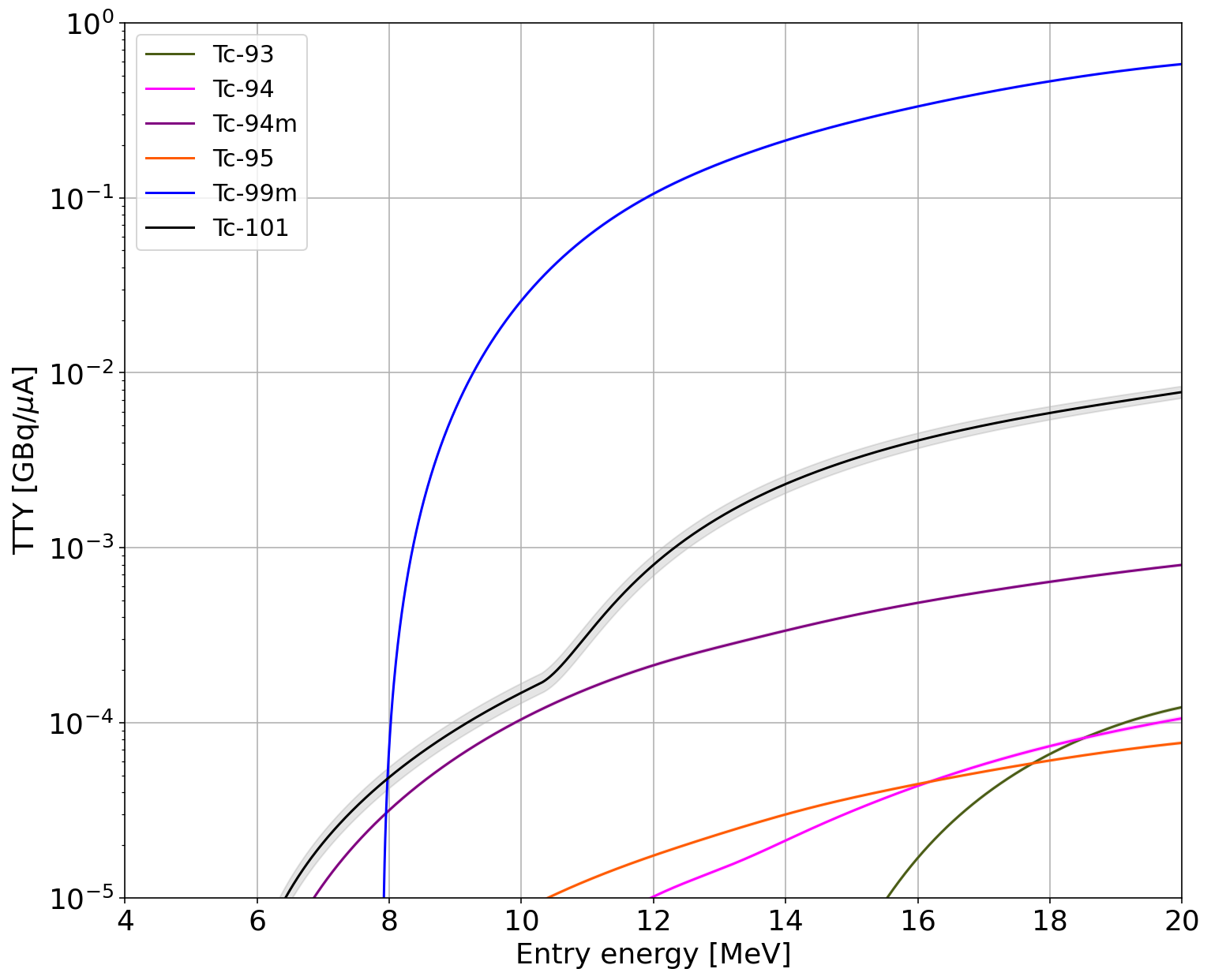}
    \caption{Thick target yields of Tc radioisotopes as a function of the proton entry energy for a 99.73\% enriched $^{100}$Mo target. The irradiation time is set to 1~h.}
    \label{fig:TTY_vs_E_pelletA_1h}
\end{figure}

\begin{figure}
    \centering
    \includegraphics[width=0.7\textwidth]{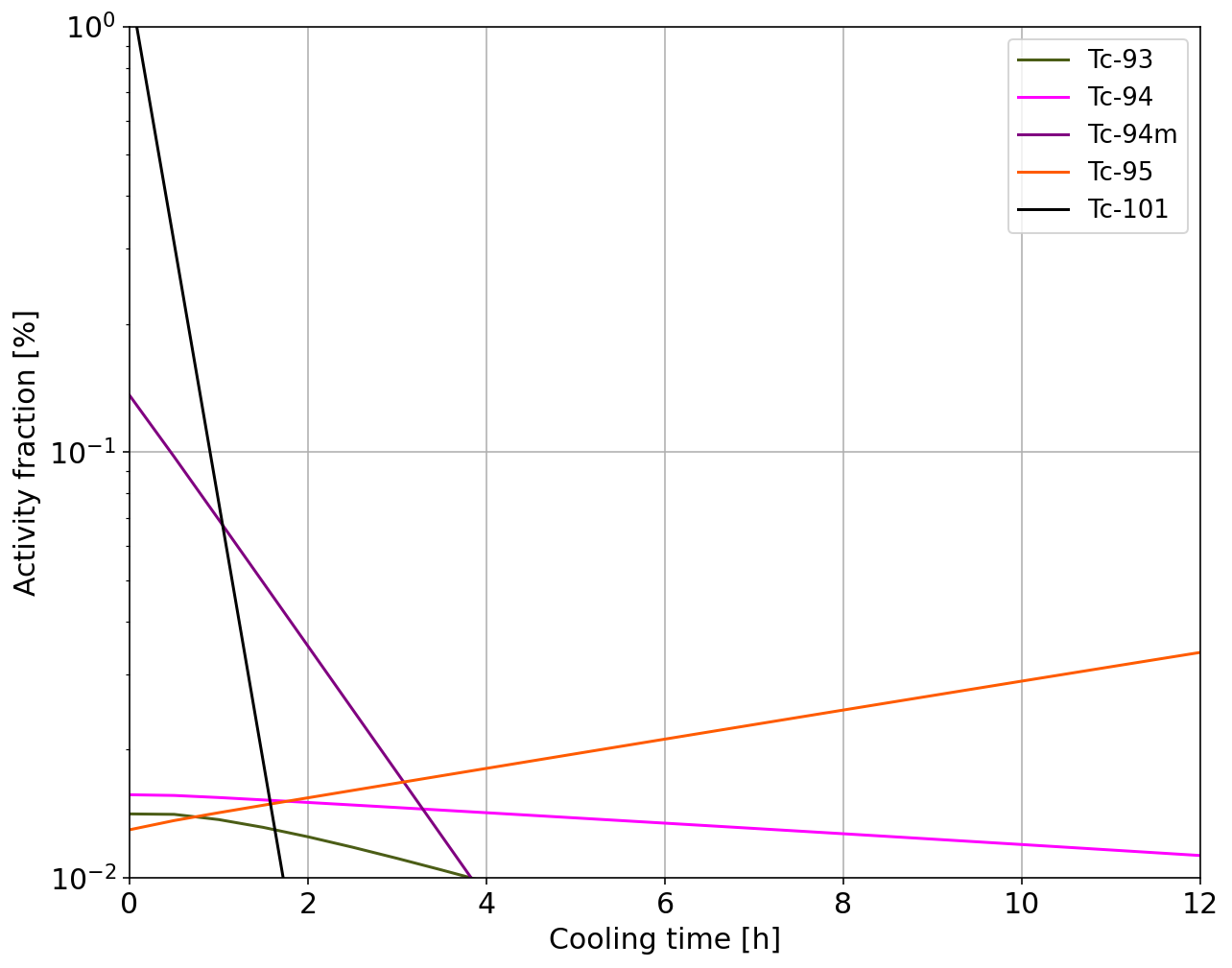}
    \caption{Activity ratio of Tc impurities as a function of time after EOB. The proton entry energy is set to 18~MeV and the irradiation time to 1~h.}
    \label{fig:A_vs_coolT_pelletA_1h}
\end{figure}

\begin{figure}
    \centering
    \includegraphics[width=0.7\textwidth]{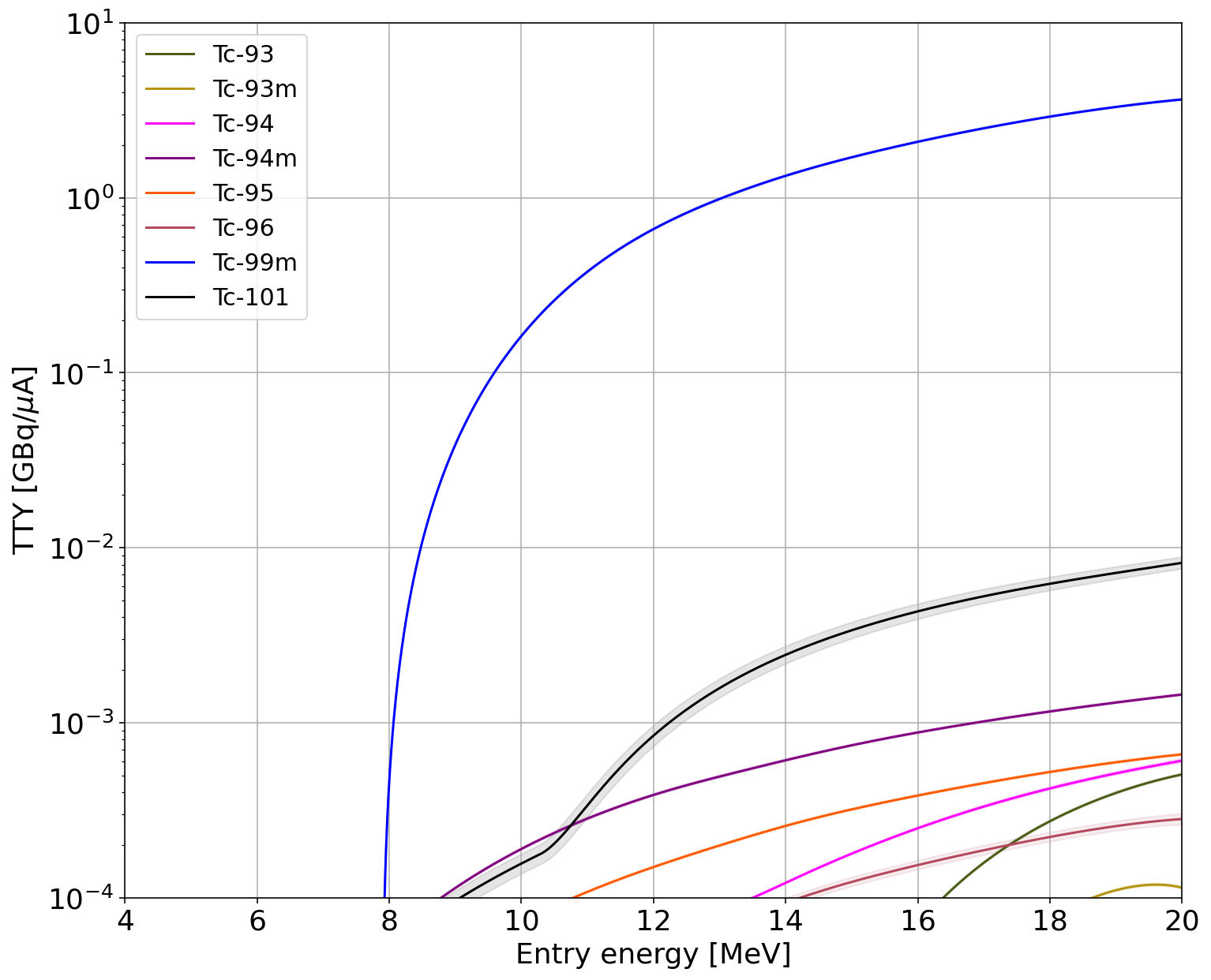}
    \caption{Thick target yields of Tc radioisotopes as a function of the proton entry energy for a 99.73\% enriched $^{100}$Mo target. The irradiation time is set to 10~h.}
    \label{fig:TTY_vs_E_pelletA_10h}
\end{figure}

\begin{figure}
    \centering
    \includegraphics[width=0.7\textwidth]{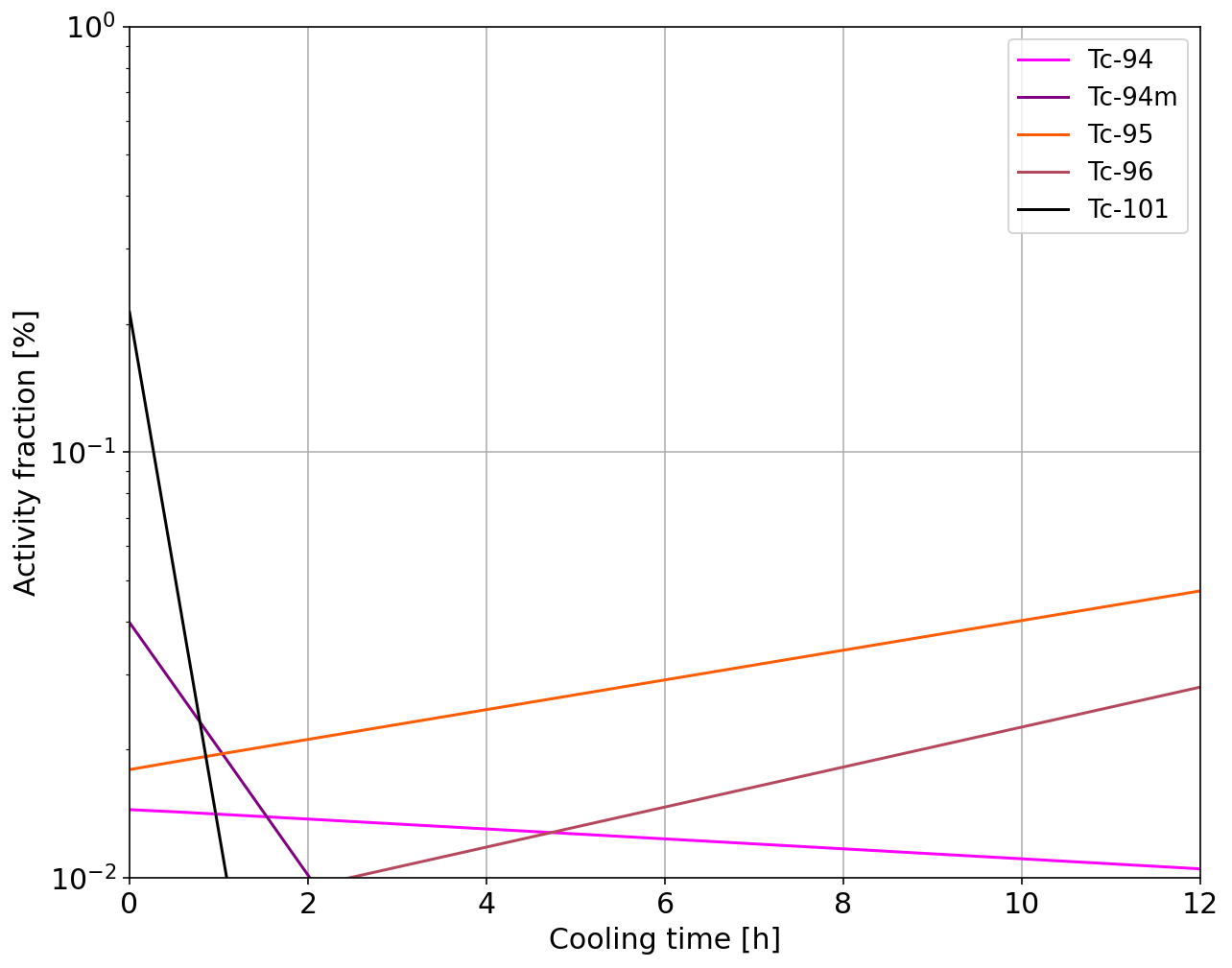}
    \caption{Activity ratio of Tc impurities as a function of time after EOB. The proton entry energy is set to 18~MeV and the irradiation time to 10~h.}
    \label{fig:A_vs_coolT_pelletA_10h}
\end{figure}

\begin{figure}
    \centering
    \includegraphics[width=0.7\textwidth]{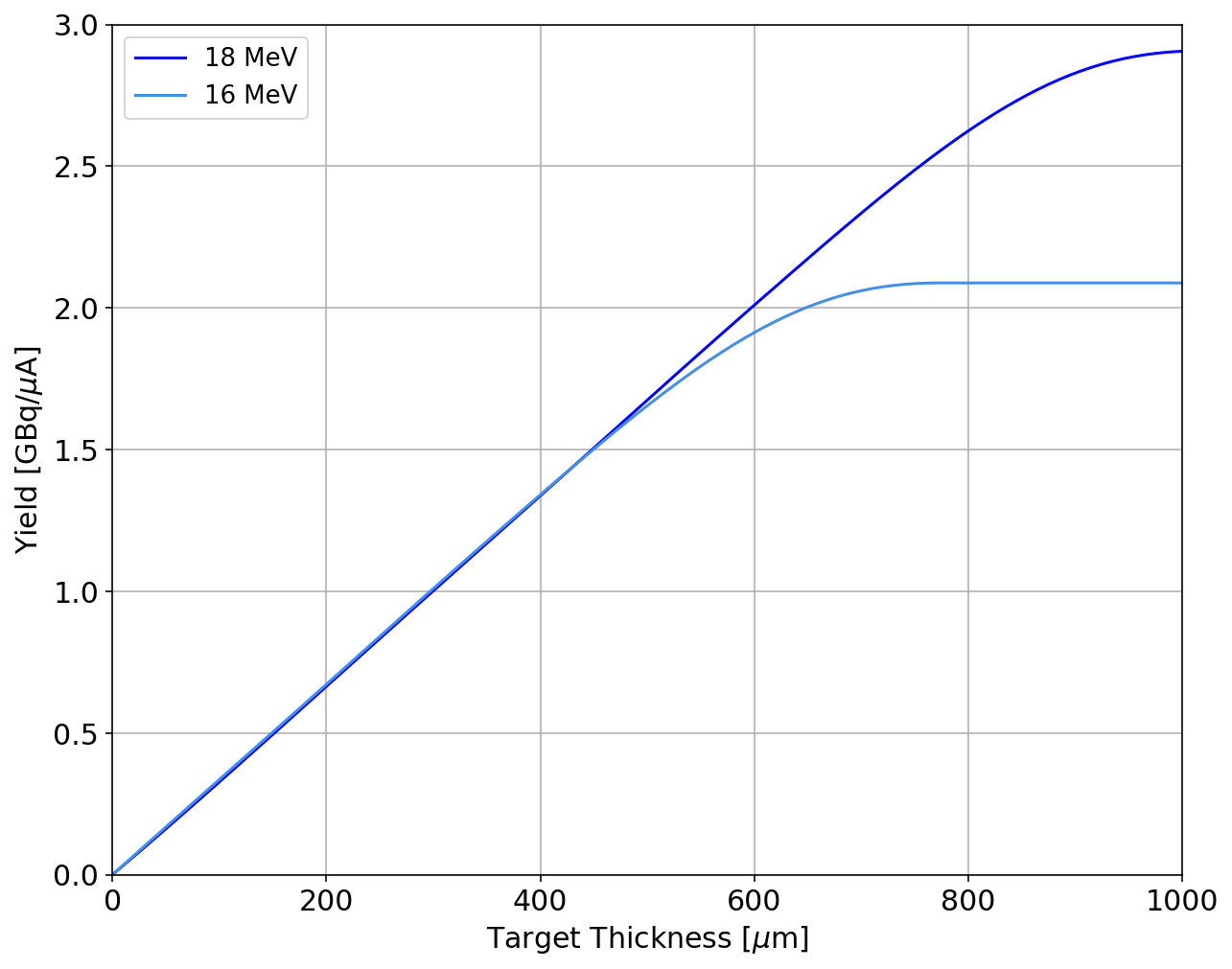}
    \caption{$^{99m}$Tc yield as a function of the target thickness for a 10-h irradiation.}
    \label{fig:Y_vs_T_pelletA_1h}
\end{figure}

\begin{figure}
    \centering
    \includegraphics[width=0.7\textwidth]{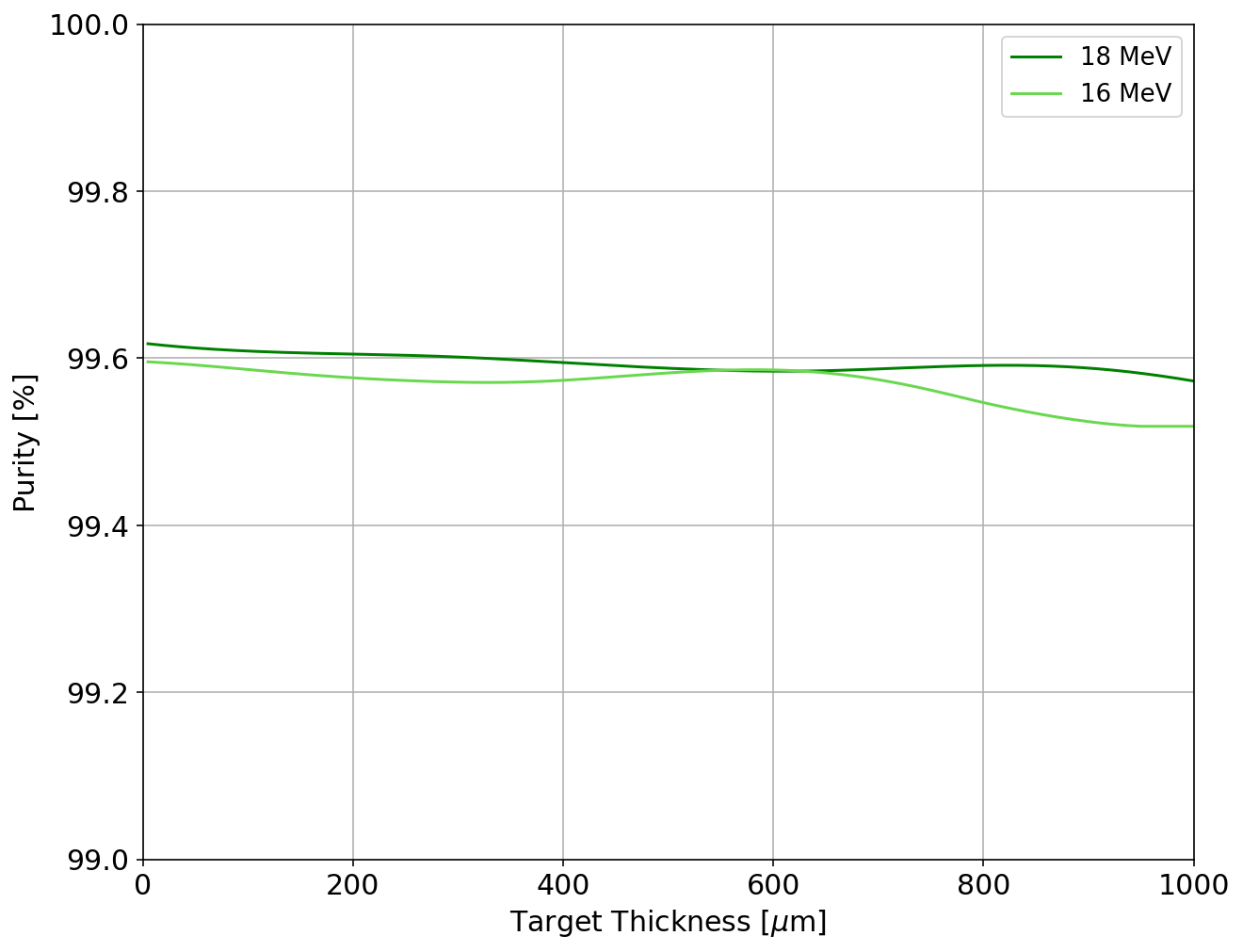}
    \caption{$^{99m}$Tc radionuclidic purity as a function of the target thickness. The irradiation time is set to 10~h.}
    \label{fig:P_vs_T_pelletA_1h}
\end{figure}

The measured cross sections of the nuclear reactions producing Tc radioisotopes up to 18~MeV were used to calculate the corresponding production cross sections for a 99.73\% enriched $^{100}$Mo target (isotopic composition reported as $^{100}$Mo (a) in Table~\ref{tab:Mo_mat}). The results are shown in Figure~\ref{fig:prod_cs_pelletA}.\\
Based on these results, the TTYs of the produced Tc radioisotopes were evaluated as a function of the proton entry energy using Eq.~\ref{eq:tty}, assuming an irradiation time of 1~h, as shown in Figure~\ref{fig:TTY_vs_E_pelletA_1h}. \\
The maximum $^{99m}$Tc yield is obtained at the highest available proton energy of 18~MeV, corresponding to a TTY of 0.46~GBq/$\mu$A. Under these conditions, the main co-produced Tc radioisotope is $^{101}$Tc. However, due to its half-life shorter than 15~min, its contribution decreases rapidly after the EOB and does not represent a limiting impurity for clinical application. The other relevant radioisotopic impurity is $^{94m}$Tc. According to the European Pharmacopoeia requirements for cyclotron-produced $^{99m}$Tc, the activity of $^{94m}$Tc at the time of injection must remain below 0.02\% of the total technetium activity. The evolution of the activity fractions of the Tc radioisotopes was therefore evaluated as a function of the cooling time (Figure~\ref{fig:A_vs_coolT_pelletA_1h}). The contribution of $^{94m}$Tc decreases below the pharmacopoeial limit approximately 3~h after EOB. The other impurities with non-negligible half-lives, $^{93}$Tc and $^{94}$Tc, are already below their respective limits of 0.04\% immediately after irradiation. Conversely, the relative contribution of $^{95}$Tc increases during the cooling period due to its longer half-life, but remains below the maximum allowed value of 0.07\% for more than 12 h after the EOB. \\
Owing to the 6-h $^{99m}$Tc half-life, extending the irradiation time can significantly increase its production yield. Figure~\ref{fig:TTY_vs_E_pelletA_10h} shows the TTYs calculated for a 10-h irradiation. At an incident energy of 18 MeV, the $^{99m}$Tc TTY increases to 2.90~GBq/$\mu$A, while the yields of the short-lived impurities $^{101}$Tc and $^{94m}$Tc increase only marginally, as these radionuclides approach saturation during irradiation. Consequently, their activities at the EOB already satisfy the radionuclidic purity limits specified by the European Pharmacopoeia (Figure~\ref{fig:A_vs_coolT_pelletA_10h}). The production of $^{95}$Tc increases by almost one order of magnitude, from 0.045~MBq/$\mu$A after a 1-h irradiation to 0.38~MBq/$\mu$A after 10~h. Nevertheless, its relative activity remains below the European Pharmacopoeia limit of 0.07\% for more than 12 h after the EOB. A similar behaviour is observed for $^{96}$Tc, whose yield increases from 0.016 to 0.16~MBq/$\mu$A. Although it remains well below the corresponding radionuclidic purity limit over the considered cooling period, its production could be further reduced, if required, by selecting a lower incident proton energy.\\
The long-lived radionuclides $^{95m}$Tc and $^{97m}$Tc are produced in negligible amounts throughout the irradiation. Their activities correspond to only 9.9$\cdot$10$^{-5}$\% and 6.5$\cdot$10$^{-4}$\% of the total technetium activity, respectively, remaining more than one order of magnitude below the European Pharmacopoeia limits of 0.005\% for $^{95m}$Tc and 0.01\% for $^{97m}$Tc.\\
To define the optimal target thickness, the $^{99m}$Tc physical yield was calculated using Eq.~\ref{eq:prod_yield} as a function of the $^{100}$Mo target thickness for a 10-h irradiation. As shown in Figure~\ref{fig:Y_vs_T_pelletA_1h}, the yield increases with target thickness and, for a proton beam energy of 18~MeV, reaches a plateau for thicknesses above approximately 900~$\mu$m. However, the additional activity gained with increasing target thickness must be balanced against the larger amount of highly enriched $^{100}$Mo required, which represents one of the main cost drivers of cyclotron-based $^{99m}$Tc production. \\
For a 500~$\mu$m-thick target, the $^{99m}$Tc yield decreases by about 40\%, from 2.83 GBq/$\mu$A at 900~$\mu$m to 1.68 GBq/$\mu$A at 500~$\mu$m, while the required amount of enriched material decreases by about 44\%, from around 131~mg to 73~mg. Consequently, the target cost decreases from approximately 1117~USD to 620~USD, assuming a material cost of 8.50~USD/mg~\cite{isoflex}. Despite the lower absolute yield, the specific yield per unit cost remains comparable and slightly improves, from $2.53\cdot10^{-3}$ to $2.70\cdot10^{-3}$ GBq/($\mu$A$\cdot$USD). This indicates that the additional amount of enriched material required for the thicker target provides only a limited benefit in terms of production efficiency. Moreover, thinner targets are advantageous for high-current operation because of their reduced heat deposition and improved thermal management.\\
For a 500~$\mu$m-thick target, the use of the maximum available proton energy is also not required. Indeed, reducing the incident proton energy from 18 to 16~MeV results in a comparable $^{99m}$Tc yield, with calculated values of 1.68 and 1.66~GBq/$\mu$A, respectively. This behaviour arises because the thinner target already integrates over the most favourable energy range of the $^{100}$Mo(p,2n)$^{99m}$Tc excitation function. From the radionuclidic purity point of view, no significant differences are observed for the target thicknesses and incident energies considered (Figure~\ref{fig:P_vs_T_pelletA_1h}). In addition, an incident energy of 16~MeV provides a practical advantage for cyclotron-based production systems with a nominal extraction energy of 18~MeV, as it allows for energy losses associated with beam transport elements, including extraction and target entrance windows, while maintaining the desired energy range at the target surface. Therefore, a proton energy of 16~MeV and a target thickness of approximately 500~$\mu$m were selected for the experimental irradiation tests.\\
\subsection{$^{99m}$Tc production tests}
\begin{figure}
    \centering
    \includegraphics[width=0.7\textwidth]{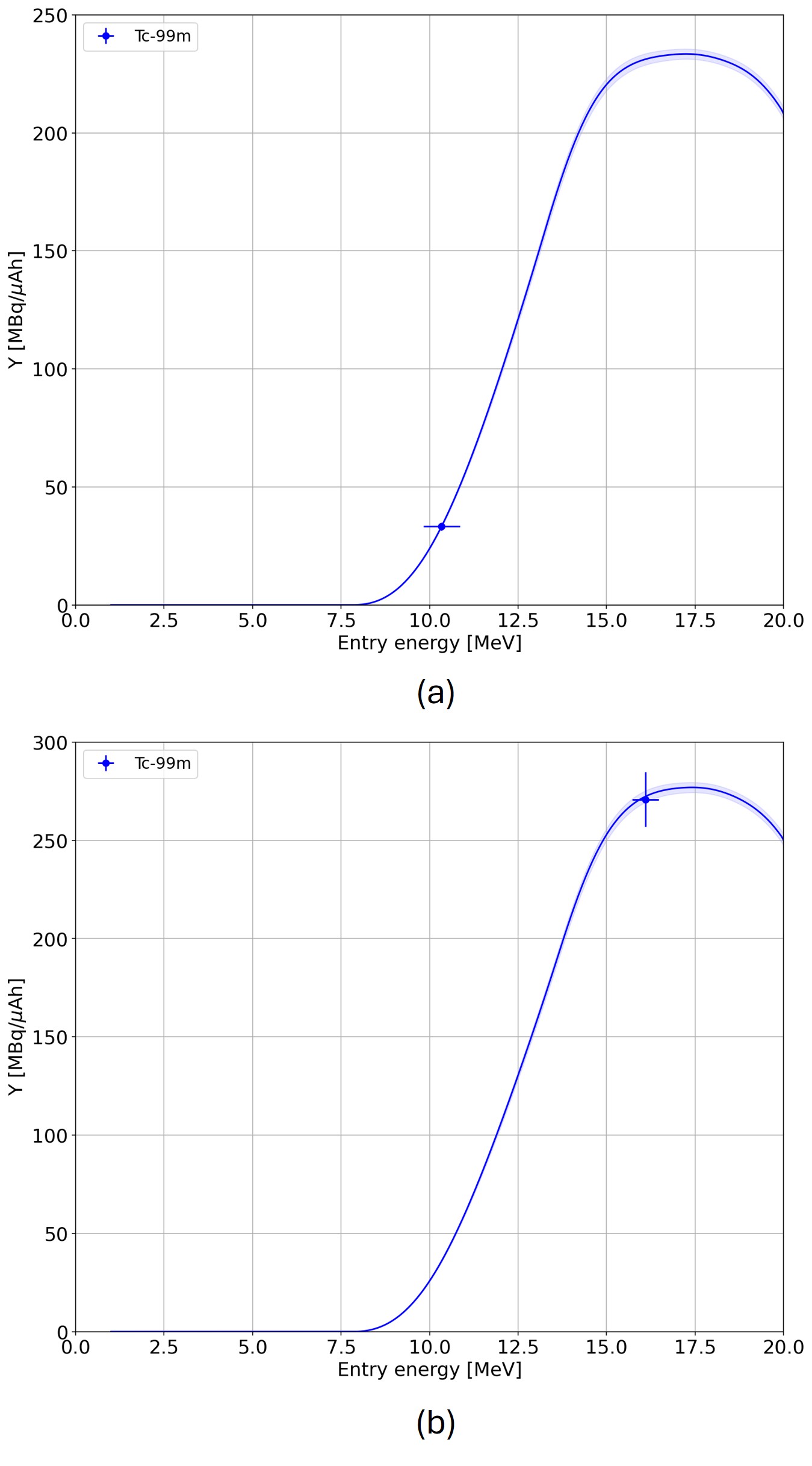}
    \caption{$^{99m}$Tc yield calculated from the measured cross sections in our irradiation conditions compared with the experimental results for the 0.50~mm-thick 92.43\% enriched $^{100}$Mo pellet (a) and the 0.48~mm-thick 99.73\% enriched $^{100}$Mo pellet (b).}
    \label{fig:prod_tests}
\end{figure}

Two irradiation experiments were performed to validate the measured excitation functions and the proposed production strategy. The first irradiation was carried out using a 0.50~mm-thick $^{100}$Mo target enriched to 92.43\% ($^{100}$Mo (b) in Table~\ref{tab:Mo_mat}) at an incident proton energy of 10.4~MeV. This experiment was primarily aimed at validating the measured cross sections for the different Tc radionuclides. The second experiment was performed using a 0.48~mm-thick $^{100}$Mo target enriched to 99.73\% ($^{100}$Mo (a) in Table~\ref{tab:Mo_mat}) at an incident proton energy of 16.1~MeV. These conditions were selected according to the optimization study presented in the previous section, which identified approximately 16~MeV as a suitable incident energy for $^{99m}$Tc production with a target thickness of about 500~$\mu$m. \\
The incident beam energies, irradiation parameters, and the activities measured by HPGe $\gamma$-ray spectrometry are summarized in Table~\ref{tab:prod_results}, together with the corresponding expected activities calculated from the measured cross sections. Overall, good agreement was observed within the uncertainty between the experimental and calculated activities. This agreement provides an independent validation of the cross-section measurements and of their use for predicting the production of different Tc radionuclides under irradiation conditions relevant to cyclotron-based $^{99m}$Tc production.\\
For $^{99m}$Tc, the two production tests resulted in yields of 33(2)~MBq/$\mu$Ah and 271(14)~MBq/$\mu$Ah, respectively. The corresponding radionuclidic purities measured 2~h after EOB were 89.6(3)\% and 99.941(4)\%. These values are in good agreement with the calculated yields of 33~MBq/$\mu$Ah and 272~MBq/$\mu$Ah (Figure~\ref{fig:prod_tests}) and the calculated radionuclidic purities of 89.9\% and 99.908\%. 
\begin{table}[]
    \centering
    \caption{Irradiation parameters and Tc activities at EOB measured for the 0.50-mm-thick 92.43\% enriched $^{100}$Mo target (Test 1) and the 0.48-mm-thick 99.73\% enriched $^{100}$Mo target (Test 2). Values in italics are the corresponding activities calculated from the measured cross sections.}  
    \resizebox{0.5\textwidth}{!}{
    \begin{tabular}{l c c}
        \hline
        ~ & Test 1 & Test 2 \\
        \hline
        E$_{in}$ [MeV] & 10.4(5) & 16.1(4) \\
        t$_{i}$ [min] & 2.89  & 3.83\\
        I [nA] & 4.81 & 3.82 \\
        \hline
        \multicolumn{3}{c}{Activity [kBq]}\\
        \hline
        \multirow{2}*{$^{93}$Tc} & 0.0125(9)  & Not Detected  \\
                                        & \textit{0.0127(3)} & \textit{0.0048(1)} \\
        \noalign{\smallskip}                                
        \multirow{2}*{$^{93m}$Tc} & 0.034(3)  & Not Detected \\
                                        & \textit{0.031(1)} & \textit{0.0036(1)} \\
        \noalign{\smallskip}
        \multirow{2}*{$^{94}$Tc} & 0.0557(4) & Not Detected\\
                                        & \textit{0.0574(9)} & \textit{0.0096(2)} \\
        \noalign{\smallskip}
        \multirow{2}*{$^{94m}$Tc}& 2.17(6)  & 0.088(7) \\ 
                                        & \textit{2.1(4)} & \textit{0.098(1)} \\
        \noalign{\smallskip}
        \multirow{2}*{$^{95}$Tc} & 0.189(1) & Not Detected \\
                                        & \textit{0.187(3)} & \textit{0.0076(1)} \\
        \noalign{\smallskip}
        \multirow{2}*{$^{95m}$Tc} & 0.0015(3) & Not Detected \\      
                                        & \textit{0.0013(1)} & \textit{4.7e-5(1)} \\
        \noalign{\smallskip}
        \multirow{2}*{$^{96}$Tc} & 0.0497(3) & 0.0021(1)\\       
                                         & \textit{0.0492(29)} & \textit{0.0023(1)}\\
        \noalign{\smallskip}
        \multirow{2}*{$^{97m}$Tc} & Not Detected & Not Detected \\       
                                        & \textit{0.00041(6)} & \textit{0.00030(3)} \\
        \noalign{\smallskip}
        \multirow{2}*{$^{99m}$Tc} & 7.69(5) & 66.1(8) \\       
                                        & \textit{7.71(12)} & \textit{66.4(7)} \\  
        \noalign{\smallskip}
        \multirow{2}*{$^{101}$Tc}  & 1.5(3) & 3.9(1)\\       
                                         & \textit{1.3(1)} & \textit{3.6(1)}\\                                 
        \noalign{\smallskip}
        \multirow{2}*{$^{99}$Mo}  & Not Detected & 0.43(1)\\       
                                         & \textit{0} & \textit{0.40(1)}\\                                        
    \hline
    \multicolumn{3}{c}{Purity [\%]}\\
    \hline
    \multirow{2}*{2h after EOB}  & 89.6(3) & 99.941(4) \\
                                 & \textit{89.9(2)} & \textit{99.908(2)}\\
    \hline
    \end{tabular}
    }
    \label{tab:prod_results}
\end{table}

\section{Conclusions and outlook}
In this work, a comprehensive experimental study of proton-induced nuclear reactions on molybdenum targets was carried out with the aim of optimizing the accelerator-based production of $^{99m}$Tc, the most widely used radionuclide in nuclear medicine. Cross sections were measured up to 18~MeV for the nuclear reactions contributing to the production of technetium radioisotopes and $^{99}$Mo, using natural Mo as well as enriched $^{97}$Mo and $^{100}$Mo targets. This data significantly extended the existing experimental database, which was found to be incomplete or inconsistent for several reaction channels, and highlighted the importance of high-quality experimental data. \\
The measured cross sections were subsequently used to evaluate thick target yields and radionuclidic purity under realistic irradiation conditions, allowing the identification of an optimized production route. A 99.73\% enriched $^{100}$Mo target with a thickness of approximately 0.50~mm irradiated with a 16-MeV proton beam was identified as an optimal configuration. For a 10-h irradiation, this configuration provides a 
$^{99m}$Tc TTY of approximately 1.7~GBq/$\mu$A while maintaining radionuclidic purity within the limits specified by the European Pharmacopoeia.\\
Two production tests were then performed using $^{100}$Mo targets with different enrichment levels. The irradiations were carried out under low-current conditions to confirm the reliability of the calculated yields and purity estimates. These results demonstrate the feasibility of producing high-purity $^{99m}$Tc using a medical cyclotron and provide a solid experimental basis for further studies aimed at scaling the production to clinically relevant beam currents.

\section{Acknowledgments}
We acknowledge the valuable support provided by the LHEP engineering and technical staff and by the Interfacial Electrochemistry group of DCBP. This research project was partially funded by the Swiss National Science Foundation (SNSF), Switzerland (grant: IZURZ2$\_$224901) and by the Tera-Care Foundation, Switzerland.

\clearpage
\section{Appendix}

\begin{table}[h]
    \centering
    \caption{Production cross-section data of $^{93}$Tc and $^{93m}$Tc measured from natural Mo and enriched $^{97}$Mo targets, whose isotopic composition is reported in Table~\ref{tab:Mo_mat}.}
    \resizebox{10cm}{!}{
    \begin{tabular}{c c c c c}
    \hline																			
    E			&	nat. Mo(p,x)$^{93}$Tc			&	enr. $^{97}$Mo(p,x)$^{93}$Tc			&	nat. Mo(p,x)$^{93m}$Tc			&	enr. $^{97}$Mo(p,x)$^{93m}$Tc			\\
    ~[MeV]			&	[mb]			&	[mb]			&	[mb]			&	[mb]			\\
    \hline																			
    5.5(9)	&	0.144(4)	&	-			&	0.110(2)	&	-			\\
    6.5(8)	&	0.317(8)	&	-			&	0.228(5)	&	-			\\
    7.9(7)	&	0.55(1)	&	-			&	0.398(7)	&	-			\\
    9.2(6)	&	0.65(1)	&	-			&	0.412(8)	&	-			\\
    10.4(5)	&	0.69(2)	&	-			&	0.279(8)	&	-			\\
    12.0(4)	&	0.48(2)	&	Not detected			&	0.21(1)	&	Not detected			\\
    12.8(4)	&	0.33(1)	&	Not detected			&	0.180(6)	&	Not detected			\\
    13.9(4)	&	0.39(1)	&	Not detected			&	0.122(7)	&	Not detected			\\
    14.9(4)	&	11.4(2)	&	0.52(1)	&	2.00(3)	&	0.090(3)	\\
    15.9(3)	&	27.0(4)	&	1.25(2)	&	5.65(9)	&	0.260(5)	\\
    17.0(3)	&	39.7(6)	&	1.84(3)	&	10.8(2)	&	0.502(9)	\\
    18.2(3)	&	44.9(8)	&	2.09(4)	&	12.7(2)	&	0.59(1)	\\
    \hline																																
    \end{tabular}
    }
    \label{appx:prod_cs_Tc93Tc93m}
\end{table}

\begin{table}[h]
    \centering
    \caption{Production cross-section data of $^{94}$Tc and $^{94m}$Tc measured from natural Mo and enriched $^{100}$Mo targets, whose isotopic composition is reported in Table~\ref{tab:Mo_mat}.}
    \resizebox{10cm}{!}{
    \begin{tabular}{c c c c c}
    \hline																			
    E			&	nat. Mo(p,x)$^{94}$Tc			&	enr. $^{100}$Mo(p,x)$^{94}$Tc			&	nat. Mo(p,x)$^{94m}$Tc			&	enr. $^{100}$Mo(p,x)$^{94m}$Tc			\\
    ~[MeV]			&	[mb]			&	[mb]			&	[mb]			&	[mb]			\\
    \hline																			
    5.5(9)	&	0.076(2)	&	-			&	2.25(5)	&	-			\\
    6.5(8)	&	0.65(1)	&	-			&	11.8(2)	&	-			\\
    7.9(7)	&	2.93(4)	&	-			&	26.6(4)	&	-			\\
    9.2(6)	&	6.6(1)	&	0.37(4)	&	38.3(7)	&	2.7(3)	\\
    10.4(5)	&	9.9(2)	&	-			&	45.6(9)	&	-			\\
    12.0(4)	&	14.3(3)	&	0.93(2)	&	47(1)	&	3.1(1)	\\
    12.8(4)	&	15.4(2)	&	1.01(2)	&	47.9(9)	&	3.14(6)	\\
    13.9(4)	&	28.7(4)	&	1.94(2)	&	58.3(8)	&	3.94(7)	\\
    14.9(4)	&	39.1(5)	&	2.70(4)	&	62.4(9)	&	4.28(8)	\\
    15.9(3)	&	46.4(7)	&	3.23(4)	&	63(1)	&	4.37(7)	\\
    17.0(3)	&	51.9(8)	&	3.65(4)	&	63(1)	&	4.40(7)	\\
    18.2(3)	&	53.4(9)	&	3.77(5)	&	63(1)	&	4.44(9)	\\
    \hline																			
						
    \end{tabular}
    }
    \label{appx:prod_cs_Tc94Tc94m}
\end{table}

\begin{table}[h]
    \centering
    \caption{Cross-section data of the $^{94}$Mo(p,n)$^{94,94m}$Tc and  $^{95}$Mo(p,2n)$^{94,94m}$Tc nuclear reactions.}
    \resizebox{10cm}{!}{
    \begin{tabular}{c c c c c}
\hline																			
E			&	$^{94}$Mo(p,n)$^{94}$Tc			&	$^{95}$Mo(p,2n)$^{94}$Tc			&	$^{94}$Mo(p,n)$^{94m}$Tc			&	$^{95}$Mo(p,2n)$^{94m}$Tc			\\
~[MeV]			&	[mb]			&	[mb]			&	[mb]			&	[mb]			\\
\hline																			
5.5(9)	&	0.92(3)	&	Not detected			&	27.0(6)	&	Not detected			\\
6.5(8)	&	7.74(14)	&	Not detected			&	142(3)	&	Not detected			\\
7.9(7)	&	35.06(49)	&	Not detected			&	319(5)	&	Not detected			\\
9.2(6)	&	79(1)	&	Not detected			&	459(8)	&	Not detected			\\
10.4(5)	&	118(2)	&	Not detected			&	547(11)	&	Not detected			\\
12.0(4)	&	171(3)	&	Not detected			&	567(13)	&	Not detected			\\
12.8(4)	&	160(4)	&	13.6(4)	&	484(6)	&	50(3)	\\
13.9(4)	&	194(1)	&	83(2)	&	386(10)	&	173(11)	\\
14.9(4)	&	166(2)	&	167(2)	&	309(19)	&	242(17)	\\
15.9(3)	&	152(22)	&	224(8)	&	215(6)	&	297(4)	\\
17.0(3)	&	109(31)	&	284(12)	&	160(18)	&	328(2)	\\
18.2(3)	&	86(28)	&	306(9)	&	118(1)	&	352(8)	\\
\hline																							\end{tabular}
    }
    \label{appx:nucl_cs_Tc94Tc94m}
\end{table}

\begin{table}[h]
    \centering
    \caption{Production cross-section data of $^{95}$Tc and $^{95m}$Tc measured from natural Mo and enriched $^{100}$Mo targets, whose isotopic composition is reported in Table~\ref{tab:Mo_mat}.}
    \resizebox{10cm}{!}{
    \begin{tabular}{c c c c c}
\hline																			
E			&	nat. Mo(p,x)$^{95}$Tc			&	enr. $^{100}$Mo(p,x)$^{95}$Tc			&	nat. Mo(p,x)$^{95m}$Tc			&	enr. $^{100}$Mo(p,x)$^{95m}$Tc			\\
~[MeV]			&	[mb]			&	[mb]			&	[mb]			&	[mb]			\\
\hline																			
5.5(9)	&	7.2(1)	&	-			&	4.1(1)	&	-			\\
6.5(8)	&	22.5(4)	&	-			&	14.2(3)	&	-			\\
7.9(7)	&	45.6(6)	&	-			&	22.2(4)	&	-			\\
9.2(6)	&	69(1)	&	3.7(2)	&	34.3(7)	&	Not detected			\\
10.4(5)	&	83(1)	&	-			&	37.1(8)	&	-			\\
12.0(4)	&	97(2)	&	6.9(1)	&	42(1)	&	3.01(8)	\\
12.8(4)	&	107(2)	&	7.8(1)	&	52(1)	&	3.76(8)	\\
13.9(4)	&	119(1)	&	8.8(1)	&	53(1)	&	3.94(6)	\\
14.9(4)	&	116(2)	&	8.6(1)	&	54(1)	&	4.01(8)	\\
15.9(3)	&	116(2)	&	8.7(1)	&	54(1)	&	4.05(7)	\\
17.0(3)	&	118(2)	&	8.9(1)	&	55(1)	&	4.10(7)	\\
18.2(3)	&	116(2)	&	8.7(1)	&	49(1)	&	3.69(9)	\\
\hline																			
			
    \end{tabular}
    }
    \label{appx:prod_cs_Tc95Tc95m}
\end{table}

\begin{table}[h]
    \centering
    \caption{Cross-section data of the $^{95}$Mo(p,n)$^{95,95m}$Tc and  $^{96}$Mo(p,2n)$^{95,95m}$Tc nuclear reactions.}
    \resizebox{10cm}{!}{
    \begin{tabular}{c c c c c}
\hline																			
E			&	$^{95}$Mo(p,n)$^{95}$Tc			&	$^{96}$Mo(p,2n)$^{95}$Tc			&	$^{95}$Mo(p,n)$^{95m}$Tc			&	$^{96}$Mo(p,2n)$^{95m}$Tc			\\
~[MeV]			&	[mb]			&	[mb]			&	[mb]			&	[mb]			\\
\hline																			
5.5(9)	&	47.4(8)	&	Not detected			&	27.1(6)	&	Not detected			\\
6.5(8)	&	149(2)	&	Not detected			&	94(2)	&	Not detected			\\
7.9(7)	&	302(4)	&	Not detected			&	147(3)	&	Not detected			\\
9.2(6)	&	458(7)	&	Not detected			&	227(5)	&	Not detected			\\
10.4(5)	&	550(9)	&	Not detected			&	246(5)	&	Not detected			\\
12.0(4)	&	580(9)	&	55(19)	&	244(14)	&	30(18)	\\
12.8(4)	&	488(23)	&	200(10)	&	213(5)	&	117(2)	\\
13.9(4)	&	389(1)	&	366(8)	&	134(4)	&	199(2)	\\
14.9(4)	&	174(4)	&	540(6)	&	57(16)	&	272(20)	\\
15.9(3)	&	120(43)	&	591(28)	&	43(20)	&	287(12)	\\
17.0(3)	&	92(42)	&	630(27)	&	29(17)	&	303(9)	\\
18.2(3)	&	66(30)	&	639(16)	&	21(15)	&	277(19)	\\
\hline																			
\end{tabular}
    }
    \label{appx:nucl_cs_Tc95Tc95m}
\end{table}

\begin{table}[h]
    \centering
    \caption{Production cross-section data of $^{96}$Tc, $^{96g}$Tc and $^{96m}$Tc measured from natural Mo and enriched $^{97}$Mo targets, whose isotopic composition is reported in Table~\ref{tab:Mo_mat}.}
    \resizebox{12cm}{!}{
    \begin{tabular}{c c c c c c c}
\hline																											
E			&	nat. Mo(p,x)$^{96}$Tc			&	enr. $^{97}$Mo(p,x)$^{96}$Tc			&	nat. Mo(p,x)$^{96g}$Tc			&	enr. $^{97}$Mo(p,x)$^{96g}$Tc			&	nat. Mo(p,x)$^{96m}$Tc			&	enr. $^{97}$Mo(p,x)$^{96m}$Tc			\\
~[MeV]			&	[mb]			&	[mb]			&	[mb]			&	[mb]			&	[mb]			&	[mb]			\\
\hline																											
5.5(9)	&	10.0(6)	&	-			&	0.4(2)	&	-			&	10.4(8)	&	-			\\
6.5(8)	&	30(2)	&	-			&	2.2(9)	&	-			&	30(2)	&	-			\\
7.9(7)	&	59(3)	&	-			&	5(1)	&	-			&	57(4)	&	-			\\
9.2(6)	&	86(5)	&	-			&	10(2)	&	-			&	80(5)	&	-			\\
10.4(5)	&	107(6)	&	-			&	16(3)	&	-			&	95(6)	&	-			\\
12.0(4)	&	146(9)	&	162(10)	&	29(3)	&	41(3)	&	123(8)	&	133(8)	\\
12.8(4)	&	139(8)	&	244(14)	&	34(3)	&	59(4)	&	110(7)	&	200(12)	\\
13.9(4)	&	120(7)	&	509(31)	&	37(4)	&	157(10)	&	87(7)	&	385(24)	\\
14.9(4)	&	100(7)	&	577(36)	&	35(4)	&	201(13)	&	69(5)	&	412(26)	\\
15.9(3)	&	95(7)	&	603(41)	&	35(4)	&	213(15)	&	63(6)	&	411(28)	\\
17.0(3)	&	93(6)	&	659(39)	&	35(4)	&	253(15)	&	61(5)	&	439(26)	\\
18.2(3)	&	85(6)	&	667(40)	&	36(4)	&	277(17)	&	53(5)	&	421(26)	\\
\hline																								
    \end{tabular}
    }
    \label{appx:prod_cs_Tc96Tc96gTc96m}
\end{table}

\begin{table}[h]
    \centering
    \caption{Cross-section data of the $^{96}$Mo(p,n)$^{96,96g,96m}$Tc and  $^{97}$Mo(p,2n)$^{96,96g,96m}$Tc nuclear reactions.}
    \resizebox{12cm}{!}{
    \begin{tabular}{c c c c c c c}
\hline																											
E			&	$^{96}$Mo(p,n)$^{96}$Tc			&	$^{97}$Mo(p,2n)$^{96}$Tc			&	$^{96}$Mo(p,n)$^{96g}$Tc			&	$^{97}$Mo(p,2n)$^{96g}$Tc			&	$^{96}$Mo(p,n)$^{96m}$Tc			&	$^{97}$Mo(p,2n)$^{96m}$Tc			\\
~[MeV]			&	[mb]			&	[mb]			&	[mb]			&	[mb]			&	[mb]			&	[mb]			\\
\hline																											
5.5(9)	&	60(4)	&	Not detected			&	2.6(9)	&	Not detected			&	63(5)	&	Not detected			\\
6.5(8)	&	178(11)	&	Not detected			&	13(5)	&	Not detected			&	178(12)	&	Not detected			\\
7.9(7)	&	353(21)	&	Not detected			&	28(7)	&	Not detected			&	342(22)	&	Not detected			\\
9.2(6)	&	518(31)	&	Not detected			&	59(11)	&	Not detected			&	484(31)	&	Not detected			\\
10.4(5)	&	644(37)	&	Not detected			&	97(17)	&	Not detected			&	574(38)	&	Not detected			\\
12.0(4)	&	771(47)	&	180(11)	&	146(19)	&	46(3)	&	652(44)	&	149(9)	\\
12.8(4)	&	672(39)	&	272(16)	&	164(16)	&	66(4)	&	528(34)	&	223(13)	\\
13.9(4)	&	379(22)	&	567(34)	&	119(20)	&	175(11)	&	267(24)	&	429(26)	\\
14.9(4)	&	216(15)	&	643(40)	&	73(13)	&	225(14)	&	138(15)	&	460(29)	\\
15.9(3)	&	166(13)	&	673(45)	&	67(14)	&	238(16)	&	105(14)	&	458(32)	\\
17.0(3)	&	119(10)	&	735(43)	&	41(11)	&	282(17)	&	74(12)	&	490(29)	\\
18.2(3)	&	66(10)	&	744(44)	&	27(11)	&	309(19)	&	36(10)	&	470(29)	\\
\hline																																									
    \end{tabular}
    }
    \label{appx:nucl_cs_Tc96Tc96gTc96m}
\end{table}

\begin{table}[h]
    \centering
    \caption{Production cross-section data of $^{97m}$Tc measured from natural Mo and enriched $^{97}$Mo targets, whose isotopic composition is reported in Table~\ref{tab:Mo_mat}.}    
    \resizebox{7cm}{!}{
    \begin{tabular}{c c c}
\hline											
E			&	nat. Mo(p,x)$^{97m}$Tc			&	enr. $^{97}$Mo(p,x)$^{97m}$Tc			\\
~[MeV]			&	[mb]			&	[mb]			\\
\hline											
5.5(9)	&	-			&	18(3)	\\
7.9(7)	&	-			&	82(11)	\\
10.4(5)	&	18(3)	&	129(12)	\\
12.0(4)	&	54(6)	&	95(6)	\\
13.9(4)	&	63(6)	&	46(4)	\\
15.9(3)	&	59(4)	&	25(4)	\\
18.2(3)	&	53(3)	&	20(2)	\\
\hline											
    \end{tabular}
    }
    \label{appx:prod_cs_Tc97m}
\end{table}

\begin{table}[h]
    \centering
    \caption{Cross-section data of the $^{97}$Mo(p,n)$^{97m}$Tc and  $^{98}$Mo(p,2n)$^{97m}$Tc nuclear reactions.}
    \resizebox{7cm}{!}{
    \begin{tabular}{c c c}
\hline											
E			&	$^{97}$Mo(p,n)$^{97m}$Tc			&	$^{98}$Mo(p,2n)$^{97m}$Tc			\\
~[MeV]			&	[mb]			&	[mb]			\\
\hline											
5.5(9)	&	20(4)	&	Not detected			\\
7.9(7)	&	92(12)	&	Not detected			\\
10.4(5)	&	143(13)	&	15(7)	\\
12.0(4)	&	96(5)	&	168(20)	\\
13.9(4)	&	38(3)	&	223(21)	\\
15.9(3)	&	15(4)	&	218(15)	\\
18.2(3)	&	10(2)	&	196(9)	\\
\hline											
\end{tabular}
    }

    \label{appx:nucl_cs_Tc97m}
\end{table}

\begin{table}[h]
    \centering
\caption{$^{100}$Mo(p,2n)$^{99m}$Tc reaction cross-section data measured using natural Mo and enriched $^{100}$Mo targets, whose isotopic compositions are reported in Table~\ref{tab:Mo_mat}, together with the final values obtained as the weighted average of the two measurements.}
    \resizebox{10cm}{!}{
    \begin{tabular}{c c c c}
\hline															
E			&	$^{100}$Mo(p,2n)$^{99m}$Tc (nat. Mo)			&	$^{100}$Mo(p,2n)$^{99m}$Tc (enr.  100Mo)			&	$^{100}$Mo(p,2n)$^{99m}$Tc			\\
~[MeV]			&	[mb]			&	[mb]			&	[mb]			\\
\hline															
5.5(9)	&	Not detected			&	-			&	Not detected			\\
6.5(8)	&	Not detected			&	-			&	Not detected			\\
7.9(7)	&	3.79(6)	&	-			&	3.79(6)	\\
9.2(6)	&	91(1)	&	93(5)	&	91(1)	\\
10.4(5)	&	180(3)	&	-			&	180(3)	\\
12.0(4)	&	254(5)	&	230(5)	&	244(3)	\\
12.8(4)	&	244(4)	&	255(4)	&	250(3)	\\
13.9(4)	&	252(3)	&	260(3)	&	256(2)	\\
14.9(4)	&	246(3)	&	259(3)	&	252(2)	\\
15.9(3)	&	250(4)	&	258(3)	&	255(2)	\\
17.0(3)	&	257(4)	&	252(3)	&	253(2)	\\
18.2(3)	&	229(4)	&	243(3)	&	237(2)	\\
\hline															
    \end{tabular}
    }
    \label{appx:prod_cs_Tc99m}
\end{table}

\begin{table}[h]
    \centering
    \caption{Cross-section data of the $^{100}$Mo(p,$\gamma$)$^{101}$Tc nuclear reaction.}
    \resizebox{4.5cm}{!}{
    \begin{tabular}{c c}
\hline											
E			&	$^{100}$Mo(p,$\gamma$)$^{101}$Tc			\\				
~[MeV]			&	[mb]			\\				
\hline											
5.5(9)	&	-			\\				
6.5(8)	&	-			\\				
7.9(7)	&	-			\\				
9.2(6)	&	Not detected			\\				
10.4(5)	&	-			\\				
12.0(4)	&	0.54(3)	\\				
12.8(4)	&	0.62(1)	\\				
13.9(4)	&	0.64(1)	\\				
14.9(4)	&	0.59(1)	\\				
15.9(3)	&	0.53(1)	\\				
17.0(3)	&	0.50(1)	\\				
18.2(3)	&	0.41(1)	\\				
\hline											

    \end{tabular}
    }

    \label{appx:prod_cs_Tc101}
\end{table}

\begin{table}[h]
    \centering
\caption{$^{100}$Mo(p,pn)$^{99}$Mo reaction cross-section data measured using natural Mo and enriched $^{100}$Mo targets, whose isotopic compositions are reported in Table~\ref{tab:Mo_mat}, together with the final values obtained as the weighted average of the two measurements.}
    \resizebox{10cm}{!}{
    \begin{tabular}{c c c c}
\hline															
E			&	$^{100}$Mo(p,pn)$^{99}$Mo (nat. Mo)			&	$^{100}$Mo(p,pn)$^{99}$Mo (enr.  100Mo)			&	$^{100}$Mo(p,pn)$^{99}$Mo			\\
~[MeV]			&	[mb]			&	[mb]			&	[mb]			\\
\hline															
5.5(9)	&	Not detected			&	-			&	Not detected			\\
6.5(8)	&	Not detected			&	-			&	Not detected			\\
7.9(7)	&	Not detected			&	-			&	Not detected			\\
9.2(6)	&	Not detected			&	Not detected			&	Not detected			\\
10.4(5)	&	Not detected			&	-			&	Not detected			\\
12.0(4)	&	Not detected			&	2.8(1)	&	2.8(1)	\\
12.8(4)	&	9.1(7)	&	8.4(1)	&	8.4(1)	\\
13.9(4)	&	13.8(4)	&	14.6(2)	&	14.5(1)	\\
14.9(4)	&	21.6(6)	&	23.9(3)	&	23.4(3)	\\
15.9(3)	&	33.7(7)	&	36.0(4)	&	35.4(3)	\\
17.0(3)	&	50(1)	&	47.9(5)	&	48.5(5)	\\
18.2(3)	&	83(3)	&	83(1)	&	83(1)	\\
\hline															

    \end{tabular}
    }
    \label{appx:prod_cs_Mo99}
\end{table}

\clearpage
\bibliographystyle{elsarticle-num}
\bibliography{mybibfile}
\end{document}